\documentclass[journal]{IEEEtran}

\usepackage{amsmath,amstext,amsfonts,amssymb,eucal,graphicx}
\usepackage{subfigure}
\usepackage{epsfig}

\newtheorem{lemma}{Lemma}
\newtheorem{corollary}{Corollary}
\newtheorem{proposition}{Proposition}
\newtheorem{theorem}{Theorem}
\newtheorem{definition}{Definition}

\newcommand{\sluttlinje}{\hspace*{1cm}\hspace*{-1cm}~\hfill}
\newcommand{\sluttmerke}{\sluttlinje\raisebox{-1mm}{\rule{2.5mm}{2.5mm}}}

\newcommand{\Z}{\mathbb{Z}}

\newcommand{\E}{\mathbb{E}}
\newcommand{\R}{\mathbb{R}}

\begin{document}
\bibliographystyle{IEEEtran}
\title{Asymptotic Behaviour of Random Vandermonde Matrices with Entries on the Unit Circle}
\author{
  \IEEEauthorblockN{\O yvind Ryan,~\IEEEmembership{Member,~IEEE} and M{\'e}rouane~Debbah,~\IEEEmembership{Senior Member,~IEEE}\\}
  \thanks{This project is partially sponsored by the project BIONET (INRIA)}
  \thanks{This work was supported by Alcatel-Lucent within the Alcatel-Lucent Chair on flexible radio at SUPELEC as well as the ANR project SESAME}
  \thanks{This paper was presented in part at the 1st Workshop on Physics-Inspired Paradigms in Wireless Communications and Networks, 2008, Berlin, Germany}
  \thanks{\O yvind~Ryan is with the Centre of Mathematics for Applications, University of Oslo, P.O. Box 1053 Blindern, NO-0316 Oslo, NORWAY, oyvindry@ifi.uio.no}
  \thanks{M{\'e}rouane~Debbah is with SUPELEC, Gif-sur-Yvette, France, merouane.debbah@supelec.fr}
}

\markboth{IEEE Transactions on Information Theory,~Vol.~1,
No.~1,~January~2009}{Shell \MakeLowercase{\textit{et al.}}: Bare
Demo of IEEEtran.cls for Journals}

\maketitle
\begin{abstract}
Analytical methods for finding moments of random
Vandermonde matrices with entries on the unit circle are developed.
Vandermonde Matrices play an
important role in signal processing and wireless applications
such as direction of arrival estimation, precoding, and sparse
sampling theory, just to name a few. Within this framework, we extend
classical freeness results on random matrices with 
independent, identically distributed (i.i.d.) entries and
show that Vandermonde structured matrices can be treated in the same
vein with different tools. We focus on various types of matrices, 
such as Vandermonde matrices with and without uniform
phase distributions, as well as generalized Vandermonde matrices. 
In each case, we provide
explicit expressions of the moments of the associated Gram matrix,
as well as more advanced models involving the Vandermonde matrix.
Comparisons with classical i.i.d. random matrix theory are provided,
and deconvolution results are discussed.
We review some applications of the results to the fields of signal processing and wireless
communications.
\end{abstract}


\section{Introduction}
Vandermonde matrices have for a long time had a central position in
signal processing due to their connections with important
tools in the field such as the FFT \cite{book.burrus85} or
Hadamard \cite{book.golub83} transforms, to name a few. 
Vandermonde matrices occur frequently in many applications, such as
finance~\cite{norberg}, signal array
processing~\cite{paper:schmidt2,paper:waxkailath,book:jonshon,paper:roykailath,paper:poratfriedlander},
ARMA processes~\cite{paper:kleinspreij}, 
cognitive radio~\cite{paper:sampaiokobayashi2}, security~\cite{paper:sampaiokobayashi1}, wireless
communications~\cite{paper:wangscaglione}, and
biology~\cite{paper:fergal}, and have been much studied.
The applied research  has been somewhat tempered by the fact that
very few theoretical results have been available.

A Vandermonde matrix with entries on the unit circle has the following form:
\begin{equation} \label{vandermonde}
  {\bf V} = \frac{1}{\sqrt{N}}
               \left( \begin{array}{lll}
                        1                    & \cdots & 1 \\
                        e^{-j \omega_1}      & \cdots & e^{-j \omega_L} \\
                        \vdots               & \ddots & \vdots \\
                        e^{-j (N-1)\omega_1} & \cdots & e^{-j (N-1)\omega_L}
                      \end{array}
               \right)
\end{equation}
We will consider the case  where $\omega_1$,...,$\omega_L$ are i.i.d., taking values in $[ 0,2\pi )$.
Throughout the paper, the $\omega_i$ will be called {\em phase distributions}.
${\bf V}$ will be used only to denote Vandermonde matrices with a given phase distribution,
and the dimensions of the Vandermonde matrices will always be $N\times L$.

Known results on Vandermonde matrices are related to the distribution of the determinant~\cite{book:girkodet}. 
The large majority of known results on the eigenvalues of the associated Gram matrix
concern Gaussian matrices~\cite{book:mehta} or matrices with
independent entries. Very few results are available in the literature on
matrices whose structure is strongly related to the Vandermonde case~\cite{paper:nordio1,paper:bordenave}.
Known results depend heavily on the distribution
of the entries, and do not give any hint on the asymptotic behaviour
as the matrices become large. In the realm of wireless channel
modeling,~\cite{mullermodelcomm} has provided some insight on the
behaviour of the eigenvalues of random Vandermonde matrices for a
specific case, without any formal proof. 

In many applications,  $N$ and $L$ are quite large, and we may be
interested in studying the case where both go to $\infty$ at a given
ratio, $\frac{L}{N}\rightarrow c$. Results in the literature
say very little on the asymptotic behaviour of (\ref{vandermonde})
under this growth condition. 
The results, however, are well known
for other models. The factor $\frac{1}{\sqrt{N}}$, as well as the
assumption that the Vandermonde entries $e^{-j\omega_i}$ lie on the
unit circle, are included in (\ref{vandermonde}) to ensure that the
analysis will give limiting asymptotic behaviour. Without this
assumption, the problem at hand is more involved, since the rows of
the Vandermonde matrix with the highest powers would dominate in the
calculations of the moments for large matrices, and also
grow faster to infinity than the $\frac{1}{\sqrt{N}}$ factor in
(\ref{vandermonde}), making asymptotic analysis difficult. In
general, often the moments, not the moments of the determinants, are
the quantities we seek. Results in the literature  say also very
little on the moments of Vandermonde matrices (however, see~\cite{paper:nordio1}), 
and also on the mixed moments of Vandermonde matrices and
matrices independent from them. This is in contrast to Gaussian
matrices, where exact expressions~\cite{paper:thorbjornsen1} and
their asymptotic behaviour~\cite{book:hiaipetz} are known through the
concept of freeness~\cite{book:hiaipetz}, which is central for
describing the mixed moments.

The framework and results presented in this paper are reminiscent
of similar results concerning i.i.d. random matrices
\cite{paper.marchenkor67} which have shed light on the design of
many important wireless communication problems such as CDMA
\cite{paper:shamai01}, MIMO \cite{paper:telatar99}, or OFDM
\cite{paper.debbah03}. This contribution aims to do the same.
We will show that, asymptotically, the moments of the Vandermonde matrices
depend only on the ratio $c$ and the phase distribution, and have explicit expressions.
The expressions are more involved than what was claimed in~\cite{mullermodelcomm}. 
Moments are useful for performing deconvolution. 
Deconvolution for our purposes will mean retrieving
"moments" $tr_L ( ({\bf D}(N))^i ),...,tr_L ( ({\bf D}(N))^i )$ (where ${\bf D}(N)$ are unknown matrices),
from "mixed moments" of ${\bf D}(N)$ and matrices on the form (\ref{vandermonde}), with $L$ and $N$ large. 
We are only able to perform such deconvolution when the ${\bf D}(N)$ are square diagonal matrices independent from ${\bf V}$.
We will see that such deconvolution can be very useful in many applications,
since the retrieved moments can give useful information about the system under study.
Deconvolution has previously been handled in cases where ${\bf V}$ is replaced with
a Gaussian matrix~\cite{paper:anderson,paper:loubatonmoulines,paper:thorbjornsen1,eurecom:freedeconvinftheory}.
As will be seen, the way the phase distribution influences these moments can be split into several cases.
Uniform phase distribution plays a central role in that it minimizes the moments.
When the phase distribution has a "nice" density (for our purposes this means that the density of the phase distribution is continuous), 
a nice connection with the moments for uniform phase distribution can be given.
When the density of the phase distribution has singularities, for instance when it has point masses,
it turns out that the asymptotics of the moments change drastically.

We will also extend our results to generalized
Vandermonde matrices, i.e. matrices where the columns do not consist
of uniformly distributed powers. Such matrices are important for
applications to finance~\cite{norberg}. 
The tools used for standard
Vandermonde matrices in this paper will allow us to find the
asymptotic behaviour of many generalized Vandermonde matrices as well.

While we provide the full computation of lower order moments, we
also describe how the higher order moments can be computed. Tedious
evaluation of many integrals is needed for this. It turns out that the
first three limit moments coincide with those of the
Mar\u{c}henko Pastur law~\cite{book:hiaipetz,book:tulinoverdu}. For
higher order moments this is not the case, although we state an
interesting inequality involving the Vandermonde limit moments and
the moments of the classical Poisson distribution and the
Mar\u{c}henko Pastur law. 

The paper is organized as follows. 
Section~\ref{section:essentials} provides background essentials on 
random matrix theory needed to state the main results. 
Section~\ref{section:theorems} states the main results of the paper.
It starts with a general result for the
mixed moments of Vandermonde matrices and matrices independent from
them. Results for the uniform phase distribution are
stated next, both for the asymptotic moments, and the lower order moments.
After this, the nice connection between uniform phase distribution and other phase distributions is stated. 
The case where the density of $\omega$ has singularities is then handled.
The section ends with results on generalized
Vandermonde matrices, and mixed moments of (more than one) independent Vandermonde matrices.
Section~\ref{section:deconvolutionperspective} discusses our results
and puts them in a general deconvolution perspective, comparing with
other deconvolution results, such as those for Gaussian
deconvolution.
Section~\ref{simulations} presents some simulations and useful
applications showing the implications of the presented results in various
applied fields, and discusses the validity of the asymptotic claims in the finite regime.
First we apply the presented Vandermonde deconvolution framework to wireless systems,
where we estimate the number of paths, the transmissions powers of the users,
the number of sources, and what is commonly referred to as wavelength.
Finally we apply the results on Vandermonde matrices to the
very active field of sparse signal reconstruction.

\section{Random matrix background essentials} \label{section:essentials}
In the following, upper (lower boldface) symbols will be used for
matrices (column vectors), whereas lower symbols will represent
scalar values, $(.)^T$ will denote transpose operator, $(.)^\star$
conjugation, and $(.)^H=\left((.)^T\right)^\star$ hermitian
transpose. ${\bf I}_L$ will represent the $L\times L$ identity matrix.
We let $Tr$ be the (non-normalized) trace for square matrices, defined by,
\[
  Tr({\bf A}) = \sum_{i=1}^L a_{ii},
\]
where $a_{ii}$ are the diagonal elements of the $L\times L$ matrix ${\bf A}$.
We also let $tr_L$ be the normalized trace, defined by $tr_L({\bf A}) = \frac{1}{L}Tr({\bf A})$.

Results in random matrix theory often refer to the empirical eigenvalue distribution of matrices:
\begin{definition}
With the empirical eigenvalue distribution of an $L\times L$ hermitian random matrix ${\bf T}$ we mean the (random) function
\begin{equation} \label{edfdef}
  F^L_{ {\bf T} }(\lambda) = \frac{\#\{ i | \lambda_i \leq \lambda \}}{L} ,
\end{equation}
where $\lambda_i$ are the (random) eigenvalues of ${\bf T}$.
\end{definition}

In the following, ${\bf D}_r(N), 1\leq r\leq n$ will denote non-random diagonal $L\times L$ matrices,
where we implicitly assume that $\frac{L}{N}\rightarrow c$.
We will assume that the ${\bf D}_r(N)$ have a joint limit
distribution as $N\rightarrow\infty$ in the following sense:
\begin{definition} \label{ddef}
  We will say that the $\{ {\bf D}_r(N) \}_{1\leq r\leq n}$ have a joint limit distribution as $N\rightarrow\infty$ if the limit
  \begin{equation} \label{alphadef}
    D_{i_1,...,i_s} = \lim_{N\rightarrow\infty} tr_L\left( {\bf D}_{i_1}(N)\cdots {\bf D}_{i_s}(N)\right)
  \end{equation}
  exists for all choices of $i_1,...,i_s\in \{ 1,..,n\}$.
\end{definition}

The matrices ${\bf D}_i(N)$ are assumed to be non-random throughout the paper. 
However, all presented formulas extend naturally to the case when ${\bf D}_i(N)$ are random and independent from the Vandermonde matrices.
The difference when the ${\bf D}_i(N)$ are random is that expectations of products of traces also come into play, 
in the sense that, similarly to expressions on the form (\ref{alphadef}), expressions of the form
\begin{eqnarray*}
  \lim_{N\rightarrow\infty} & E [ & tr_L\left( {\bf D}_{i_1}(N)\cdots {\bf D}_{i_s}(N)\right) \cdots \nonumber \\
                            &     & tr_L\left( {\bf D}_{j_1}(N)\cdots {\bf D}_{j_r}(N)\right) ]      \label{alphadef2}
\end{eqnarray*}
also enter the picture. 
Our framework can also be extended naturally to compute the covariance of traces, defined in the following way:
\begin{definition}
By the covariance $C_{i,j}$ of two traces $tr_L({\bf A}^i)$ and $tr_L({\bf A}^j)$ of an $L\times L$ random matrix ${\bf A}$, we mean the quantity
\begin{equation} \label{covtwotraces}
\begin{array}{lll}
  C_{i,j}({\bf A}) &=& E\left[ tr_L \left( {\bf A}^i \right) tr_L \left( {\bf A}^j \right) \right] \\
                   & & - E\left[ tr_L \left( {\bf A}^i \right) \right] E\left[ tr_L \left( {\bf A}^j \right) \right].
\end{array}
\end{equation}
When ${\bf A}$ is replaced with an ensemble of matrices, ${\bf A}_L$, 
the limits $\lim_{L\rightarrow\infty} LC_{i,j}({\bf A}_L)$ are also called second order moments. 
\end{definition}

The normalizing factor $L$ is included in order to obtain a limit. 
It will be explained later why this is the correct normalizing factor for the matrices we consider. 
The term second order moment is taken from~\cite{secondorderfreeness3}, 
where different matrix ensembles were considered. 
For these matrices, the second order moments were instead defined as $\lim_{L\rightarrow\infty} L^2C_{i,j}({\bf A}_L)$
(i.e. a higher order normalizing factor was used),
since these matrices displayed faster convergence to a limit.
We will present expressions for the second order moments $\lim_{N\rightarrow\infty} L C_{i,j}({\bf D}(N) {\bf V}^H {\bf V})$. 

Most theorems in this paper will present expressions for various mixed moments, defined in the following way:
\begin{definition}
By a mixed moment we mean the limit
\begin{equation} \label{computethis}
\begin{array}{ll}
  M_n = \lim_{N\rightarrow\infty} E[ tr_L ( & {\bf D}_1(N) {\bf V}^H {\bf V} {\bf D}_2(N) {\bf V}^H {\bf V} \\
                                                      & \cdots \times {\bf D}_n(N) {\bf V}^H {\bf V} ) ],
\end{array}
\end{equation}
whenever this exists.
\end{definition}

A joint limit distribution of $\{ {\bf D}_r(N) \}_{1\leq r\leq n}$ is always assumed in the presented results on mixed moments.
Note that when ${\bf D}_1(N) = \cdots = {\bf D}_n(N) = {\bf I}_L$,
the $M_n$ compute to the asymptotic moments of the Vandermonde matrices themselves,
defined by
\begin{eqnarray*}
  V_n &=& \lim_{N\rightarrow\infty} E \left[ tr_L \left( \left( {\bf V}^H {\bf V} \right)^n \right) \right] \\
      &=& \lim_{N\rightarrow\infty} E \left[\int \lambda^n dF^L_{ {\bf V}^H {\bf V} }(\lambda) \right].
\end{eqnarray*}
Similarly, when ${\bf D}_1(N) = \cdots = {\bf D}_n(N) = {\bf D}(N)$, we will also write
\begin{equation} \label{dmomdef}
  D_n = \lim_{N\rightarrow\infty} tr_L ({\bf D}(N)^n).
\end{equation}
Note that this is in conflict with the notation $D_{i_1,...,i_s}$, but the name of the index will resolve such conflicts.

To prove the results of this paper, the random matrix concepts presented up to now need to be extended using concepts from partition theory. 
We denote by ${\cal P}(n)$ the set of all partitions of $\{ 1,...,n\}$, 
and use $\rho$ as notation for a partition in ${\cal P}(n)$.
Also, we will write $\rho = \{ W_1 ,..., W_k\}$, 
where $W_j$ will be used repeatedly to denote the blocks of $\rho$, 
$|\rho |=k$ will denote the number of blocks in $\rho$, 
and $|W_j|$ will denote the number of elements in a given block. 
Definition~\ref{ddef} can now be extended as follows. 
\begin{definition} \label{ddef2}
  For $\rho = \{ W_1,...,W_k \}$, with $W_i = \{ w_{i1},...,w_{i|W_i|} \}$,
  we define
  \begin{eqnarray}
    D_{W_i}  &=& D_{i_{w_{i1}},...,i_{w_{i|W_i|}}} \label{dblockdef} \\
    D_{\rho} &=& \prod_{i=1}^k D_{W_i}. \label{dpartdef}
  \end{eqnarray}
\end{definition}

To better understand the presented expressions for mixed moments, the notion of free cumulants will be helpful. 
They are defined in terms of noncrossing partitions~\cite{book:comblect}.
\begin{definition}
  A partition $\rho$ is called noncrossing if, whenever we have $i<j<k<l$ with $i\sim k$, $j\sim l$ ($\sim$ meaning belonging to the same block),
  we also have $i\sim j\sim k\sim l$ (i.e. $i,j,k,l$ are all in the same block).
  The set of noncrossing partitions of $\{ 1,,,.,n\}$ is denoted $NC(n)$.
\end{definition}

The noncrossing partitions have already shown their usefulness
in expressing what is called the freeness relation in a particularly nice way~\cite{book:comblect}.

\begin{definition} \label{cumulantdef}
  Assume that ${\bf A}_1,...,{\bf A}_n$ are $L\times L$-random matrices.
  By the free cumulants of ${\bf A}_1,...,{\bf A}_n$ we mean the unique set of multilinear functionals $\kappa_{r}$ ($r\geq 1$) which satisfy
  \begin{equation} \label{momcumformula}
    E \left[ tr_L \left( {\bf A}_{i_1} \cdots {\bf A}_{i_n} \right) \right]
    =
    \sum_{\rho\in NC(n)} \kappa_{\rho}[ {\bf A}_{i_1},...,{\bf A}_{i_n}]
  \end{equation}
  for all choices of $i_1,...,i_n$, where
  \begin{eqnarray*}
    \kappa_{\rho}[ {\bf A}_{i_1},...,{\bf A}_{i_n}] &=& \prod_{j=1}^ k\kappa_{W_j}[ {\bf A}_{i_1},...,{\bf A}_{i_n}] \\
    \kappa_{W_i}[ {\bf A}_{i_1},...,{\bf A}_{i_n}]  &=& \kappa_{|W_i|}[{\bf A}_{i_{w_{i1}}},...,{\bf A}_{i_{w_{i|W_i|}}}],
  \end{eqnarray*}
  where $\rho = \{ W_1,...,W_k \}$, with $W_i = \{ w_{i1},...,w_{i|W_i|} \}$.
  By the classical cumulants of ${\bf A}_1,...,{\bf A}_n$ we mean the unique set of multilinear functionals
  which satisfy (\ref{momcumformula}) with $NC(n)$ replaced by the set of all partitions ${\cal P}(n)$.
\end{definition}

We have restricted our definition of cumulants to random matrices, although their general definition 
is in terms of more general probability spaces (Lecture 11 of~\cite{book:comblect}).
(\ref{momcumformula}) is also called the (free or classical) moment-cumulant formula.
The importance of the free moment-cumulant formula comes from the fact that,
had we replaced Vandermonde matrices with Gaussian matrices, it could help us perform deconvolution.
For this, the cumulants of the Gaussian matrices are needed, which asymptotically have a very nice form.
For Vandermonde matrices, it is not known what a useful definition of cumulants would be.
However, from the calculations in Appendix~\ref{appendixteo0},
it will turn out that the following quantities are helpful.
\begin{definition} \label{expansiondef}
  For $\rho\in {\cal P}(n)$, define
  \begin{equation} \label{kpindef}
  \begin{array}{ll}
    K_{\rho , \omega , N} =& \frac{1}{N^{n+1-|\rho |}} \times \\
                           & \int_{(0,2\pi)^{|\rho |}} \prod_{k=1}^n \frac{1-e^{j N (\omega_{b(k-1)} -\omega_{b(k)})}}{1-e^{j (\omega_{b(k-1)} -\omega_{b(k)})}} \\
                           & d\omega_1\cdots d\omega_{|\rho |},
  \end{array}
  \end{equation}
  where $\omega_{W_1},...,\omega_{W_{|\rho |}}$ are i.i.d. (indexed by the blocks of $\rho$), all with the same distribution as $\omega$,
  and where $b(k)$ is the block of $\rho$ which contains $k$ (notation is cyclic, i.e. $b(0) = b(n)$).
  If the limit
  \[
      K_{\rho , \omega} = \lim_{N\rightarrow\infty} K_{\rho , \omega , N}
  \]
  exists, then it is called a {\em Vandermonde mixed moment expansion coefficient}.
\end{definition}

These quantities do not behave exactly as cumulants, but rather as weights which tell us how a partition in the moment formula we present should be weighted.
In this respect our formulas for the moments are different from classical or free moment-cumulant formulas, since these do not perform this weighting.
The limits $K_{\rho , \omega}$ may not always exist, and necessary and sufficient conditions for their existence seem to be hard to find.
However, it is easy to prove from their definition that they do not exist if the density of $\omega$ has singularities
(for instance when the density has point masses).
On the other hand, Theorem~\ref{teo:generaldist} will show that they exist when the same density is continuous.

${\cal P}(n)$ is equipped with the refinement order $\leq$~\cite{book:comblect}, 
i.e. $\rho_1\leq\rho_2$ if and only if any block of $\rho_1$ is
contained within a block of $\rho_2$. 
The partition with $n$ blocks, denoted $0_n$, is the smallest partition within this order,
while the partition with $1$ block, denoted $1_n$, is the largest partition within this order. 
In the following sections, we will encounter the complementation map of Kreweras (p. 147 of~\cite{book:comblect}),
which is an order-reversing isomorphism of $NC(n)$ onto itself.
To define this we need the circular representation of  a partition:
We mark $n$ equidistant points $1,...,n$ (numbered clockwise) on the circle,
and form the convex hull of points lying in the same block of the partition.
This gives us a number of convex sets $H_i$, equally many as there are blocks in the partition, which do not intersect if and only if the partition is noncrossing.
Put names $\bar{1},...,\bar{n}$ on the midpoints of the $1,...,n$ (so that $\bar{i}$ is the midpoint of the segment from $i$ to $i+1$).
The complement of the set $\cup_i H_i$ is again a union of disjoint convex sets $\tilde{H}_i$.

\begin{definition}
The Kreweras complement of $\rho$, denoted $K(\rho)$, is defined as the partition on $\{ \bar{1},...,\bar{n}\}$ determined by
\[
  i\sim j \mbox{ in } K(\rho) \iff \bar{i},\bar{j} \mbox{ belong to the same convex set } ~\tilde{H}_k.
\]
\end{definition}

An important property of the Kreweras complement is that (p. 148 of~\cite{book:comblect})
\begin{equation} \label{krewerasprop}
  |\rho | + |K(\rho)| = n+1.
\end{equation}

\section{Statement of main results} \label{section:theorems}
We first state the main result of the paper, which applies to Vandermonde matrices with any phase distribution.
It restricts to the case when the expansion coefficients $K_{\rho , \omega}$ exist.
Different versions of it adapted to different Vandermonde matrices will be stated in succeeding sections.

\begin{theorem} \label{teo0}
  Assume that the $\{ {\bf D}_r(N) \}_{1\leq r\leq n}$ have a joint limit distribution as $N\rightarrow\infty$.
  Assume also that all Vandermonde mixed moment expansion coefficients $K_{\rho , \omega}$ exist.
  Then the limit
  \begin{equation} \label{cumequation0}
  \begin{array}{ll}
    M_n = \lim_{N\rightarrow\infty} E[ tr_L ( & {\bf D}_1(N) {\bf V}^H {\bf V} {\bf D}_2(N) {\bf V}^H {\bf V} \\
                                              & \cdots \times {\bf D}_n(N) {\bf V}^H {\bf V} ) ]
  \end{array}
  \end{equation}
  also exists when $\frac{L}{N}\rightarrow c$, and equals
  \begin{equation} \label{cumequation}
    \sum_{\rho\in{\cal P}(n)} K_{\rho , \omega} c^{|\rho |-1} D_{\rho}.
  \end{equation}
\end{theorem}

The proof of Theorem~\ref{teo0} can be found in Appendix~\ref{appendixteo0}.
Theorem~\ref{teo0} explains how "convolution" with Vandermonde matrices can be performed, and also provides us with an extension
of the concept of free convolution to Vandermonde matrices.
It also gives us means for performing deconvolution.
Indeed, suppose ${\bf D}_1(N) = \cdots = {\bf D}_n(N) = {\bf D}(N)$, and that one knows all the moments $M_n$.
One can then infer on the moments $D_n$ by inspecting (\ref{cumequation}) for increasing values of $n$.
For instance, the first two equations can also be written
\begin{eqnarray*}
  D_{1_1} &=& \frac{M_1}{K_{1_1,\omega}} \\
  D_{1_2} &=& \frac{M_2 - c K_{0_2,\omega} D_{0_2} }{K_{1_2,\omega}},
\end{eqnarray*}
where we have used  (\ref{dpartdef}), 
that the one-block partition $0_1=1_1$ is the only partition of length $1$, 
and that the two-block partition $0_2$ and the one-block partition $1_2$ are the only partitions of length $2$. 
This gives us the first moments $D_1$ and $D_2$ defined by (\ref{dmomdef}), since $D_{1_1}=D_1$, $D_{0_2}=D_1^2$, and $D_{1_2}=D_2$.

\subsection{Uniformly distributed $\omega$}
For the case of Vandermonde matrices with uniform phase distribution,
it turns out that the noncrossing partitions play a central role.
The role is somewhat different than the relation for freeness.
Let $u$ denote the uniform distribution on $[0,2\pi)$.

\begin{proposition} \label{teo1}
  The Vandermonde mixed moment expansion coefficient
  \[
    K_{\rho , u} = \lim_{N\rightarrow\infty} K_{\rho , u , N}
  \]
  exists for all $\rho$.
  Moreover, $0 < K_{\rho , u} \leq 1$, the $K_{\rho , u}$ are rational numbers for all $\rho$,
  and $K_{\rho , u}=1$ if and only if $\rho$ is noncrossing.
\end{proposition}

The proof of Proposition~\ref{teo1} can be found in Appendix~\ref{appendixteo1}. 
The same result is proved in~\cite{paper:nordio1}, where the $K_{\rho,u}$ are given an equivalent description. 
The proof in the appendix only translates the result in~\cite{paper:nordio1} to the current notation. 
Due to Proposition~\ref{teo0}, Theorem~\ref{teo1} guarantees that the mixed moments (\ref{cumequation0}) exist in the limit
for the uniform phase distribution, and are given by (\ref{cumequation}).
The $K_{\rho , u}$ are in general hard to compute for higher order $\rho$ with crossings.
It turns out that the following computations suffice to obtain the $7$ first moments.
\begin{proposition} \label{lemma:kcompute}
  The following holds:
  \begin{eqnarray*}
    K_{ \{ \{ 1,3\} , \{ 2,4\} \} , u}                       &=& \frac{2}{3} \\
    K_{ \{ \{ 1,4\} , \{ 2,5\} , \{ 3,6\} \} , u}            &=& \frac{1}{2} \\
    K_{ \{ \{ 1,4\} , \{ 2,6\} , \{ 3,5\} \} , u}            &=& \frac{1}{2} \\
    K_{ \{ \{ 1,3,5\} , \{ 2,4,6\} \} , u}                   &=& \frac{11}{20} \\
    K_{ \{ \{ 1,5\} , \{ 3,7\} , \{ 2,4,6\} \} , u}          &=& \frac{9}{20} \\
    K_{ \{ \{ 1,6\} , \{ 2,4\} , \{ 3,5,7\} \} , u}          &=& \frac{9}{20}.
  \end{eqnarray*}
\end{proposition}

The proof of Proposition~\ref{lemma:kcompute} is given in Appendix~\ref{appendixkcompute}.
Combining Proposition~\ref{teo1} and Proposition~\ref{lemma:kcompute} one can prove the following:
\begin{proposition} \label{teo:first7moments}
  Assume ${\bf D}_1(N) = \cdots = {\bf D}_n(N) = {\bf D}(N)$, and that the limits
  \begin{eqnarray}
    m_n &=& (cM)_n = c \lim_{N\rightarrow\infty} E\left[ tr_L \left( \left( {\bf D}(N) {\bf V}^H {\bf V} \right)^n \right) \right] \label{substequations1} \\
    d_n &=& (cD)_n = c \lim_{N\rightarrow\infty} tr_L\left( {\bf D}^n(N) \right). \label{substequations2}
  \end{eqnarray}
  exist. When $\omega = u$, we have that
  \begin{eqnarray*}
    m_1 &=& d_1 \\
    m_2 &=& d_2 + d_1^2 \\
    m_3 &=& d_3 + 3d_2d_1 + d_1^3\\
    m_4 &=& d_4 + 4d_3d_1 + \frac{8}{3}d_2^2 + 6d_2d_1^2 + d_1^4\\
    m_5 &=& d_5 + 5d_4d_1 + \frac{25}{3}d_3d_2 + 10 d_3d_1^2 + \\
        & & \frac{40}{3}d_2^2d_1 + 10 d_2d_1^3 + d_1^5\\
    m_6 &=& d_6 + 6d_5d_1 + 12 d_4d_2 + 15 d_4d_1^2 + \\
        & & \frac{151}{20} d_3^2 + 50 d_3d_2d_1 + 20 d_3d_1^3 + \\
        & & 11 d_2^3 + 40 d_2^2d_1^2 + 15 d_2d_1^4 + d_1^6\\
    m_7 &=& d_7 + 7d_6d_1 + \frac{49}{3}  d_5d_2 + 21 d_5d_1^2 + \\
        & & \frac{497}{20} d_4d_3 + 84 d_4 d_2 d_1 + 35 d_4 d_1^3 + \\
        & & \frac{1057}{20} d_3^2 d_1 + \frac{693}{10} d_3d_2^2 + 175 d_3 d_2 d_1^2 + \\
        & & 35 d_3d_1^4 + 77 d_2^3 d_1 + \frac{280}{3} d_2^2 d_1^3 + \\
        & & 21 d_2 d_1^5 + d_1^7.
  \end{eqnarray*}
\end{proposition}

Proposition~\ref{teo:first7moments} is proved in Appendix~\ref{appendixkcounting}.
Several of the following theorems will also be stated in terms of the scaled moments (\ref{substequations1})-(\ref{substequations2}), rather than $M_n,D_n$.
The reason for this is that the dependency on the matrix aspect ratio $c$ can be absorbed in $m_n,d_n$,
so that the result itself can be expressed independently of $c$, as in the equations of Proposition~\ref{teo:first7moments}.
The same usage of scaled moments has been applied for large Wishart matrices~\cite{eurecom:freedeconvinftheory}. 
Similar computations to those in the proof of Proposition~\ref{teo:first7moments} are performed in~\cite{paper:nordio1}, 
although the computations there do not go up as high as the first seven mixed moments. 
To compute higher order moments, $K_{\rho , u}$ must be computed for partitions of higher order also.
The computations performed in Appendix~\ref{appendixkcompute} and~\ref{appendixkcounting} should convince the reader that this can be done,
but that it is very tedious.

Following the proof of Proposition~\ref{teo1}, we can also obtain formulas for the second order moments of Vandermonde matrices.
Since it is easily seen that $C_{1,n}({\bf D}(N) {\bf V}^H {\bf V})=C_{n,1}({\bf D}(N) {\bf V}^H {\bf V})=0$, 
the first nontrivial second order moment is the following:
\begin{proposition} \label{secondorderexps}
Assume that ${\bf V}$ has uniform phase distribution, let $d_n$ be as in (\ref{substequations2}), and define
\begin{equation} \label{substequations3}
  m_{i,j} = c \lim_{L\rightarrow\infty} LC_{i,j}\left( {\bf D}(N) {\bf V}^H {\bf V}) \right).
\end{equation}
Then we have that
\begin{equation} \label{vfluctuations2}
  m_{2,2} = d_4 + 4d_3d_1  \frac{4}{3}d_2^2 + 4d_2d_1^2.
\end{equation}
\end{proposition}

Proposition~\ref{secondorderexps} is proved in Appendix~\ref{appendixsecondorderexps}, 
and relies on the same type of calculations as those in Appendix~\ref{appendixkcompute}.
Following the proof of Proposition~\ref{teo1} again, we can also obtain exact
expressions for moments of lower order random Vandermonde matrices with uniform phase distribution,
not only the limit. We state these only for the first four moments.
\begin{theorem} \label{teo:exact4moments}
  Assume ${\bf D}_1(N) = {\bf D}_2(N) = \cdots = {\bf D}_n(N)$, set $c=\frac{L}{N}$, and define
  \begin{eqnarray}
    m_n^{(N,L)} &=& c E\left[ tr_L \left( \left( {\bf D}(N) {\bf V}^H {\bf V} \right)^n \right) \right] \\
    d_n^{(N,L)} &=& c tr_L\left( {\bf D}^n(N) \right).
  \end{eqnarray}
  When $\omega = u$ we have that
  \begin{eqnarray*}
    m_1^{(N,L)} &=& d_1^{(N,L)} \\
    m_2^{(N,L)} &=& \left( 1 - N^{-1} \right) d_2^{(N,L)} + (d_1^{(N,L)})^2 \\
    m_3^{(N,L)} &=& \left( 1  - 3N^{-1}  + 2 N^{-2} \right) d_3^{(N,L)} \\
                & & + 3 \left( 1 - N^{-1} \right) d_1^{(N,L)} d_2^{(N,L)} + (d_1^{(N,L)})^3 \\
    m_4^{(N,L)} &=& \left( 1 - \frac{20}{3} N^{-1} + 12 N^{-2} - \frac{19}{3} N^{-3} \right) d_4^{(N,L)} \\
                & & + \left( 4 - 12 N^{-1} + 8 N^{-2} \right) d_3^{(N,L)} d_1^{(N,L)} \\
                & & + \left( \frac{8}{3} - 6 N^{-1} + \frac{10}{3} N^{-2} \right) (d_2^{(N,L)})^2 \\
                & & + 6 \left( 1 - N^{-1} \right) d_2^{(N,L)} (d_1^{(N,L)})^2 + (d_1^{(N,L)})^4.
  \end{eqnarray*}
\end{theorem}

Theorem~\ref{teo:exact4moments} is proved in Appendix~\ref{appendixteoexactdist}.
Exact formulas for the higher order moments also exist, but they become increasingly complex,
as higher order terms $N^{-k}$ also enter the picture.
These formulas are also harder to prove for higher order moments.
In many cases, exact expressions are not what we need:
first order approximations (i.e. expressions where only the $N^{-1}$-terms are included) can suffice for many purposes.
In Appendix~\ref{appendixteoexactdist}, we explain how the simpler case of these first order approximations can be computed.
It seems much harder to prove a similar result when the phase distribution is not uniform.

An important result building on the results we present is the following, 
which provides a major difference from the limit eigenvalue distributions of Gaussian matrices. 
\begin{proposition} \label{propunbounded}
  The asymptotic mean eigenvalue distribution of a Vandermonde matrix with uniform phase distribution has unbounded support.
\end{proposition}

Proposition~\ref{propunbounded} is proved in Appendix~\ref{appendixpropunbounded}.

\subsection{$\omega$ with continuous density} \label{section:generaldistfinite}
The following result tells us that the limit $K_{\rho ,\omega}$ exists for many $\omega$,
and also gives a useful expression for them in terms of $K_{\rho , u}$ and the density of $\omega$.
\begin{theorem} \label{teo:generaldist}
  The Vandermonde mixed moment expansion coefficients $K_{\rho ,\omega} = \lim_{N\rightarrow\infty} K_{\rho , \omega , N}$
  exist whenever the density $p_{\omega}$ of $\omega$ is continuous on $[0,2\pi )$.
  If this is fulfilled, then
  \begin{equation} \label{generalintegral}
    K_{\rho , \omega} = K_{\rho , u} (2\pi)^{|\rho |-1} \left( \int_0^{2\pi} p_{\omega}(x)^{|\rho |} dx \right) .
  \end{equation}
\end{theorem}

The proof is given in Appendix~\ref{appendixgeneraldist}. 
Although the proof assumes a continuous density, we remark that it can be generalized to cases where the density contains a finite set of jump discontinuities also. 
In Section~\ref{simulations}, several examples are provided where the integrals
(\ref{generalintegral}) are computed.
An important consequence of Theorem~\ref{teo:generaldist} is the following, which gives the uniform phase distribution an important role.
\begin{proposition} \label{teo:uniformmin}
Let ${\bf V}_{\omega}$ denote a Vandermonde matrix with phase distribution $\omega$, and set
\[ V_{\omega,n} = \lim_{N\rightarrow\infty} E \left[ tr_L \left( \left( {\bf V}_{\omega}^H {\bf V}_{\omega} \right)^n \right) \right]. \]
Then we have that
\[ V_{u,n} \leq V_{\omega,n}. \]
\end{proposition}

The proof is given in Appendix~\ref{appendixuniformmin}.
An immediate consequence of this and Proposition~\ref{propunbounded} is that all phase distributions, not only uniform phase distribution,
give Vandermonde matrices with unbounded mean eigenvalue distributions in the limit.
Besides providing us with a deconvolution method for finding the
mixed moments of the $\{ {\bf D}_r(N) \}_{1\leq r\leq n}$,
Theorem~\ref{teo:generaldist} also provides us with a way of
inspecting the phase distribution $\omega$, by first finding the
moments of the density, i.e. $\int_0^{2\pi} p_{\omega}(x)^k dx$.
However, note that we can not expect to find the density of $\omega$
itself, only the density of the density of $\omega$. 
This follows immediately by noting that $\int_0^{2\pi} p_{\omega}(x)^k dx$ remains unchanged when the phase distribution $\omega$ is cyclically shifted.

\subsection{$\omega$ with density singularities} \label{section:generaldistinfinite}
The asymptotics of Vandermonde matrices are different when the
density of $\omega$ has singularities, and depends on the density
growth rates near the singular points. It will be clear from the following 
results that one can not perform deconvolution for such $\omega$ to
obtain the higher order moments of the $\{ {\bf D}_r(N) \}_{1\leq
r\leq n}$, as only their first moment can be obtained. The asymptotics
are first described for $\omega$ with atomic density singularities,
as this is the simplest case to prove. After this, densities with
polynomic growth rates near the singularities are addressed.

\begin{theorem} \label{generalinfinitedensity1}
Assume that $p_{\omega}=\sum_{i=1}^r p_i \delta_{\alpha_i}$ is
atomic (where $\delta_{\alpha_i}(x)$ is dirac measure (point mass)
at $\alpha_i$), and denote by $p^{(n)} = \sum_{i=1}^r p_i^n$. Then
\begin{eqnarray*}
  \lim_{N\rightarrow\infty} & E[ Tr ( & {\bf D}_1(N) \frac{1}{N} {\bf V}^H {\bf V} {\bf D}_2(N) \frac{1}{N} {\bf V}^H {\bf V} \\
                            &         & \cdots \times {\bf D}_n(N) \frac{1}{N} {\bf V}^H {\bf V} )] \\
  =                         &         & c^{n-1} p^{(n)} \lim_{N\rightarrow\infty} \prod_{i=1}^n tr_L\left( {\bf D}_i(N) \right) .
\end{eqnarray*}
Note here that the non-normalized trace is used.
\end{theorem}

The proof can be found in Appendix~\ref{appendixgeneralinfinitedensity1}.
In particular, Theorem~\ref{generalinfinitedensity1} states that the asymptotic moments of $\frac{1}{N} {\bf V}^H {\bf V}$ can be computed from $p^{(n)}$.
The theorem is of great importance for the estimation of the point masses $p_i$.
In blind seismic and telecommunication applications, one would like to detect the locations $\alpha_i$.
Unfortunately, Theorem~\ref{generalinfinitedensity1} tells us that this is impossible with our deconvolution framework, 
since the $p^{(n)}$, which are the quantities we can find through deconvolution,
have no dependency to them. This parallels Theorem~\ref{teo:generaldist}, since also there we could not recover the density $p_{\omega}$ itself.
Having found the $p^{(n)}$ through deconvolution, one can find the point masses $p_i$, by solving
for $p_1,p_2,...$ in the Vandermonde equation
\[
  \left( \begin{array}{llll} p_1 & p_2 & \cdots & p_r \\ p_1^2 & p_2^2 & \cdots & p_r^2 \\ \vdots & \vdots & \vdots & \vdots \end{array} \right)
  \left( \begin{array}{l} 1 \\ 1 \\ \vdots \end{array} \right)
  =
  \left( \begin{array}{l} p^{(1)} \\ p^{(2)} \\ \vdots \end{array} \right) .
\]

The case when the density has non-atomic singularities is more complicated.
We provide only the following result, which addresses the case when the density has polynomic growth rate near the singularities.

\begin{theorem} \label{generalinfinitedensity2}
  Assume that
  \[ \lim_{x\rightarrow\alpha_i} |x-\alpha_i|^s p_{\omega}(x) = p_i \mbox{ for some } 0<s<1 \]
  for a set of points $\alpha_1,...,\alpha_r$, with $p_{\omega}$ continuous for $\omega\neq \alpha_1,...,\alpha_r$. Then
\begin{eqnarray*}
  \lim_{N\rightarrow\infty} & E[Tr( & {\bf D}_1(N) \frac{1}{N^s} {\bf V}^H {\bf V} {\bf D}_2(N) \frac{1}{N^s} {\bf V}^H {\bf V} \\
                            &         & \cdots \times {\bf D}_n(N) \frac{1}{N^s} {\bf V}^H {\bf V} )] \\
  =                         &         & c^{n-1} q^{(n)} \lim_{N\rightarrow\infty} \prod_{i=1}^n tr_L\left( {\bf D}_i(N) \right)
\end{eqnarray*}
where
\begin{equation} \label{pmintegral}
\begin{array}{lll}
  q^{(n)} &=& \left( 2\Gamma(1-s) \cos\left( \frac{(1-s)\pi}{2} \right) \right)^n p^{(n)} \times \\
          & & \int_{[0,1]^n} \prod_{k=1}^{n} \frac{1}{ \left| x_{k-1} - x_k \right|^{1-s}} dx_1\cdots dx_n,
\end{array}
\end{equation}
and $p^{(n)} = \sum_i p_i^n$.
Note here that the non-normalized trace is used.
\end{theorem}

The proof can be found in
Appendix~\ref{appendixgeneralinfinitedensity2}. Also in this case it
is only the point masses $p_i$ which can be found through
deconvolution, not the locations $\alpha_i$. Note that
the integral in (\ref{pmintegral}) can also be written as an
$m$-fold convolution. Similarly, the definition of $K_{\rho , \omega
, N}$ given by (\ref{kpindef}) can also be viewed as a $2$-fold
convolution when $\rho$ has two blocks, and as a $3$-fold
convolution when $\rho$ has three blocks (but not for $\rho$ with
more than $3$ blocks).

A useful application of Theorem~\ref{generalinfinitedensity2} we will return to is the case when
$\omega = k\sin(\theta)$ for some constant $k$ (see (\ref{vandermonde2})), with $\theta$ uniformly distributed on some interval. 
This case is simulated in Section~\ref{firstscenario}. 
It is apparent from (\ref{sindensity2}) that the density goes to infinity near $\omega = \pm k$, with rate $x^{-1/2}$. 
Theorem~\ref{generalinfinitedensity2} thus applies with $s=1/2$.
For this case, however, the "edges" at $\pm \pi/2$ are never reached in practice. 
Indeed, in array processing \cite{paper:krimviberg}, 
the antenna array is a sector antenna which scans an angle interval which never includes the edges. 
We can therefore restrict $\omega$ in our analysis to clusters of intervals $[\alpha_i,\beta_i]$ not containing $\pm 1$,
for which the results of Section~\ref{section:generaldistfinite} suffice.
In this way, we also avoid the computation of the cumbersome integral (\ref{pmintegral}).

\subsection{Generalized Vandermonde matrices} \label{section:nongeometric}
We will consider generalized Vandermonde matrices on the form
\begin{equation} \label{vandermondegeneralized1}
  {\bf V} = \frac{1}{\sqrt{N}}
               \left( \begin{array}{lll}
                        e^{-j \lfloor Nf(0) \rfloor \omega_1}             & \cdots & e^{-j \lfloor  Nf(0) \rfloor \omega_L} \\
                        e^{-j \lfloor Nf(\frac{1}{N}) \rfloor  \omega_1}  & \cdots & e^{-j \lfloor  Nf(\frac{1}{N}) \rfloor \omega_L} \\
                        \vdots                                            & \ddots & \vdots \\
                        e^{-j \lfloor Nf(\frac{N-1}{N}) \rfloor \omega_1} & \cdots & e^{-j \lfloor  Nf(\frac{N-1}{N}) \rfloor \omega_L}
                      \end{array}
               \right),
\end{equation}
where $f$ is called the {\em power distribution}, and is a function from $[0,1)$ to $[0,1)$.
We will also consider the more general case when $f$ is replaced with a random variable $\lambda$, i.e.
\begin{equation} \label{vandermondegeneralized2}
  {\bf V} = \frac{1}{\sqrt{N}}
               \left( \begin{array}{lll}
                        e^{-j N\lambda_1 \omega_1} & \cdots & e^{-j N\lambda_1 \omega_L} \\
                        e^{-j N\lambda_2 \omega_1} & \cdots & e^{-j N\lambda_2 \omega_L} \\
                        \vdots                     & \ddots & \vdots \\
                        e^{-j N\lambda_N \omega_1} & \cdots & e^{-j N\lambda_N \omega_L}
                      \end{array}
               \right),
\end{equation}
with the $\lambda_i$ i.i.d. and distributed as $\lambda$, defined and taking values in $[0,1)$,
and also independent from the $\omega_j$. 

We will define mixed moment expansion coefficients for generalized Vandermonde matrices also.
The difference is that, while we in Definition~\ref{expansiondef} simplified using the geometric sum formula, we
can not do this now since we do not assume uniform power distribution anymore. 
To define expansion coefficients for generalized Vandermonde matrices of the form (\ref{vandermondegeneralized1}),
define first integer functions $f_N$ from $[0,N-1]$ to $[0,N-1]$ by $f_N(r)=\lfloor N f\left(\frac{r}{N}\right) \rfloor$.
Let $p_{f_N}$ be the corresponding density for $f_N$.
The procedure is similar for matrices of the form (\ref{vandermondegeneralized2}).
The following definition captures both cases:

\begin{definition}
  For (\ref{vandermondegeneralized1}) and (\ref{vandermondegeneralized2}), define
  \begin{equation} \label{kpindef2}
  \begin{array}{ll}
    K_{\rho , \omega , f, N}       =& \frac{1}{N^{1-|\rho |}} \times \\
                                    & \int_{(0,2\pi)^{|\rho |}} \\
                                    & \prod_{k=1}^n \left( \sum_{r=0}^{N-1} p_{f_N}(r) e^{j r (\omega_{b(k-1)} -\omega_{b(k)})} \right) \\
                                    & d\omega_1\cdots d\omega_{|\rho |} \\
    K_{\rho , \omega , \lambda, N} =& \frac{1}{N^{1-|\rho |}} \times \\
                                    & \int_{(0,2\pi)^{|\rho |}} \prod_{k=1}^n \left( \int_0^1 Ne^{j N\lambda (\omega_{b(k-1)} -\omega_{b(k)})} d\lambda \right) \\
                                    & d\omega_1\cdots d\omega_{|\rho |},
  \end{array}
  \end{equation}
  where $\omega_{W_1},...,\omega_{W_{|\rho |}}$ are as in Definition~\ref{expansiondef}.
  If the limits
  \begin{eqnarray*}
      K_{\rho , \omega , f}       &=& \lim_{N\rightarrow\infty} K_{\rho , \omega , f, N} \\
      K_{\rho , \omega , \lambda} &=& \lim_{N\rightarrow\infty} K_{\rho , \omega , \lambda, N},
  \end{eqnarray*}
  exist, then they are called {\em Vandermonde mixed moment expansion coefficients}.
\end{definition}

Note that (\ref{vandermonde}) corresponds to (\ref{vandermondegeneralized1}) with $f(x)=x$.
The following result holds:
\begin{theorem} \label{teo0gen}
  Theorem~\ref{teo0} holds also with Vandermonde matrices (\ref{vandermonde}) replaced with generalized Vandermonde matrices on either form
  (\ref{vandermondegeneralized1}) or (\ref{vandermondegeneralized2}), and with $K_{\rho , \omega}$ replaced with either $K_{\rho , \omega , f}$
  or $K_{\rho , \omega , \lambda}$.
\end{theorem}

The proof follows the same lines as those in Appendix~\ref{appendixteo0},
and is therefore only explained briefly at the end of that appendix.
As for matrices of the form (\ref{vandermonde}),
it is the case of uniform phase distribution which is most easily described how to compute for generalized Vandermonde matrices also.
Appendix~\ref{appendixteo1} shows how the computation of $K_{\rho,u}$ boils down to computing certain integrals.
The same comments are valid for matrices of the form (\ref{vandermondegeneralized1}) or (\ref{vandermondegeneralized2}) in order to compute
$K_{\rho , \omega , f}$ and $K_{\rho , \omega , \lambda}$. This is further commented at the end of that appendix.

We will not consider generalized Vandermonde matrices with density singularities.

\subsection{The joint distribution of independent Vandermonde matrices} \label{section:manymatrices}
When many independent random Vandermonde matrices are involved, the following holds:
\begin{theorem} \label{teo2}
  Assume that the $\{ {\bf D}_r(N) \}_{1\leq r\leq n}$ have a joint limit distribution as $N\rightarrow\infty$.
  Assume also that ${\bf V}_1,{\bf V}_2,...$ are independent Vandermonde matrices with the same phase distribution $\omega$,
  and that the density of $\omega$ is continuous.
  Then the limit
  \[
  \begin{array}{ll}
    \lim_{N\rightarrow\infty} E[ tr_L ( & {\bf D}_1(N) {\bf V}_{i_1}^H {\bf V}_{i_2} {\bf D}_2(N) {\bf V}_{i_2}^H {\bf V}_{i_3} \\
                                                  & \cdots \times {\bf D}_n(N) {\bf V}_{i_n}^H {\bf V}_{i_1} ) ]
  \end{array}
  \]
  also exists when $\frac{L}{N}\rightarrow c$. The limit is $0$ when $n$ is odd, and equals
  \begin{equation} \label{generalequation}
    \sum_{\rho\leq\sigma\in{\cal P}(n)} K_{\rho , \omega} c^{|\rho |-1} D_{\rho},
  \end{equation}
  where $\sigma=\{\sigma_1,\sigma_2\}=\{\{ 1,3,5,...,\},\{ 2,4,6,...\}\}$ is the partition 
  where the two blocks are the even numbers, and the odd numbers. 
\end{theorem}

The proof of Theorem~\ref{teo2} can be found in Appendix~\ref{appendix:manymatrices}.
That appendix also contains some remarks on the case when the matrices ${\bf D}_i(N)$ are placed at different positions relative to the Vandermonde matrices.
From Theorem~\ref{teo2}, the following corollary is immediate:
\begin{corollary} \label{kornew}
  The first three mixed moments
  \[
    V^{(2)}_n = \lim_{N\rightarrow\infty} E\left[ tr_L \left( \left( {\bf V}_1^H {\bf V}_2 {\bf V}_2^H {\bf V}_1 \right)^n \right) \right]
  \]
  of independent Vandermonde matrices ${\bf V}_1, {\bf V}_2$ are given by
  \begin{eqnarray*}
    V^{(2)}_1 &=& I_2 \\
    V^{(2)}_2 &=& \frac{2}{3}I_2 + 2I_3 + I_4 \\
    V^{(2)}_3 &=& \frac{11}{20}I_2 + 4I_3 + 9I_4 + 6I_5 + I_6,
  \end{eqnarray*}
  where
  \[
    I_k = (2\pi)^{k-1} \left( \int_0^{2\pi} p_{\omega}(x)^{k} dx \right).
  \]
  In particular, when the phase distribution is uniform, the first three mixed moments are given by
  \begin{eqnarray*}
    V^{(2)}_1 &=& 1 \\
    V^{(2)}_2 &=& \frac{11}{3} \\
    V^{(2)}_3 &=& \frac{411}{20}
  \end{eqnarray*}
\end{corollary}

The results here can also be extended to the case with independent Vandermonde matrices with different phase distributions:
\begin{theorem} \label{notproved}
  Assume that $\{ {\bf V}_i \}_{1\leq i\leq s}$ are independent Vandermonde matrices,
  where ${\bf V}_i$ has continuous phase distribution $\omega_i$.
  Denote by $p_{\omega_i}$ the density of $\omega_i$.
  Then Equation (\ref{generalequation}) still holds, with $K_{\rho , \omega}$ replaced by
  \[
    K_{\rho , u} (2\pi)^{|\rho |-1} \int_0^{2\pi} \prod_{i=1}^{s}  p_{\omega_i}(x)^{|\rho_i |} dx ,
  \]
  where $\rho_i$ consists of all numbers $k$ such that $i_k=i$. 
\end{theorem}

The proof is omitted, as it is a straightforward extension of the
proofs of Theorem~\ref{teo:generaldist} and Theorem~\ref{teo2}.

\section{Discussion} \label{section:deconvolutionperspective}
In the recent work \cite{paper:nordio1}, the Vandermonde model (\ref{vandermonde}) is encountered 
in reconstruction of multidimensional signals in wireless sensor networks.
The authors also recognize a similar expression for the Vandermonde mixed moment expansion coefficient as in Definition~\ref{expansiondef}.
They also state that, for the case of uniform phase distribution, closed form expressions for the moments can be found, building on
an analysis of partitions and calculation of volumes of convex polytopes described by certain constraints.
This is very similar to what is done in this paper.

We will in the following discuss some differences and similarities between Gaussian and Vandermonde matrices.

\subsection{Convergence rates}
\begin{figure}
  \subfigure[$80$ samples]{\epsfig{figure=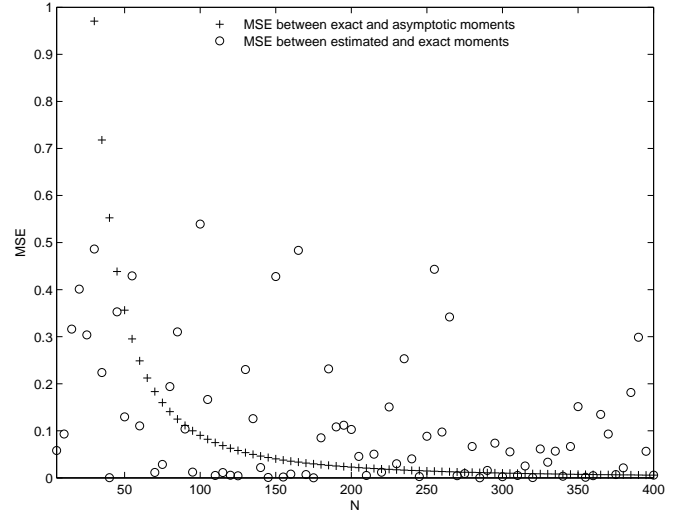,width=0.99\columnwidth}}
  \subfigure[$320$ samples]{\epsfig{figure=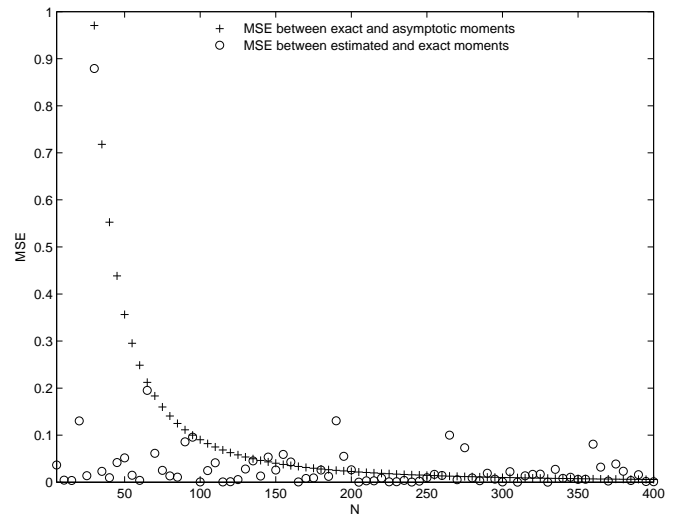,width=0.99\columnwidth}}
  \caption{MSE of the first $4$ estimated moments from the exact moments for $80$ and $320$ samples for varying matrix sizes, with $N=L$.
    Matrices are on the form ${\bf V}^H {\bf V}$ with ${\bf V}$ a Vandermonde matrix with uniform phase distribution.
    The MSE of the first $4$ exact moments from the asymptotic moments is also shown.} \label{fig:80320it}
\end{figure}

\begin{figure}
  \begin{center}
    \epsfig{figure=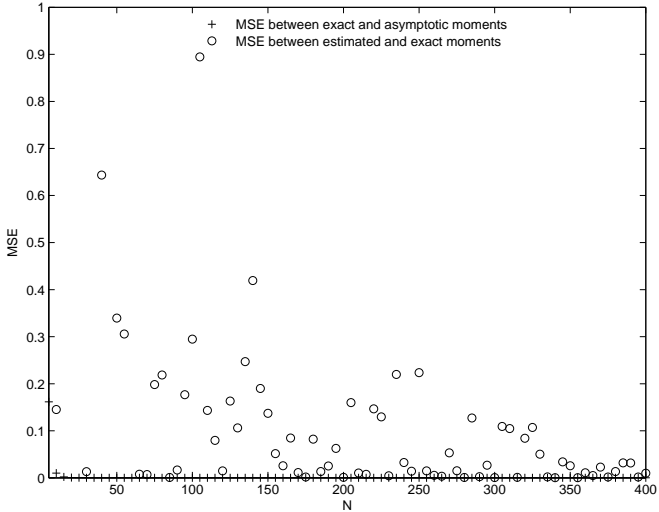,width=0.99\columnwidth}
  \end{center}
  \caption{MSE of the first $4$ estimated moments from the exact moments for $5$ samples for varying matrix sizes, with $N=L$.
    Matrices are on the form $\frac{1}{N} {\bf X} {\bf X}^H$ with ${\bf X}$ a complex standard Gaussian matrix.
    The MSE of the first $4$ exact moments from the asymptotic moments is also shown.} \label{fig:5itg}
\end{figure}

In~\cite{paper:thorbjornsen1}, almost sure convergence of Gaussian matrices was shown by proving exact formulas for the distribution
of lower order Gaussian matrices. These deviated from their limits by terms of order $1/N^2$.
In Theorem~\ref{teo:exact4moments}, we see that terms of order $1/N$ are involved.
This slower rate of convergence may not be enough to make a statement on whether we have almost sure convergence for Vandermonde matrices.
However,~\cite{paper:brycdembojiang} shows some almost sure convergence properties for certain Hankel and Toeplitz matrices.
These matrices are seen in that paper to have similar combinatorial descriptions for the moments, when compared to Vandermonde matrices in this paper.
Therefore, it may be the case that the techniques in~\cite{paper:brycdembojiang} can be generalized to address almost sure convergence of Vandermonde matrices also.
Figure~\ref{fig:80320it} shows the speed of convergence of the moments of Vandermonde matrices
(with uniform phase distribution) towards the asymptotic moments as the
matrix dimensions grow, and as the number of samples grow.
The differences between the asymptotic moments and the exact moments are also shown.
To be more precise, the MSE values in Figure~\ref{fig:80320it} are computed as follows:
\begin{enumerate}
  \item $K$ samples ${\bf V}_i$ are independently generated using (\ref{vandermonde}).
  \item The $4$ first sample moments $\hat{v}_{ji} = \frac{1}{L} tr_n\left( \left( {\bf V}_i^H {\bf V}_i \right)^j \right)$ ($1\leq j\leq 4$)
    are computed from the samples.
  \item The $4$ first estimated moments $\hat{V}_j$ are computed as the mean of the sample moments,
    i.e. $\hat{V}_j = \frac{1}{K} \sum_{i=1}^K \hat{m}_{ji}$.
  \item The $4$ first exact moments $E_j$ are computed using Theorem~\ref{teo:exact4moments}.
  \item The $4$ first asymptotic moments $A_j$ are computed using Proposition~\ref{teo:first7moments}.
  \item The mean squared error (MSE) of the first $4$ estimated  moments from the exact moments is computed as $\sum_{j=1}^4 \left( \hat{V}_j - E_j \right)^2$.
  \item The MSE of the first $4$ exact moments from the asymptotic moments is computed as $\sum_{j=1}^4 \left( E_j - A_j \right)^2$.
\end{enumerate}
Figure~\ref{fig:80320it} is in sharp contrast with Gaussian matrices,
as shown in Figure~\ref{fig:5itg}.
First of all, it is seen that the asymptotic moments can be used just as well instead of the exact moments
(for which expressions can be found in~\cite{eurecom:channelcapacity}), due to the $O(1/N^2)$ convergence of the moments.
Secondly, it is seen that only $5$ samples were needed to get a reliable estimate for the moments.

\subsection{Inequalities between moments of Vandermonde matrices and moments of known distributions}
We will state an inequality involving the moments of Vandermonde matrices, and the moments of known distributions.
The classical Poisson distribution with rate $\lambda$ and jump size  $\alpha$ is defined as the limit of
\[
  \left( \left( 1 -\frac{\lambda}{n} \right) \delta_0 + \frac{\lambda}{n} \delta_{\alpha} \right)^{\ast n}
\]
as $n\rightarrow\infty$~\cite{book:comblect},
where $\ast$ denotes classical (additive) convolution, and $\ast n$ denotes $n$-fold convolution with itself.
For our analysis, we will only need the classical Poisson distribution with rate $c$ and jump size $1$, denoted $\nu_c$.
The free Poisson distribution with rate $\lambda$ and jump size $\alpha$ is defined similarly as the limit of
\[
  \left( \left( 1 -\frac{\lambda}{n} \right) \delta_0 + \frac{\lambda}{n} \delta_{\alpha} \right)^{\boxplus n}
\]
as $n\rightarrow\infty$, where $\boxplus$ is the free probability counterpart of $\ast$~\cite{book:comblect,book:hiaipetz},
and where $\boxplus n$ denotes $n$-fold free convolution with itself.
For our analysis, we will only need the free Poisson distribution with rate $\frac{1}{c}$ and jump size $c$, denoted $\mu_c$.
$\mu_c$ is the same as the better known Mar\u{c}henko Pastur law, i.e. it has the density~\cite{book:hiaipetz}
\begin{equation} \label{mpdensity}
  f^{\mu_c}(x) = (1-\frac{1}{c})^+ \delta_0(x) + \frac{\sqrt{(x-a)^+(b-x)^+}}{2\pi cx},
\end{equation}
where $(z)^+ =\mbox{max}(0,z)$, $a=(1-\sqrt{c})^2$, $b=(1+\sqrt{c})^2$.
Since the classical (free) cumulants of the classical (free) Poisson distribution are $\lambda\alpha^n$~\cite{book:comblect},
we see that the (classical) cumulants of $\nu_c$ are $c,c,c,c,...$, and that the (free) cumulants of $\mu_c$ are $1,c,c^2,c^3,...$.
In other words, if $a_1$ has the distribution $\mu_c$, then
\begin{equation} \label{freepo}
\begin{array}{lll}
  \phi(a_1^n) &=& \sum_{\rho\in NC(n)} c^{n-|\rho |} = \sum_{\rho\in NC(n)} c^{|K(\rho)|-1} \\
              &=& \sum_{\rho\in NC(n)} c^{|\rho|-1}.
\end{array}
\end{equation}
Here we have used the Kreweras complementation map and (\ref{krewerasprop}), 
with $\phi$ denoting the expectation in a non-commutative probability space~\cite{book:hiaipetz}.
Also, if $a_2$ has the distribution $\nu_c$, then
\begin{equation} \label{po}
  E(a_2^n) = \sum_{\rho\in{\cal P}(n)} c^{|\rho |}.
\end{equation}
We immediately recognize the $c^{|\rho |-1}$-entry of Theorem~\ref{teo0} in (\ref{freepo}) and (\ref{po})
(with an additional power of $c$ in (\ref{po})).
Combining Proposition~\ref{teo1} with ${\bf D}_1(N) = \cdots = {\bf D}_n(N) = {\bf I}_L$,
(\ref{freepo}), and (\ref{po}), we thus get the following corollary to Proposition~\ref{teo1}:

\begin{corollary} \label{freekor}
  Assume that ${\bf V}$ has uniform phase distribution.
  Then the limit moment
  \[ V_n = \lim_{N\rightarrow\infty} E\left[ tr_L \left( \left( {\bf V}^H {\bf V} \right)^n \right) \right] \]
  satisfies the inequality
  \[ \phi(a_1^n) \leq V_n \leq \frac{1}{c} E(a_2^n) , \]
  where $a_1$ has the distribution $\mu_c$ of the Mar\u{c}henko Pastur law,
  and $a_2$ has the Poisson distribution $\nu_c$.
  In particular, equality occurs for $m=1,2,3$ and $c=1$ (since all partitions are noncrossing for $m=1,2,3$).
\end{corollary}

Corollary~\ref{freekor} thus states that the moments of Vandermonde matrices with uniform phase distribution are bounded above and below by the
moments of the classical and free Poisson distributions, respectively.
The left part of the inequality in Corollary~\ref{freekor} was also observed in Section VI in~\cite{paper:nordio1}.
The different Poisson distributions enter here because their
(free and classical) cumulants resemble the $c^{|\rho |-1}$-entry in Theorem~\ref{teo0},
where we also can use that $K_{\rho , u} = 1$ if and only if $\rho$ is noncrossing to get a connection with the Mar\u{c}henko Pastur law.
To see how close the asymptotic Vandermonde moments are to these upper and lower bounds, the following corollary to Proposition~\ref{teo:first7moments}
contains the first moments:
\begin{corollary} \label{momkor}
  When $c=1$, the limit moments
  \[ V_n = \lim_{N\rightarrow\infty} E\left[ tr_L \left( \left( {\bf V}^H {\bf V} \right)^n \right) \right] ,\]
  the moments $fp_n$ of the Mar\u{c}henko Pastur law $\mu_1$, and the moments $p_n$ of the Poisson distribution $\nu_1$ satisfy
  \[
  \begin{array}{lllll}
    fp_4 = 14  &\leq &  V_4 = \frac{44}{3}     \approx 14.67  &\leq & p_4 = 15 \\
    fp_5 = 42  &\leq &  V_5 = \frac{146}{3}    \approx 48.67  &\leq & p_5 = 52 \\
    fp_6 = 132 &\leq &  V_6 = \frac{3571}{20}  \approx 178.55 &\leq & p_6 = 203\\
    fp_7 = 429 &\leq &  V_7 = \frac{2141}{3}   \approx 713.67 &\leq & p_7 = 877.
  \end{array}
  \]
  The first three moments coincide for the three distributions, and are $1,2$, and $5$, respectively.
\end{corollary}

\begin{figure}
  \begin{center}
    \epsfig{figure=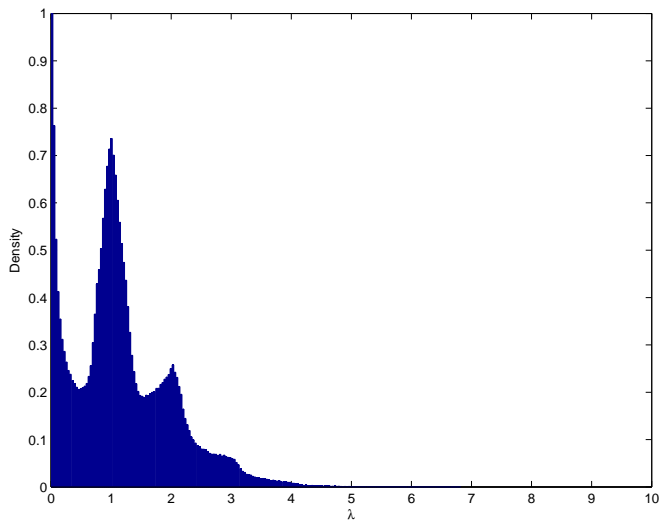,width=0.99\columnwidth}
  \end{center}
  \caption{Histogram of the mean eigenvalue distribution of $640$ samples of ${\bf V}^H {\bf V}$,
    with ${\bf V}$ a $1600\times 1200$ Vandermonde matrix with uniform phase distribution.} \label{fig:640iteig1600x1200}
\end{figure}

\begin{figure}
  \begin{center}
    \epsfig{figure=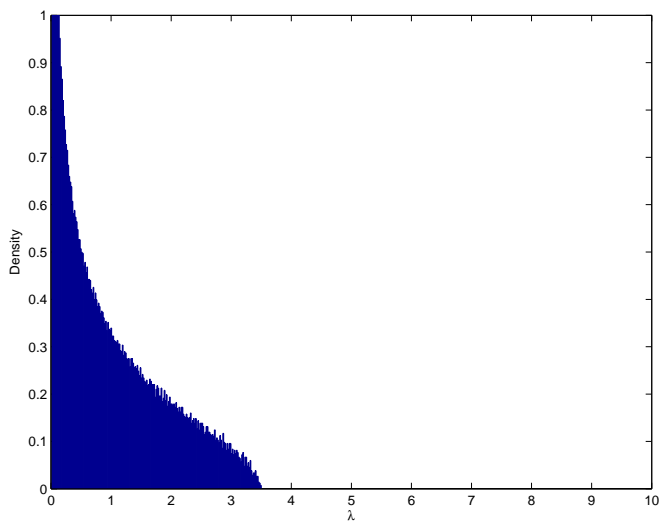,width=0.99\columnwidth}
  \end{center}
  \caption{Histogram of the mean eigenvalue distribution of $20$ samples of $\frac{1}{N} {\bf X} {\bf X}^H$,
    with ${\bf X}$ an $L\times N = 1200\times 1600$ complex, standard, Gaussian matrix.} \label{fig:20iteigg1600x1200}
\end{figure}

\begin{figure}
  \begin{center}
    \epsfig{figure=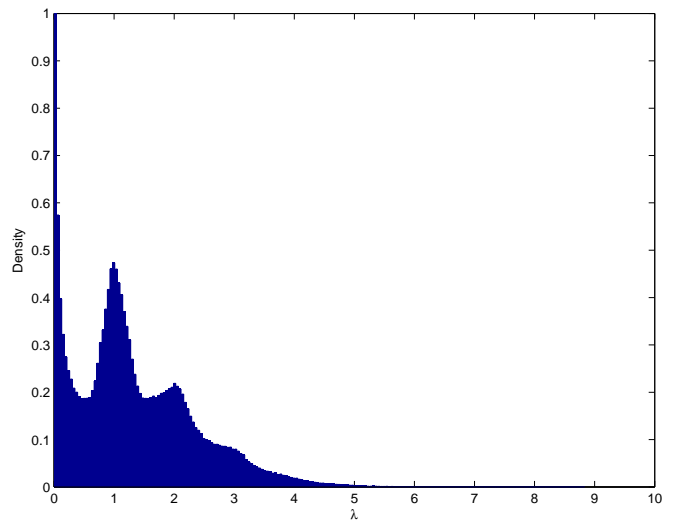,width=0.99\columnwidth}
  \end{center}
  \caption{Histogram of the mean eigenvalue distribution of $640$ samples of ${\bf V}^H {\bf V}$,
    with ${\bf V}$ a $1600\times 1200$ Vandermonde matrix with phase distribution $p_{\omega}$ defined in (\ref{sindensity2}) with $\lambda=2d,\alpha=\frac{\pi}{4}$.} \label{2part640iteig1600x1200}
\end{figure}

\begin{figure}
  \begin{center}
    \epsfig{figure=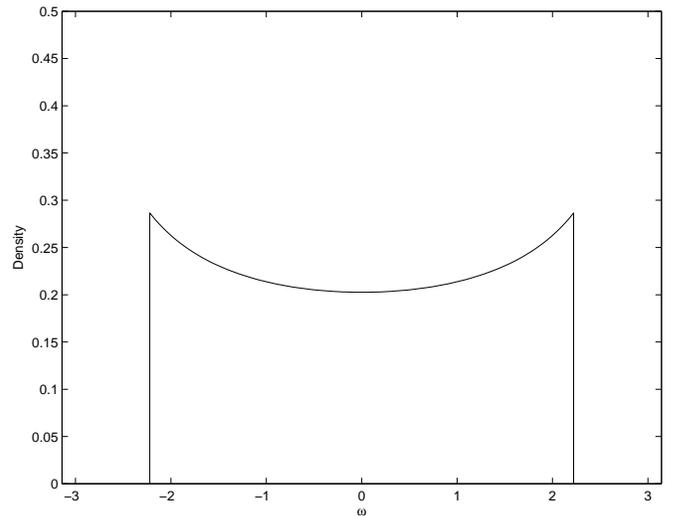,width=0.99\columnwidth}
  \end{center}
  \caption{The density $p_{\omega}(x)$ given by (\ref{sindensity2}), with $\lambda = 2d,\alpha = \frac{\pi}{4}$.} \label{fig:sindensity}
\end{figure}

The numbers $fp_n$ and $p_n$ are simply the number of partitions in $NC(n)$ and ${\cal P}(n)$, respectively.
The number of partitions in $NC(n)$ equals the Catalan number $C_n=\frac{1}{n+1}\binom{2n}{n}$~\cite{book:comblect}, and are easily computed.
The number of partitions of ${\cal P}(n)$ are also known as the Bell numbers $B_n$~\cite{book:comblect}.
They can easily be computed from the recurrence relation
\[
  B_{n+1} = \sum_{k=0}^n B_k \binom{n}{k}.
\]
In Figure~\ref{fig:640iteig1600x1200},
the mean eigenvalue distribution of $640$ samples of a $1600\times 1200$ (i.e. $c=0.75$) Vandermonde matrix with uniform phase distribution
is shown. While the Poisson distribution $\nu_1$ is purely atomic
and has masses at $0$, $1$, $2$, and $3$ which are $e^{-1}$, $e^{-1}$, $e^{-1}/2$, and $e^{-1}/6$
(the atoms consist of all integer multiples), the Vandermonde histogram shows a more continuous eigenvalue distribution,
with the peaks which the Poisson distribution has at integer multiples clearly visible, although not as sharp.
We remark that the support of ${\bf V}^H {\bf V}$ for a fixed $N$ goes all the way up to $N$, but lies within $[0, N]$.
It is unknown whether the peaks at integer multiples in the Vandermonde histogram grow to infinity as we let $N\rightarrow\infty$.
From the histogram, only the peak at $0$ seems to be of atomic nature.
The effect of decreasing $c$ amounts to stretching the eigenvalue density vertically, and compressing it horizontally,
just as the case for the different Mar\u{c}henko Pastur laws.
An eigenvalue histogram for Gaussian matrices which in the limit give the corresponding (in the sense of Corollary~\ref{freekor}) Mar\u{c}henko Pastur law
for Figure~\ref{fig:640iteig1600x1200} (i.e. $\mu_{0.75}$)
is shown in Figure~\ref{fig:20iteigg1600x1200}.
Figure~\ref{2part640iteig1600x1200} shows an eigenvalue histogram in the case of a non-uniform phase distribution.
Here we have taken $640$ samples of a $1600\times 1200$
Vandermonde matrix with phase distribution with density (\ref{sindensity2}), 
with $\lambda=2d,\alpha=\frac{\pi}{4}$. 
This density, also shown in Figure~\ref{fig:sindensity}, is used in the applications of Section~\ref{firstscenario}. 
Experiments show that the eigenvalue histogram tends to flatten when the phase distribution becomes "less uniform", 
with a higher concentration of larger eigenvalues. 

\begin{figure}
  \subfigure[Realizations of $\frac{1}{N}\log_2\det\left( {\bf I}_N + \rho \frac{1}{N} {\bf X}{\bf X}^H \right)$
             when ${\bf X}$ is standard, complex, Gaussian. The asymptotic capacity (\ref{asymptoticcapacity}) is also shown.]{\epsfig{figure=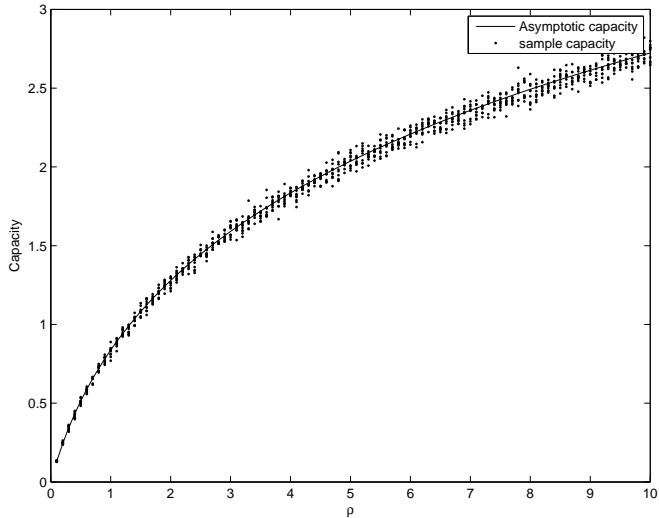,width=0.99\columnwidth}}
  \subfigure[Realizations of $\frac{1}{N}\log_2\det\left( {\bf I}_N + \rho {\bf V}{\bf V}^H \right)$ when $\omega$ has uniform phase distribution.]{\epsfig{figure=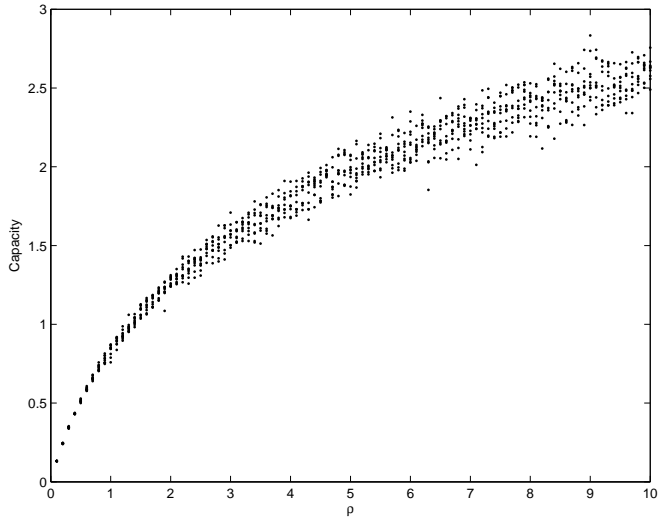,width=0.99\columnwidth}}
  \caption{Realizations of the capacity for Gaussian and Vandermonde matrices of size $36\times 36$.}\label{gvsamples}
\end{figure}

It is unknown whether the inequalities for the moments can be extended to inequalities for the associated capacity. 
If ${\bf X}$ is an $N\times N$ standard, complex, Gaussian matrix,
then an explicit expression for the asymptotic capacity
exists~\cite{book:tulinoverdu}:
\begin{equation} \label{asymptoticcapacity}
\begin{array}{l}
  \lim_{N\rightarrow\infty} \frac{1}{N} \log_2\det\left( {\bf I}_N + \rho\left( \frac{1}{N}{\bf X}{\bf X}^H \right) \right) = \\
   \hspace{1cm} 2\log_2\left( 1+\rho -\frac{1}{4}\left( \sqrt{4\rho + 1} - 1 \right)^2 \right) \\
   \hspace{1cm} - \frac{\log_2 e}{4\rho} \left( \sqrt{4\rho + 1} - 1 \right)^2.
\end{array}
\end{equation}
In Figure~\ref{gvsamples}(a), several realizations of the capacity are computed for Gaussian matrix samples of size $36\times 36$.
The asymptotic capacity (\ref{asymptoticcapacity}) is also shown.
In Figure~\ref{gvsamples}(b), several realizations of the capacity are computed for Vandermonde matrix samples of the same size,
for the case of uniform phase distribution.
It is seen that the variance of the Vandermonde capacities is higher
than for the Gaussian counterparts. This should come as no surprise,
due to the slower convergence to the asymptotic limits for
Vandermonde matrices. Although the
capacities of Vandermonde matrices with uniform phase distribution
and Gaussian matrices seem to be close, we have no proof
that the capacities of Vandermonde matrices are even finite due to the unboundedness of its support.

\subsection{Deconvolution}
Deconvolution with Vandermonde
matrices (as stated in (\ref{cumequation}) in Theorem~\ref{teo0})
differs from the Gaussian deconvolution
counterpart~\cite{book:comblect} in the sense that there is no
multiplicative~\cite{book:comblect} structure involved, since
$K_{\rho , \omega}$ is not multiplicative in $\rho$.
The Gaussian equivalent of Proposition~\ref{teo:first7moments}
(i.e. ${\bf V}^H {\bf V}$ replaced with $\frac{1}{N} {\bf X}{\bf X}^H$, with ${\bf X}$ an $L\times N$ complex, standard, Gaussian matrix) is
\begin{eqnarray}
    m_1 &=& d_1 \label{gaussianeqv1} \\
    m_2 &=& d_2 + d_1^2 \label{gaussianeqv2} \\
    m_3 &=& d_3 + 3d_2d_1 + d_1^3\\
    m_4 &=& d_4 + 4d_3d_1 + 2d_2^2 + 6d_2d_1^2 + d_1^4\\
    m_5 &=& d_5 + 5d_4d_1 + 5d_3d_2 + 10 d_3d_1^2 + \nonumber \\
        & & 10d_2^2d_1 + 10 d_2d_1^3 + d_1^5\\
    m_6 &=& d_6 + 6d_5d_1 + 6 d_4d_2 + 15 d_4d_1^2 + \nonumber \\
        & & 3 d_3^2 + 30 d_3d_2d_1 + 20 d_3d_1^3 + \nonumber \\
        & & 5 d_2^3 + 10 d_2^2d_1^2 + 15 d_2d_1^4 + d_1^6\\
    m_7 &=& d_7 + 7d_6d_1 + 7 d_5d_2 + 21 d_5d_1^2 + \nonumber \\
        & & 7 d_4d_3 + 42 d_4 d_2 d_1 + 35 d_4 d_1^3 + \nonumber \\
        & & 21 d_3^2 d_1 + 21 d_3d_2^2 + 105 d_3 d_2 d_1^2 + \nonumber \\
        & & 35 d_3d_1^4 + 35 d_2^3 d_1 + 70 d_2^2 d_1^3 + \nonumber \\
        & & 21 d_2 d_1^5 + d_1^7 \label{gaussianeqv7},
\end{eqnarray}
where the $m_i$ and the $d_i$ are computed as in (\ref{substequations1})-(\ref{substequations2}).
This follows immediately from asymptotic freeness~\cite{book:hiaipetz}, and from the fact that $\frac{1}{N} {\bf X}{\bf X}^H$
converges to the Mar\u{c}henko Pastur law $\mu_c$.
In particular, when all ${\bf D}_i(N)={\bf I}_L$ and $c=1$, we obtain the limit moments $1,2,5,14,42,132,429$,
which also were listed in Corollary~\ref{momkor}.
One can also write down Gaussian equivalents to the second order moments of Vandermonde matrices 
(\ref{vfluctuations2}) using techniques from~\cite{secondorderfreeness3}.
However the formulas look quite different, and the asymptotic behaviour is different. 
We have for instance 
\begin{equation} \label{gfluctuations1}
  \lim_{L\rightarrow\infty} L^2 D_{1,1}\left( {\bf D}(N) \frac{1}{N} {\bf X}{\bf X}^H \right) = c d_2, 
\end{equation}
where it is not needed that the matrices ${\bf D}(N)$ are diagonal.
Similarly, one can write down an equivalent to Theorem~\ref{teo:exact4moments} for the exact moments.
For the first three moments (the fourth moment is dropped, since this is more involved), these are
\begin{eqnarray*}
  m_1 &=& d_1 \\
  m_2 &=& d_2 + d_1^2 \\
  m_3 &=& \left( 1 + N^{-2} \right) d_3 + 3 d_1 d_2 + d_1^3.
\end{eqnarray*}
This follows from a careful count of all possibilities after the matrices have been multiplied together
(see also~\cite{eurecom:channelcapacity},
where one can see that the restriction that the matrices ${\bf D}_i(N)$ are diagonal can be dropped in the Gaussian case).
It is seen, contrary to Theorem~\ref{teo:exact4moments} for Vandermonde matrices,
that the second exact moment equals the second asymptotic moment (\ref{gaussianeqv2}),
and also that the convergence is faster (i.e. $O(N^{-2})$) for the third moment (this will also be the case for higher moments).

The two types of (de)convolution also differ in how they can be computed in practice.
In~\cite{eurecom:freedeconvinftheory}, an algorithm for
free convolution with the Mar\u{c}henko Pastur law was sketched. A similar algorithm may not exist
for Vandermonde convolution. However, Vandermonde
convolution can be subject to numerical approximation: To see
this, note first that Theorem~\ref{teo:generaldist} splits the
numerics into two parts: The approximation of the integrals $\int
p_{\omega}(x)^{|\rho |}dx$, and the approximation of the $K_{\rho , u}$.
A strategy for obtaining the latter quantities could be to
randomly generate many numbers between $0$ and $1$ and estimate the
volume as the ratio of the solutions which satisfy
(\ref{xequations}) in Appendix~\ref{appendixteo1}.
Implementations of the various Vandermonde convolution variants given in this paper can be found in~\cite{eurecom:vandermondeimpl}.

In practice, one often has a random matrix model where independent Gaussian and Vandermonde matrices are both present.
In such cases, it is possible to combine the individual results for both of them.
In Section~\ref{simulations}, examples on how this can be done are presented.

\section{Applications} \label{simulations}
The applications presented here all use the deconvolution framework for Vandermonde matrices.
Since additive, white, Gaussian noise also is taken into account, Vandermonde deconvolution is combined with Gaussian deconvolution.
Matlab code for running the different simulations can be found in~\cite{eurecom:vandermondeimpl}.

In the eigenvalue histograms for Vandermonde matrices shown in figures~\ref{fig:640iteig1600x1200} and~\ref{2part640iteig1600x1200},
large matrices were used in order to obtain something close to the asymptotic limit.
In practical scenarios, and in the applications we present, $N$ and $L$ are much smaller than what was used in these figures,
which partially explains the uncertainty in some of the simulations.
In particular, the uncertainty for non-uniform phase distributions such as those in Section~\ref{firstscenario} is high,
since exact expressions for the lower order moments are not known,
contrary to the case of uniform phase distribution. In all the following, $d$ is the distance between the antennas whereas $\lambda$
is the wavelength.  The ratio $\frac{d}{\lambda}$ is a figure of the resolution with which the system will be able to separate (and therefore estimate the position of) users in space.

\subsection{Detection of the number of sources} \label{firstscenario}
Let us consider a basestation equipped with $N$ receiving antennas, and with $L$ mobiles (each
with a single antenna) in the cell. The received signal at the base
station is given by
\begin{eqnarray} \label{thismodel0}
{\bf r}_i= {\bf V} {\bf P}^{\frac{1}{2}} {\bf s}_i+ {\bf n}_i.
\end{eqnarray}
Here ${\bf r}_i$ is the $N\times 1$ received vector,
${\bf s}_i$ is the $L \times 1$ transmit vector by the $L$ users which is assumed to satisfy $\E\left[ {\bf s}_i{\bf s}_i^H \right]={\bf I}_L$,
${\bf n}_i$ is $N \times 1$  additive, white, Gaussian noise of
variance $\frac{\sigma}{\sqrt{N}}$ (all components in ${\bf s}_i$ and ${\bf n}_i$ are assumed independent).
In the case of a line of sight between the users and the base station,
and considering a Uniform Linear Array (ULA),
the matrix ${\bf V}$ has the following form:
\begin{equation} \label{vandermonde2}
  {\bf  V} = \frac{1}{\sqrt{N}}
               \left( \begin{array}{lll} 1                             & \cdots & 1 \\
                                       e^{-j2 \pi \frac{d}{\lambda} \sin(\theta_1) } & \cdots & e^{-j2 \pi \frac{d}{\lambda} \sin(\theta_L) } \\
                                       \vdots                        & \ddots & \vdots \\
                                       e^{-j2 \pi (N-1) \frac{d}{\lambda} \sin(\theta_1) } & \cdots & e^{-j2 \pi \frac{d}{\lambda} \sin(\theta_L) }
                      \end{array}
               \right)
\end{equation}
Here, $\theta_i$ is the angle of the user in the cell and is
supposed to be uniformly distributed over $[-\alpha,\alpha]$. ${\bf
P}^{\frac{1}{2}}$ is an $L \times L$ diagonal power matrix due to the
different distances from which the users emit.
In other words, we assume that the phase distribution has the form $2 \pi \frac{d}{\lambda} \sin(\theta)$
with $\theta$ uniformly distributed on $[-\alpha , \alpha]$. The fact that the phase has the form $2 \pi \frac{d}{\lambda} \sin(\theta)$
is a well known result in array processing \cite{paper:krimviberg}. 
The user's distribution can be known 
(in the case of  these simulations, the uniform distribution has been accounted for without loss of generality) 
through measurements in wireless systems up to some parameters (here, $\alpha$ typically). 
This  is usually done to have a better understanding of the user's behaviour.
It is easily seen, by taking inverse functions, that the density is, when $\frac{2d\sin\alpha}{\lambda}<1$,
\begin{equation} \label{sindensity2}
  p_{\omega}(x) = \frac{1}{2\alpha\sqrt{\frac{4\pi^2d^2}{\lambda^2}-x^2}}
\end{equation}
on $[ -\frac{2\pi d\sin\alpha}{\lambda} , \frac{2\pi d\sin\alpha}{\lambda}]$, 
and $0$ elsewhere (see Figure~\ref{fig:sindensity}).

Throughout the paper we will assume, as in Figure~\ref{2part640iteig1600x1200},
that $\lambda = 2d,\alpha = \frac{\pi}{4}$ when model (\ref{vandermonde2}) is used.
With this assumption, $\frac{2d\sin\alpha}{\lambda}<1$ is always fulfilled.

The goal  is to detect the number of sources $L$ and their
respective power based on the sample covariance matrix supposing
that we have $K$ observations, of the same order as $N$. When the
number of observation is quite higher than $N$ (and the noise
variance is known), classical subspace methods \cite{paper.comon94}
provide tools to detect the number of sources. Indeed, let ${\bf R}$
be the true covariance matrix given by
\[
  {\bf V} {\bf P}{\bf V}^H+\sigma^2 {\bf I}_N,
\]
where $\sigma^2$ is the noise variance.
This matrix has $N-L$ eigenvalues equal to $\sigma^2$ and
$L$ eigenvalues strictly superior to $\sigma^2$. One can therefore
determine the number of source by counting the number of eigenvalues
different from $\sigma^2$. However, in practice, one has only access
to the sample covariance matrix given by
\[{\bf W} = \frac{1}{K} {\bf Y}{\bf Y}^H, \]
with
\begin{equation} \label{withnoiseform}
{\bf Y}=[{\bf r}_1,...{\bf r}_K]={\cal \bf V} {\bf P}^{\frac{1}{2}} [{\bf s}_1,...,{\bf s}_K]+ [{\bf n}_1,...,{\bf n}_K].
\end{equation}
If one has only the sample covariance matrix ${\bf W}$,
we have three independent parts which must be dealt
with in order to get an estimate of ${\bf P}$: the Gaussian matrices
${\bf S}=[{\bf s}_1,...,{\bf s}_K]$ and
${\bf N}=[{\bf n}_1,...,{\bf n}_K]$,
and the Vandermonde matrix ${\bf V}$.
It should thus be possible to combine Gaussian
deconvolution~\cite{eurecom:channelcapacity} and Vandermonde deconvolution by performing the
following steps:
\begin{enumerate}
  \item Estimate the moments of $\frac{1}{K} {\bf V} {\bf P}^{\frac{1}{2}} {\bf S} {\bf S}^H {\bf P}^{\frac{1}{2}} {\bf V}^H$
    using multiplicative free convolution as described in~\cite{eurecom:freedeconvinftheory}. This is the denoising part.
  \item Estimate the moments of ${\bf P} {\bf V}^H {\bf V}$, again using multiplicative free deconvolution.
  \item Estimate the moments of ${\bf P}$ using Vandermonde deconvolution as described in this paper.
\end{enumerate}
Putting these steps together, we will prove the following:
\begin{proposition} \label{propasymptotic}
  Define
  \begin{equation} \label{needthis}
    I_n = (2\pi)^{n-1} \int_0^{2\pi} p_{\omega}(x)^n dx,
  \end{equation}
  and denote the moments of ${\bf P}$ and the sample covariance matrix, respectively, by
  \begin{eqnarray*}
    P_i &=& tr_L({\bf P}^i) \\
    W_i &=& tr_N({\bf W}^i).
  \end{eqnarray*}
  Then the equations
  \begin{eqnarray*}
    W_1 &=& c_2 P_1 + \sigma^2 \\
    W_2 &=& c_2P_2  + (c_2^2 I_2 + c_2c_3)(P_1)^2 \\
        & & + 2\sigma^2(c_2+c_3)P_1 + \sigma^4(1+c_1) \\
    W_3 &=& c_2P_3 + (3 c_2^2 I_2 + 3 c_2 c_3)P_1 P_2 \\
        & & + \left( c_2^3 I_3 + 3 c_2^2 c_3 I_2 + c_2 c_3^2 \right) (P_1)^3 \\
        & & + 3\sigma^2 (1+c_1)c_2 P_2 \\
        & & + 3\sigma^2 ( (1+c_1)c_2^2 I_2 + c_3 (c_3+2c_2) ) (P_1)^2 \\
        & & + 3\sigma^4 ( c_1^2+3c_1+1)c_2 P_1 \\
        & & + \sigma^6 (c_1^2+3c_1+1)
  \end{eqnarray*}
  provide an asymptotically unbiased estimator for the moments $P_i$ from the moments of $W_i$ (or vice versa)
  when $\lim_{N\rightarrow\infty} \frac{N}{K} = c_1$, $\lim_{N\rightarrow\infty} \frac{L}{N} = c_2$, $\lim_{N\rightarrow\infty} \frac{L}{K}=c_3$.
\end{proposition}

The proof of this can be found in Appendix~\ref{mixedvandgauss}.
Note that $c_3=c_1c_2$, so that the definition of $c_3$ is really not necessary.
We still include it however, since $c_1$, $c_2$ and $c_3$ are matrix aspect ratios which represent different deconvolution stages,
so that they all are used when these stages are implemented and combined serially.
In the simulations, Proposition~\ref{propasymptotic} is put to the test when ${\bf P}$
has three sets of powers, 0.5, 1, and 1.5, with equal probability,
with phase distribution given by (\ref{vandermonde2}). Both the
number of sources and the powers are estimated. For the phase
distribution (\ref{vandermonde2}), the integrals $I_2$ and $I_3$ can
be computed exactly (for general phase distributions they are
computed numerically), and are~\cite{rottmann}
\begin{eqnarray*}
  I_2 &=& \frac{\lambda}{4d\alpha^2} \ln\left( \frac{1+\sin\alpha}{1-\sin\alpha} \right) \\
  I_3 &=& \frac{\lambda^2\tan\alpha}{4d^2\alpha^3}.
\end{eqnarray*}
Under the assumptions $\lambda = 2d, \alpha = \frac{\pi}{4}$ used throughout this paper, the integrals above take the values
\begin{eqnarray*}
  I_2 &=& \frac{40}{\pi^2} \ln\left( \frac{2+\sqrt{2}}{2-\sqrt{2}}\right) \\
  I_3 &=& \frac{1600}{\pi^3}.
\end{eqnarray*}

For estimation of the powers,
knowing that we have only three sets of powers with equal probability,
it suffices to estimate the three lowest moments in order to get an estimate of the powers
(which are the three distinct eigenvalues of ${\bf P}$).
Therefore, in the following simulations, Proposition~\ref{propasymptotic} is first used to get an estimate of the moments of ${\bf P}$.
Then these are used to obtain an estimate of the three distinct eigenvalues of ${\bf P}$ using the Newton-Girard formulas~\cite{book:programmingmath}.
These should then lie close to the three powers of ${\bf P}$.
Power estimation for the model (\ref{vandermonde2}) is shown in
the first plot of Figure~\ref{fig:part22power}. In the plot,
$K=L=N=144$, and $\sigma=\sqrt{0.1}$. 
Experiments show that when the phase distribution becomes "less" uniform, larger matrix sizes are needed in 
order for accurate power estimation using this method. 
This will also be seen when we perform power estimation using uniform phase distribution in the next section.

For estimation of the number of users $L$, we assume that the power distribution of ${\bf P}$ is known, but not $L$ itself.
Since $L$ is unknown, in the simulations we enter different candidate values of it into the following procedure:
\begin{enumerate}
  \item Computing the moments $P_i = tr_L({\bf P}^i)$ of ${\bf P}$.
  \item The moments $tr_L({\bf P}^i)$ are fed into the formulas of Proposition~\ref{propasymptotic},
    and we thus obtain candidate moments $W_i$ of the sample covariance matrix ${\bf W}$.
  \item Compute the sum of the square errors between these candidate moments, and the moments $\hat{W_i}$ of the observed sample covariance matrix ${\bf \hat{W}}$,
    i.e. compute $\sum_{i=1}^3 |W_i-\hat{W_i}|^2$.
\end{enumerate}
The estimate $L$ for the number of users is chosen as the one which gives the minimum value for the sum of square errors after these steps.

\begin{figure}
  \subfigure[$K=1$]{\epsfig{figure=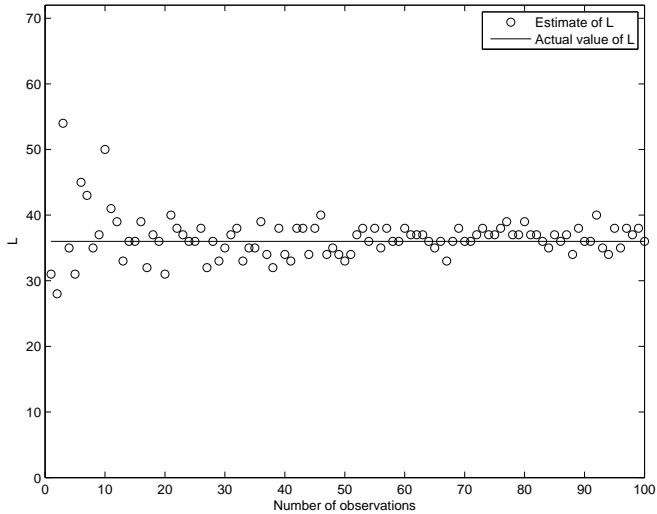,width=0.99\columnwidth}}
  \subfigure[$K=10$]{\epsfig{figure=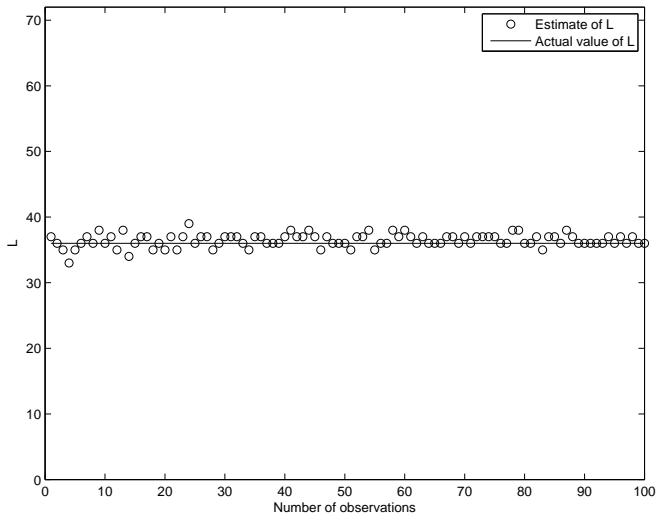,width=0.99\columnwidth}}
  \caption{Estimate for the number of users. Actual value of $L$ is $36$. Also, $\sigma =\sqrt{0.1}$, $N = 100$. The powers were $0.5$, $1$, and $1.5$, with equal probability.}\label{part22bpathssin}
\end{figure}

In Figure~\ref{part22bpathssin}, we have set $\sigma =\sqrt{0.1}$,
$N = 100$, and $L=36$. We tried the procedure described
above for $1$ all the way up to $100$ observations.
It is seen that only a small number of observations are needed in order to get an accurate estimate of $L$.
When $K=1$, it is seen that more observations are needed to get an accurate estimate of $L$, when compared to $K=10$.

\subsection{Estimation of the number of paths} \label{secondscenario}
In many channel modeling applications, one needs to determine the
number of paths in the channel \cite{book.rappaport}. For this
purpose,  consider a multi-path channel of the form:
\begin{eqnarray*}
h(\tau)=\sum_{i=1}^L s_i \delta(\tau -\tau_i)
\end{eqnarray*}
Here, $s_i$ are i.d. Gaussian random variables with power $P_i$
and $\tau_i$ are uniformly distributed delays over $[0,T]$. The $s_i$ represent the attenuation factors 
due to the different reflections. $L$ is the total number of paths.  In the frequency domain, the channel
is given by
\begin{eqnarray*}
H(f)=\sum_{i=1}^L s_i G(f) e^{-j 2 \pi f \tau_i}.
\end{eqnarray*}
Sampling the continuous frequency signal at
$f_i=i \frac{W}{N}$ where $W$ is the bandwidth, the model becomes (for a given channel realization)
$${\bf H}={\bf V} {\bf P}^{\frac{1}{2}} {\bf s}$$
where
\begin{equation} \label{vandermonde3}
  {\bf  V} = \frac{1}{\sqrt{N}}
               \left(
               \begin{array}{lll}
                 1                                                   & \cdots & 1 \\
                 e^{-j2 \pi \frac{W \tau_1}{N}  }       & \cdots & e^{-j2 \pi \frac{W \tau_L}{N}}  \\
                 \vdots                                              & \ddots & \vdots \\
                 e^{-j2 \pi (N-1) \frac{W \tau_1}{N}} & \cdots & e^{-j2 \pi (N-1) \frac{W \tau_L}{N}}
               \end{array}
               \right),
\end{equation}
We will here set $W=T=1$, which means that the
$\omega_i$  of (\ref{vandermonde}) are uniformly distributed over
$[0, 2 \pi)$. The corresponding eigenvalue histogram was shown in Figure~\ref{fig:640iteig1600x1200}. 
When additive noise (${\bf n}$) again is taken into consideration, our model again becomes that of (\ref{thismodel0}),
the only difference being that the phase distribution of the Vandermonde matrix now is uniform.
$L$ now is the number of paths,
$N$ the number of frequency samples,
and ${\bf P}$ is the unknown $L\times L$ diagonal power matrix.
Taking $K$ observations we arrive at the same form as in (\ref{withnoiseform}).
In this case with uniform phase distribution, we can do even better than Proposition~\ref{propasymptotic},
in that one can write down estimators for the moments which are unbiased for any number of observations and frequency samples:
\begin{proposition} \label{propexact}
  Assume that ${\bf V}$ has uniform phase distribution,
  and let $P_i$ be the moments of ${\bf P}$,
  and $W_i = tr_N({\bf W}^i)$
  the moments of the sample covariance matrix.
  Define also $c_1 = \frac{N}{K}$, $c_2 = \frac{L}{N}$, and $c_3 = \frac{L}{K}$.
  Then
  \begin{eqnarray*}
    E\left[ W_1 \right] &=& c_2 P_1 + \sigma^2 \\
    E\left[ W_2 \right] &=& c_2 \left( 1-\frac{1}{N} \right) P_2 + c_2(c_2+c_3)(P_1)^2 \\
                        & & + 2\sigma^2(c_2+c_3)P_1 + \sigma^4 (1+c_1) \\
    E\left[ W_3 \right] &=& c_2\left( 1+\frac{1}{K^2} \right) \left( 1-\frac{3}{N}+\frac{2}{N^2} \right)P_3 \\
                        & & + \left( 1-\frac{1}{N} \right) \left( 3 c_2^2 \left( 1+\frac{1}{K^2} \right)+3c_2c_3\right)P_1 P_2 \\
                        & & + \left( c_2^3 \left( 1+\frac{1}{K^2} \right) + 3c_2^2c_3 + c_2c_3^2 \right)(P_1)^3 \\
                        & & + 3\sigma^2 \left( (1+c_1)c_2 + \frac{c_1 c_2^2}{KL} \right) \left( 1-\frac{1}{N} \right) P_2 \\
                        & & + 3\sigma^2 \left( \frac{c_1 c_2^3}{KL} + c_2^2 + c_3^2 + 3c_2c_3 \right) (P_1)^2 \\
                        & & + 3\sigma^4 \left( c_1^2+3c_1+1+\frac{1}{K^2} \right) c_2 P_1 \\
                        & & + \sigma^6 \left( c_1^2+3c_1+1+\frac{1}{K^2} \right)
  \end{eqnarray*}
\end{proposition}

Just as Proposition~\ref{propasymptotic}, this is proved in Appendix~\ref{mixedvandgauss}.
In the following, this result is used in order to determine the number of paths as
well as the power of each path.
The different convergence rates of the approximations are clearly seen in the plots.

\begin{figure}
  \subfigure[$K=1$]{\epsfig{figure=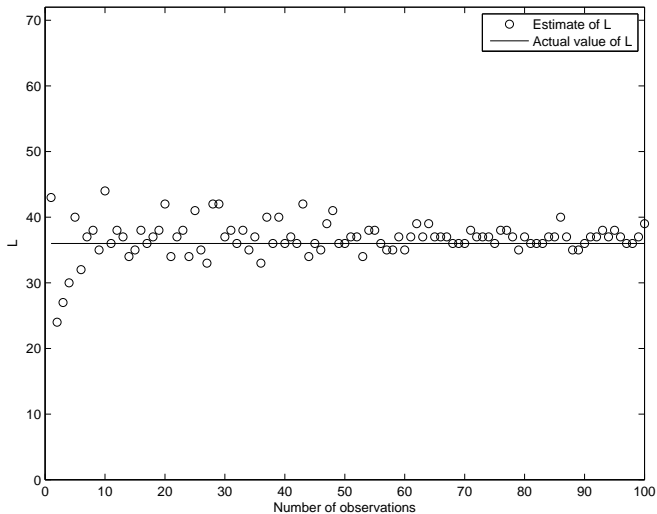,width=0.99\columnwidth}}
  \subfigure[$K=10$]{\epsfig{figure=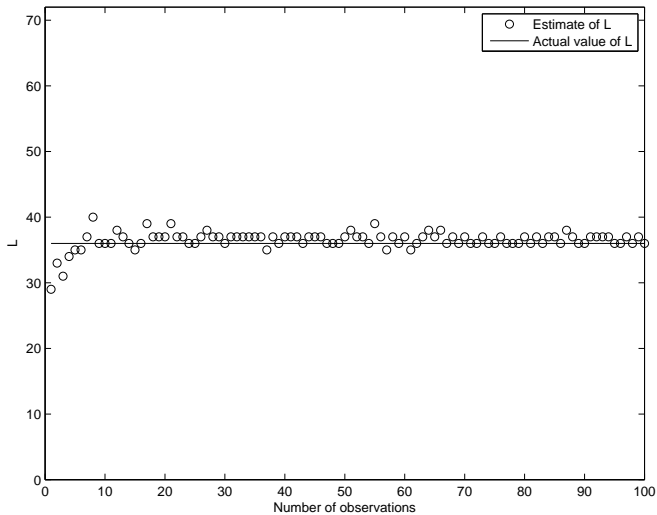,width=0.99\columnwidth}}
  \caption{Estimate for the number of paths. Actual value of $L$ is $36$. Also, $\sigma =\sqrt{0.1}$, $N = 100$.}\label{part22bpaths}
\end{figure}

In Figure~\ref{part22bpaths}, the number of paths is estimated based on
the procedure sketched above. We have set $\sigma =\sqrt{0.1}$, $N = 100$, and $L=36$.
The procedure is tried for $1$ all the way up to $100$ observations.
The plot is very similar to Figure~\ref{part22bpathssin}, in that only a small number of observations are needed in order to get an accurate estimate of $L$.
When $K=1$, it is seen that more observations are needed to get an accurate estimate of $L$, when compared to $K=10$.

\begin{figure}
  \subfigure[The model (\ref{vandermonde2}) of Section~\ref{firstscenario}.]
    {\epsfig{figure=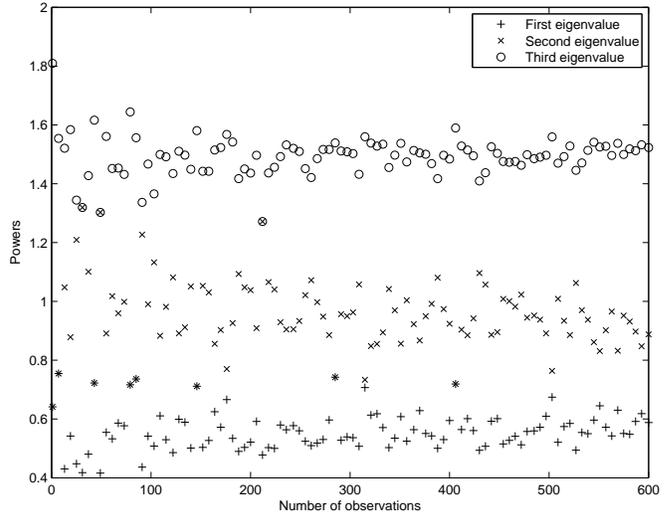,width=0.99\columnwidth}}
  \subfigure[The model (\ref{vandermonde3}) of Section~\ref{secondscenario}.]
    {\epsfig{figure=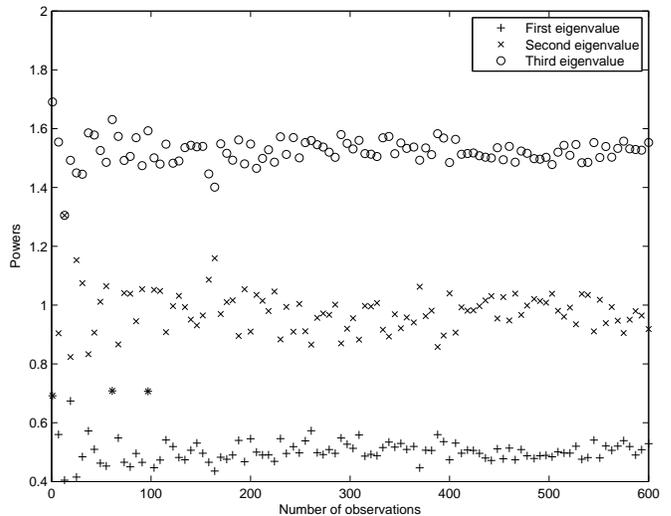,width=0.99\columnwidth}}
  \caption{Estimation of powers for the two models (\ref{vandermonde2}) and (\ref{vandermonde3}), 
           for various number of observations. $K=N=L=144$, and $\sigma=\sqrt{0.1}$. 
           The actual powers were $0.5$, $1$, and $1.5$, with equal probability.}
  \label{fig:part22power}
\end{figure}

For the estimation of powers simulation, we have set $K=N=L=144$, and $\sigma=\sqrt{0.1}$,
following the procedure also described above, up to $1000$
observations. The second plot in Figure~\ref{fig:part22power} shows the
results which confirms the usefulness of the approach. 

\subsection{Estimation of wavelength}
\begin{figure}
  \begin{center}
    \epsfig{figure=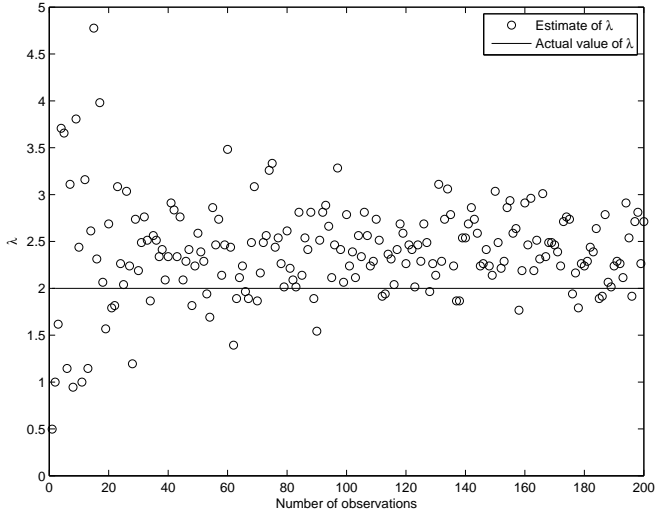,width=0.99\columnwidth}
  \end{center}
  \caption{Estimation of wavelength. Deconvolution was performed for varying number of observations, assuming different wavelengths,
           In the true model (\ref{vandermonde2}), $\lambda = 2,d=1$, $\alpha=\frac{\pi}{4}$, $K=10$, $L=36$, $N=100$, and $\sigma=\sqrt{0.1}$.} \label{fig:simwavelength}
\end{figure}

In the field of MIMO cognitive sensing
\cite{paper.cardoso08,paper.haykin05}, terminals must decide on the
band on which to transmit and in particular sense which band is
occupied. One way of doing so is to find the wavelength $\lambda$ in
(\ref{vandermonde2}), based on some realizations of the sample
covariance matrix. In our simulation, we have set $K=10$, $L=36$, $N=100$, and $\sigma=\sqrt{0.1}$, 
in addition to $\lambda = 2,d=1,\alpha = \frac{\pi}{4}$. We have
tried values between $0$ and $5$ as candidate wavelengths (to be more precise, the values $0.05,0.1,0.15,...,5$ are tried), 
and chosen the one which gives the smallest deviation (in the same sense as
above, i.e. the sum of the squared errors of the first three moments
are taken) from a different number of realizations of sample
covariance matrices. The resulting plot is shown in
Figure~\ref{fig:simwavelength}, and shows that the Vandermonde
deconvolution method can also be used for wavelength estimation. 

\subsection{Signal reconstruction and estimation of the sampling distribution} \label{section:signalreconstruction}
For signal reconstruction, one can provide a general framework where only the sampling
distribution matters asymptotically. 
The sampling distribution can be estimated with the help of the presented results. 
Several works have investigated
how irregular sampling affects the performance of signal
reconstruction in the presence of noise  in different fields, namely
sensor networks \cite{paper.ganesan04,paper.nordio2006}, image
processing \cite{paper.strohmer96,paper.early2001}, geophysics
\cite{paper.rauth97}, and compressive sampling \cite{paper.candes07}.
The usual Nyquist Theorem states that for a signal with maximum
frequency $f_{\textrm{max}}$, one needs to sample the signal at a
rate which is at least twice this number. However, in many cases,
this can not be performed, or one has an observation of a signal at
only a subset of the frequencies. Moreover, one feels that if the
signal has a sparse spectrum, one can take fewer samples and still
have the same information on the original signal. One of the central
motivations of sparse sampling is exactly to understand under which
condition one can still have less samples and recover the original
signal up to an error of $\epsilon$ \cite{paper.feichtinger95}. Let
us consider the signal of interest as a superposition of its
frequency components (this is also the case for a unidimensional
bandlimited physical signal),  i.e.
\begin{eqnarray*}
r(t)=\frac{1}{\sqrt{N}} \sum_{k=0}^{N-1} s_k e^{\frac{-j 2 \pi k
t}{N}}
\end{eqnarray*}
and suppose that the signal is sampled at various instants
$\left[t_1,...,t_L\right]$ with $t_i \in [0,1]$. This can be
identically written as
\begin{eqnarray*}
r(\omega)=\frac{1}{\sqrt{N}} \sum_{k=0}^{N-1} s_k e^{-j  k \omega},
\end{eqnarray*}
or ${\bf r} = {\bf V}^T {\bf s}$.
In the presence of noise, one can write
\begin{equation} \label{thismodelnew}
{\bf r} ={\bf V}^T {\bf s} + {\bf n},
\end{equation}
where ${\bf r} = [r(\omega_1),...r(\omega_L)]^T$,
${\bf s}$ and ${\bf n}$ are as in (\ref{thismodel0}),
and with ${\bf V}$ on the form (\ref{vandermonde}). 
A similar analysis for such cases can be found in~\cite{paper:nordio1}.

In the following, we suppose that one has $K$ observations
of the received sampled vector ${\bf r}$:
\begin{equation} \label{neweq}
{\bf Y}=[{\bf r}_1,...{\bf r}_{K}]={\cal \bf V}^T [{\bf
s}_1,...,{\bf s}_{K}]+ [{\bf n}_1,...,{\bf n}_{K}]
\end{equation}
The vector ${\bf r}$ is the discrete output of the sampled
continuous signal $r(w)$ for which the distribution is
unknown (however, $c$ is known). This case happens when one has an
observation without the knowledge of the sampling rate for example.
The difference in (\ref{neweq}) from the model (\ref{withnoiseform})
lies in that the adjoint of a Vandermonde matrix is used,
and in that there is no additional diagonal matrix ${\bf P}$ included.
The following result can now be stated and proved similarly to
Proposition~\ref{propasymptotic} and~\ref{propexact}:
\begin{proposition} \label{propsamplingdist}
\begin{eqnarray}
    E\left[ tr_n \left( {\bf W}   \right) \right] &=& 1  + \sigma^2 \label{samplingdistequations1} \\
    E\left[ tr_n \left( {\bf W}^2 \right) \right] &=& c_2 I_2 + (1+c_3)(1+\sigma^2)^2 \\
    E\left[ tr_n \left( {\bf W}^3 \right) \right] &=& 1 + 3c_2(1+c_3)I_2 \nonumber \\
                                                  & & 3c_3 + c_3^2 + c_2^2 I_3 \nonumber \\
                                                  & & 3\sigma^2(1+3c_3+c_3^2+c_2(1+c_3)I_2) \nonumber \\
                                                  & & 3\sigma^4c_2(c_3^2+3c_3+1) \nonumber \\
                                                  & & \sigma^6(c_3^2+3c_3+1) , \label{samplingdistequations3}
\end{eqnarray}
where
$\lim_{N\rightarrow\infty} \frac{N}{K} = c_1$,
$\lim_{N\rightarrow\infty} \frac{L}{N} = c_2$,
$\lim_{N\rightarrow\infty} \frac{L}{K} = c_3$,
$I_n$ is defined as in Proposition~\ref{propasymptotic},
and ${\bf W} = \frac{1}{K} {\bf Y}{\bf Y}^H$.
\end{proposition}

\begin{figure}
  \begin{center}
    \epsfig{figure=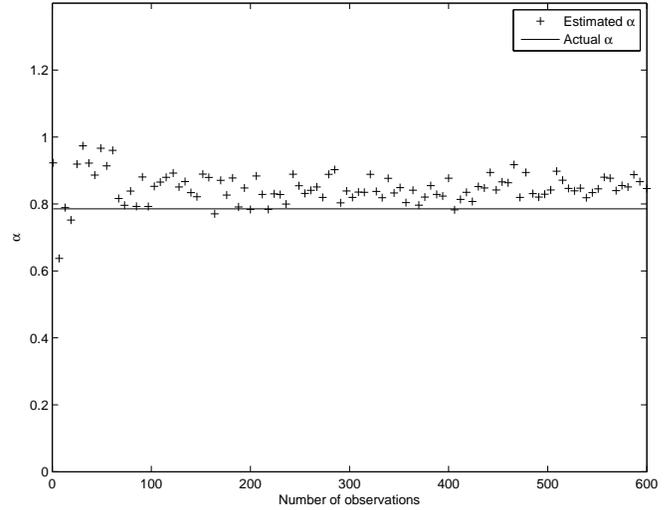,width=0.99\columnwidth}
  \end{center}
  \caption{Estimated values of $\alpha$ using (\ref{samplingdistequations1})-(\ref{samplingdistequations3}), for various number of observations, and for
    $K=10,L=36,N=100,\sigma=\sqrt{0.1}$. The actual value of $\alpha$ was $\frac{\pi}{4}$.} \label{fig:estimatedalpha}
\end{figure}

\begin{figure}
  \begin{center}
    \epsfig{figure=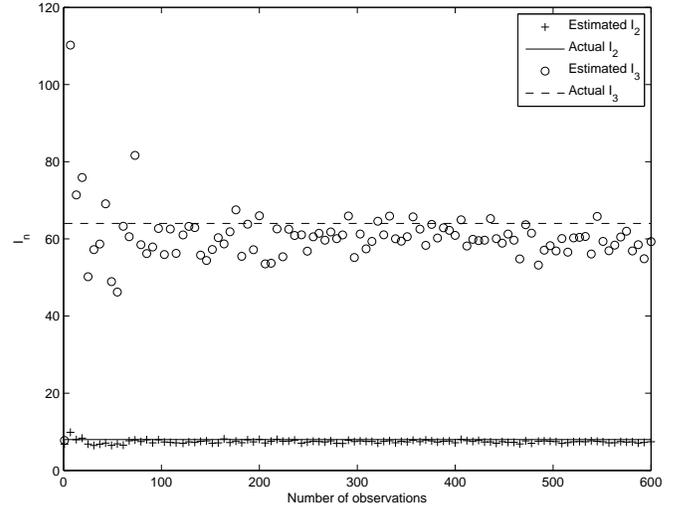,width=0.99\columnwidth}
  \end{center}
  \caption{Estimated values of $I_2$ and $I_3$ using (\ref{samplingdistequations1})-(\ref{samplingdistequations3}), for various number of observations, and for
    $K=10,L=36,N=100,\sigma=\sqrt{0.1}$. The actual value of $\alpha$ was $\frac{\pi}{4}$.} \label{fig:estimatedi2i3}
\end{figure}

The proof of Proposition~\ref{propsamplingdist} is commented in Appendix~\ref{mixedvandgauss}.
We have tested (\ref{samplingdistequations1})-(\ref{samplingdistequations3}) 
by taking a phase distribution $\omega$ which is uniform on $[0, \alpha]$, and $0$ elsewhere.
The density is thus $\frac{2\pi}{\alpha}$ on $[0, \alpha]$, and $0$ elsewhere.
In this case we can compute that
\begin{eqnarray*}
  I_2 &=& \frac{2\pi}{\alpha} \\
  I_3 &=& \left( \frac{2\pi}{\alpha} \right)^2.
\end{eqnarray*}
The first of these equations, combined with
(\ref{samplingdistequations1})-(\ref{samplingdistequations3}), enables us to estimate $\alpha$.
This is tested in Figure~\ref{fig:estimatedalpha} for various number of observations.
In Figure~\ref{fig:estimatedi2i3} we have also tested estimation of
$I_2,I_3$ from the observations using the same equations.
When one has a distribution which is not uniform, the integrals $I_3,I_4,...$
would also be needed in finding the characteristics of the
underlying phase distribution. Figure~\ref{fig:estimatedi2i3} shows
that the estimation of $I_2$ requires far fewer observations than the
estimation of $I_3$. In both figures, the values $K=10,L=36,N=100$,
and $\sigma=\sqrt{0.1}$ were used and  $\alpha$ was $\frac{\pi}{4}$.
It is seen that the estimation of $I_3$ is a bit off even for higher number of observations.
This is to be expected, since an asymptotic result is applied.

\section{Conclusion and further directions}
We have shown how asymptotic moments of random Vandermonde matrices with entries on the unit circle 
can be computed analytically, and treated many different cases.
Vandermonde matrices with uniform phase distribution proved to be
the easiest case, and it was shown
how the case with more general phases could be expressed in terms of this. 
The case where the phase
distribution has singularities was also handled, as
this case displayed different asymptotic behaviour. Also, mixed
moments of independent Vandermonde matrices were investigated, as well
as the moments of generalized Vandermonde matrices. In addition to
the general asymptotic expressions stated, exact expressions for the
first moments of Vandermonde matrices with uniform phase distribution
were also stated. We have also provided some useful applications of random
Vandermonde matrices. The applications concentrated on deconvolution and
signal sampling analysis. As shown, many useful system models use independent Vandermonde
matrices and Gaussian matrices combined in some way. The presented
examples show how random Vandermonde matrices in such
systems can be handled in practice to obtain estimates on quantities
such as the number of paths in channel modeling, the transmission
powers of the users in wireless transmission, or the sampling
distribution for signal recovery. The paper has only touched upon a
limited number of applications, but the results already provide
benchmark figures in the non-asymptotic regime.

From a theoretical perspective, it would also be interesting to find methods for obtaining 
the generalized expansion coefficients $K_{\rho,\omega,\lambda}$ from $K_{\rho,u,u}$,
similar to how we found the expansion coefficients $K_{\rho,\omega}$ from $K_{\rho,u}$.
This could also shed some light on whether uniform phase- and power distribution also minimizes moments of generalized Vandermonde matrices,
similarly to how we showed that it minimizes moments in the non-generalized case.

Throughout the paper, we assumed that only diagonal matrices were
involved in mixed moments of Vandermonde matrices. The case of
non-diagonal matrices is harder, and should be addressed
in future research. The analysis of the maximum and
minimum eigenvalue is also of importance.
The methods presented in this paper can not be used directly
to obtain explicit expressions for the p.d.f. of the asymptotic mean eigenvalue distribution,
so this is also a case for future research. A way of attacking this problem could be to develop
for Vandermonde matrices analytic counterparts to what one has in free probability, 
such as the $R$-, $S$-, and the Stieltjes transform~\cite{book:hiaipetz}. 
Interestingly, certain matrices similar to Vandermonde matrices, have analytical expressions for the moments: 
in~\cite{paper:bordenave},
analytical expressions for the moments of matrices with entries of the form $A_{i,j}=F(\omega_i-\omega_j)$ are found. 
This is interesting for the Vandermonde matrices we consider, since
\[
  \left( \frac{1}{N}{\bf V}^H{\bf V} \right)_{i,j} = \frac{\sin\left(\frac{N}{2}(\omega_i-\omega_j)\right)}{N\sin\left(\frac{1}{2}(\omega_i-\omega_j)\right)}.
\]
Unfortunately, the function $F_N(x)=\frac{\sin\left(\frac{N}{2}x\right)}{N\sin\left(\frac{1}{2}x\right)}$ depends on the matrix dimension $N$,
so that we can not find a function $F$ which fits the result from~\cite{paper:bordenave}.

Finally, another case for future research is the
asymptotic behaviour of Vandermonde matrices when the matrix entries
lie outside the unit circle.

\appendices

\section{The proof of Theorem~\ref{teo0}} \label{appendixteo0}
We can write
\begin{equation} \label{limitmoment}
  E\left[ tr_L \left( {\bf D}_1(N) {\bf V}^H {\bf V} {\bf D}_2(N) {\bf V}^H {\bf V} \cdots {\bf D}_n(N) {\bf V}^H {\bf V} \right) \right]
\end{equation}
as
\begin{equation} \label{limitmoment2}
              \begin{array}{ll}
                L^{-1} \sum_{ \stackrel{i_1,...,i_n}{j_1,...,j_n} } E( & {\bf D}_1(N)(j_1,j_1) {\bf V}^H(j_1,i_2) {\bf V}(i_2,j_2) \\
                                                                       &  {\bf D}_2(N)(j_2,j_2) {\bf V}^H(j_2,i_3) {\bf V}(i_3,j_3) \\
                                                                       & \vdots \\
                                                                       & {\bf D}_n(N)(j_n,j_n) {\bf V}^H(j_n,i_1) {\bf V}(i_1,j_1) )
              \end{array}
\end{equation}
The $(j_1,...,j_n)$ uniquely identifies a partition $\rho$ of $\{ 1,...,n\}$, 
where each block $W_j$ of $\rho$ consists of the positions of the indices which equal $j$, i.e.
\[ W_j = \{k | j_k = j\}.\]
We will also say that $(j_1,...,j_n)$ give rise to $\rho$. Write 
\[
  W_j = \{ w_{j1},w_{j2},...,w_{j|W_j |} \}.
\]
When $(j_1,...,j_n)$ give rise to $\rho$, we see that since
\[
  j_{w_{j1}}=j_{w_{j2}}= \cdots = j_{w_{j|W_j |}},
\]
we also have that
\[
  \omega_{j_{w_{j1}}}=\omega_{j_{w_{j2}}} = \cdots = \omega_{j_{w_{j|W_j |}}},
\]
and we will denote their common value by $\omega_{W_j}$ as in Definition~\ref{expansiondef}.
With this in mind, it is straightforward to verify that (\ref{limitmoment2}) can be written as
\begin{eqnarray}
  & & \sum_{\rho\in{\cal P}(n)} \sum_{(i_1,...,i_n)} \sum_{\begin{array}{c} \scriptsize (j_1,...,j_n) \\ \scriptsize\mbox{giving rise to }\rho \end{array}} \nonumber\\
  & & \hspace{1cm} N^{-n} L^{-1} \nonumber\\
  & & \hspace{1cm} \times \prod_{k=1}^{|\rho |} E \left( e^{j \left( \sum_{k\in W_j} i_{k-1} - \sum_{k\in W_j} i_k \right)  \omega_{W_k}} \right) \nonumber\\
  & & \hspace{1cm} \times {\bf D}_1(N)(j_1,j_1) \times\cdots\times {\bf D}_n(N)(j_n,j_n) \label{limitmoment3},
\end{eqnarray}
where $i_1,...,i_n$ takes values between $0$ and $N-1$. 
We will in the following switch between the form (\ref{limitmoment3}) and the form
\begin{eqnarray}
  & & \sum_{\rho\in{\cal P}(n)} \sum_{\begin{array}{c} \scriptsize (j_1,...,j_n) \\ \scriptsize\mbox{giving rise to }\rho \end{array}} \sum_{(i_1,...,i_n)} \nonumber \\
  & & \hspace{1cm} N^{|\rho |-n-1} c^{|\rho |-1} L^{-|\rho |} \nonumber \\
  & & \hspace{1cm} \times E \left( \prod_{k=1}^n \left( e^{j (\omega_{b(k-1)} -\omega_{b(k)}) i_k} \right) \right) \nonumber \\
  & & \hspace{1cm} \times {\bf D}_1(N)(j_1,j_1) \times\cdots\times {\bf D}_n(N)(j_n,j_n), \label{limitmoment4}
\end{eqnarray}
where we also have reorganized the powers of $N$ and $L$ in (\ref{limitmoment3}), 
and changed the order of summation (i.e. summed over the different $i_1,...,i_n$ first). 
Noting that 
\begin{eqnarray}
  & & \sum_{(i_1,...,i_n)} N^{|\rho |-n-1} E \left( \prod_{k=1}^n e^{j (\omega_{b(k-1)} -\omega_{b(k)}) i_k} \right) \\
  &=& N^{|\rho |-n-1} E \left( \sum_{(i_1,...,i_n)} \prod_{k=1}^n e^{j (\omega_{b(k-1)} -\omega_{b(k)}) i_k} \right) \\
  &=& N^{|\rho |-n-1} E \left( \prod_{k=1}^n \left( \sum_{i_k=0}^{N-1} e^{j (\omega_{b(k-1)} -\omega_{b(k)}) i_k} \right) \right) \label{wrongfromhere} \\
  &=& N^{|\rho |-n-1} E \left( \prod_{k=1}^n \frac{1-e^{j N (\omega_{b(k-1)} -\omega_{b(k)})}}{1-e^{j (\omega_{b(k-1)} -\omega_{b(k)})}} \right) \\
  &=& N^{|\rho |-n-1} \times \nonumber \\
  & & \int_{(0,2\pi)^{|\rho |}} \prod_{k=1}^n \frac{1-e^{j N (\omega_{b(k-1)} -\omega_{b(k)})}}{1-e^{j (\omega_{b(k-1)} -\omega_{b(k)})}} \nonumber \\
  & & d\omega_1\cdots d\omega_{|\rho |} \\
  &=& K_{\rho , \omega , N},
\end{eqnarray}
Definition~\ref{expansiondef} of the Vandermonde mixed moment expansion coefficients comes into play, 
so that (\ref{limitmoment4}) can also be written
\begin{eqnarray}
  & & \sum_{\rho\in{\cal P}(n)} \sum_{\begin{array}{c} \scriptsize (j_1,...,j_n) \\ \scriptsize\mbox{giving rise to }\rho \end{array}} \nonumber \\
  & & \hspace{1cm} c^{|\rho |-1} L^{-|\rho |} K_{\rho , \omega , N} \nonumber \\
  & & \hspace{1cm} \times {\bf D}_1(N)(j_1,j_1) \cdots\times\times {\bf D}_n(N)(j_n,j_n). \label{limitmoment5}
\end{eqnarray}
The notation for a joint limit distribution simplifies
(\ref{limitmoment4}). Indeed, add to (\ref{limitmoment4}) for each
$\rho$ the terms
\begin{eqnarray}
    & & \sum_{\rho '\in{\cal P}(n), \rho ' > \rho} \sum_{\begin{array}{c} \scriptsize (j_1,...,j_n) \\ \scriptsize\mbox{giving rise to }\rho '\end{array}} \nonumber \\
    & & \hspace{1cm} c^{|\rho |-1} L^{-|\rho |} K_{\rho , \omega , N} \nonumber \\
    & & \hspace{1cm} \times {\bf D}_1(N)(j_1,j_1) \cdots\times {\bf D}_n(N)(j_n,j_n). \label{addterms1}
\end{eqnarray}
These go to $0$ as $N\rightarrow\infty$, since they are bounded by
\[
  c^{|\rho |-1} L^{-|\rho |} K_{\rho , \omega , N} L^{| \rho '|} = K_{\rho , \omega , N} c^{|\rho |-1} L^{| \rho '| - |\rho |} = O(L^{-1}).
\]
After this addition, the limit of (\ref{limitmoment5}) can be written
\begin{equation} \label{limitform}
  \sum_{\rho\in{\cal P}(n)} c^{|\rho |-1} K_{\rho , \omega} D_{\rho},
\end{equation}
which is what we had to show.\sluttmerke

We also need to comment on the statement of Theorem~\ref{teo0gen}, where generalized Vandermonde matrices are considered. 
In this case, the derivations after (\ref{limitmoment4}) are different since the power distribution is not uniform. 
For the case of (\ref{vandermondegeneralized1}), we can in (\ref{wrongfromhere}) replace
$\sum_{i_k=1}^n e^{j (\omega_{b(k-1)} -\omega_{b(k)}) i_k}$ with 
$\sum_{r=0}^{N-1} Np_{f_N}(r) e^{j r (\omega_{b(k-1)} -\omega_{b(k)})}$, 
since the number of occurrences of the power $e^{j r (\omega_{b(k-1)} -\omega_{b(k)})}$  is $Np_{f_N}(r)$. 
The rest of the proof of Theorem~\ref{teo0gen} follows by canceling  $n$ powers of $N$ after this replacement. 
The details are similar for the case (\ref{vandermondegeneralized2}), where the law of large numbers is applied to arrive at the second formula in (\ref{kpindef2}).

\section{The proof of Proposition~\ref{teo1}} \label{appendixteo1}
Note that for each block $W_j$, 
\[
  E \left( e^{j \left( \sum_{k\in W_j} i_{k-1} - \sum_{k\in W_j} i_k \right) \omega_{W_j}} \right) = 0
\]
when
\[
  \sum_{k\in W_j} i_{k-1} \neq \sum_{k\in W_j} i_k,
\]
and $1$ if
\begin{equation} \label{iequations}
  \sum_{k\in W_j} i_{k-1} = \sum_{k\in W_j} i_k.
\end{equation}
If we denote by $S_{\rho , N}$ the set of all $n$-tuples $(i_1,...,i_n)$ ($0\leq i_k\leq N-1$, $1\leq k\leq n$) 
which solve (\ref{iequations}), and define $| S_{\rho , N} |$ to be the cardinality of $S_{\rho , N}$, it is clear that 
\[
  K_{\rho , u} = \lim_{N\rightarrow\infty} K_{\rho , u, N} = \lim_{N\rightarrow\infty} \frac{1}{N^{n+1-|\rho |}} | S_{\rho , N} |.
\]
It is straightforward to show that the solution set of (\ref{iequations}) has $n+1-|\rho |$ free variables. 
After dividing the equations (\ref{iequations}) by $N$ and letting $N$ go to infinity, $K_{\rho , u}$ can thus alternatively be 
expressed as the volume in $\R^{n+1-|\rho |}$ of the solution set of
\begin{equation} \label{xequations}
  \sum_{k\in W_j} x_{k-1} = \sum_{k\in W_j} x_k, 
\end{equation}
with $0\leq x_k \leq 1$. It is clear that the volume of this solution set computes to a rational number. 
It is the form (\ref{xequations}) which will be used in the other appendices to compute $K_{\rho , u}$ for certain lower order $\rho$. 
Appendix D of~\cite{paper:nordio1} states the same equations for finding quantities equivalent 
to Vandermonde mixed moment expansion coefficients for the uniform phase distribution. 
The fact that $K_{\rho , u} \leq 1$ follows directly from Appendix D of~\cite{paper:nordio1}.
The same applies for the fact that $K_{\rho,u}=1$ if and only if $\rho$ is noncrossing. 

For any $\rho$, we can define a partition of $\{ 1,...,n\}$ into $n+1-|\rho |$ blocks, where two elements are defined to be 
in the same block if and only if the corresponding variables in solutions to (\ref{xequations}) are linearly dependent. 
When $\rho$ is noncrossing, it is straightforward to show that two such variables are dependent if and only if they are equal, 
and also that this partition is the Kreweras complement $K(\rho)$ of $\rho$. 
This fact is used elsewhere in this paper. 

We will also briefly explain why the computations in this appendix are useful for generalized Vandermonde matrices with uniform phase distribution. 
For (\ref{vandermondegeneralized1}), the number of solutions $i_1,...,i_k$ to (\ref{iequations}) needs to be multiplied by 
\[
  Np_{f_N}(i_1) \cdots Np_{f_N}(i_k),
\]
since each $i_j$ now may occur $Np_{f_N}(i_j)$ times.
This means that $K_{\rho,\omega,f}$ can be computed as the integrals in this appendix, but that we also need to multiply with the density $p_f$ for each variable. 
The computations of these new integrals become rather involved when $f$ is not uniform, and are therefore dropped.

\section{The proof for Proposition~\ref{lemma:kcompute}} \label{appendixkcompute}
We will in the following compute the volume of the solution set of (\ref{xequations}), 
as a volume in $[0,1]^{n+1-|\rho |} \subset \R^{n+1-|\rho |}$, as explained in the proof of Proposition~\ref{teo1}. 
These integrals are very tedious to compute, and many of the details are skipped. 
The formula 
\[
  \frac{r!s!}{(r+s+1)!} = \int_0^1 x^r (1-x)^s dx
\]
can be used to simplify some of the calculations for higher values of $n$. 

\subsection{Computation of $K_{ \{ \{ 1,3\} , \{ 2,4\} \} , u}$}    
This is equivalent to finding the volume of the solution set of
\[
  x_1 + x_3 = x_2 + x_4
\]
in $\R^3$. Since this means that
\[
  x_4 = x_1 + x_3 - x_2 \mbox{ lies between $0$ and $1$,}
\]
we can set up the following integral bounds: 
When $x_1+x_3 \leq 1$, we must have that $0\leq x_2 \leq x_1+x_3$, so that we get the contribution
\[
  \int_0^1 \int_0^{1-x_1} \int_0^{x_1+x_3} dx_2 dx_3 dx_1 , 
\]
which computes to $\frac{1}{3}$. 
When $1\leq x_1+x_3$, we must have that $x_1 + x_3 - 1 \leq x_2 \leq 1$, so that we get the contribution
\[
  \int_0^1 \int_{1-x_1}^1 \int_{x_1+x_3-1}^1 dx_2 dx_3 dx_1 ,
\]
which also computes to $\frac{1}{3}$. 
Adding the contributions together we get $\frac{2}{3}$, 
which is the stated value for $K_{ \{ \{ 1,3\} , \{ 2,4\} \} , u}$. 

It turns out that when the blocks of $\rho$ are cyclic shifts of each other, the computation of $K_{\rho , u}$ can be simplified. 
Examples of such $\rho$ are $\{ \{ 1,3\} , \{ 2,4\} \}$ (for which we just computed $K_{\rho , u}$), 
$\{ \{ 1,3,5\} , \{ 2,4,6\} \}$, and $\{ \{ 1,4\} , \{ 2,5\} , \{ 3,6\} \}$. 
We will in the following describe this simplified computation. 
Let $a_l^{(m)}(x)$ be the polynomial which gives the volume in $\R^{m-1}$ of the solutions set to $x_1+\cdots +x_m = x$ 
(constrained to $0\leq x_i\leq 1$) for $l\leq x\leq l+1$. 
It is clear that these satisfy the integral equations
\begin{equation} \label{integralequations}
  a_l^{(m+1)}(x) = \int_{x-1}^l a_{l-1}^{(m)}(t)dt + \int_l^x a_l^{(m)}(t)dt, 
\end{equation}
which can be used to compute the $a_l^m(x)$ recursively. 
Note first that $a_0^{(1)}(x) = 1$. For $m=2$ we have
\begin{eqnarray*}
  a_0^{(2)}(x) &=& \int_0^x a_0^{(1)}(t)dt = x \\
  a_1^{(2)}(x) &=& \int_{x-1}^1 a_0^{(1)}(t)dt = 2-x.
\end{eqnarray*}
For $m=3$ we have 
\begin{eqnarray*}
  a_0^{(3)}(x) &=& \int_0^x a_0^{(2)}(t) dt = \frac{1}{2} x^2 \\
  a_1^{(3)}(x) &=& \int_{x-1}^1 a_0^{(2)}(t) dt + \int_1^x a_1^{(2)}(t) dt \\
               &=& 1 - \frac{1}{2} (x-1)^2 - \frac{1}{2} (2-x)^2 \\ 
  a_2^{(3)}(x) &=& \int_{x-1}^2 a_1^{(2)}(t) dt = \frac{1}{2} (3-x)^2. 
\end{eqnarray*}

\subsection{Computation of $K_{ \{ \{ 1,3,5\} , \{ 2,4,6\} \} , u}$}
For $m=3$, integration gives
\[
  \int_0^1 (a_0^{(3)})^2(t)dt + \int_1^2 (a_1^{(3)})^2(t)dt + \int_2^3 (a_2^{(3)})^2(t)dt ,
\]
which computes to $\frac{11}{20}$. This is the stated expression for $K_{ \{ \{ 1,3,5\} , \{ 2,4,6\} \} , u}$. 

\subsection{Computation of $K_{ \{ \{ 1,4\} , \{ 2,5\} , \{ 3,6\} \} , u}$} 
This is equivalent to finding the volume of the solution set of
\[
  x_1 + x_4 = x_2 + x_5 = x_3 + x_6
\]
in $\R^4$, which is computed as 
\[
  \int_0^1 (a_0^{(2)})^3(t)dt + \int_1^2 (a_1^{(2)})^3(t)dt ,
\]
which computes to $\frac{1}{2}$. 
This is the stated expression for $K_{ \{ \{ 1,4\} , \{ 2,5\} , \{ 3,6\} \} , u}$. 

\subsection{Computation of $K_{ \{ \{ 1,4\} , \{ 2,6\} , \{ 3,5\} \} , u}$}
This is equivalent to finding the volume of the solution set of 
\begin{eqnarray*}
  x_1 + x_4 &=& x_2 + x_5 \\
  x_2 + x_6 &=& x_3 + x_1
\end{eqnarray*}
in $\R^4$. Since this means that 
\begin{eqnarray*}
  x_5 &=& x_1-x_2+x_4 \mbox{ lies between $0$ and $1$, }\\
  x_6 &=& x_1-x_2+x_3 \mbox{ lies between $0$ and $1$, }
\end{eqnarray*}
we can set up the following integral bounds: 

For $x_2\geq x_1$ we must have $x_2-x_1\leq x_3,x_4\leq 1$, so that we get the contribution
\[
  \int_0^1 \int_{x_1}^1 \int_{x_2-x_1}^1 \int_{x_2-x_1}^1 dx_4dx_3dx_2dx_1 ,
\]
which computes to $\frac{1}{4}$. 
It is clear that for $x_1\geq x_2$ we get the same result by symmetry, 
so that the total contribution is $\frac{1}{4} + \frac{1}{4} = \frac{1}{2}$, which proves the claim.

\subsection{Computation of $K_{ \{ \{ 1,5\} , \{ 3,7\} , \{ 2,4,6\} \} , u}$} 
This is equivalent to finding the volume of the solution set of 
\begin{eqnarray*}
  x_1 + x_5 &=& x_2 + x_6 \\
  x_3 + x_7 &=& x_4 + x_1
\end{eqnarray*}
in $\R^5$, or 
\begin{equation} \label{twoequations}
\begin{array}{lll}
  x_6 &=& x_5 + x_1 - x_2 \mbox{ lies between $0$ and $1$, }\\
  x_7 &=& x_4 + x_1 - x_3 \mbox{ lies between $0$ and $1$  }.
\end{array}
\end{equation}
This can be split into the following volumes:
\begin{enumerate}
  \item $x_1\leq x_2 \leq x_3$, 
  \item $x_1\leq x_3 \leq x_2$, 
  \item $x_3\leq x_2 \leq x_1$, 
  \item $x_2\leq x_3 \leq x_1$, 
  \item $x_2\leq x_1 \leq x_3$, 
  \item $x_3\leq x_1 \leq x_2$.
\end{enumerate}
Each of these volumes can be computed by setting up an integral with corresponding bounds. 
Computing these integrals, we get the values $\frac{1}{15}$, $\frac{1}{15}$, $\frac{1}{15}$, $\frac{1}{15}$, $\frac{11}{120}$, $\frac{11}{120}$, respectively. 
Adding these contributions together, we get 
\[
  \frac{4}{15} + \frac{11}{60} = \frac{27}{60} = \frac{9}{20},
\]
which proves the claim.

\subsection{The computation of $K_{ \{ \{ 1,6\} , \{ 2,4\} , \{ 3,5,7\} \} , u}$}
This is equivalent to finding the volume of the solution set of 
\begin{eqnarray*}
  x_1 + x_6 &=& x_2 + x_7 \\
  x_2 + x_4 &=& x_3 + x_5
\end{eqnarray*}
in $\R^5$, or 
\begin{eqnarray*}
  x_6 &=& x_7 + x_2 - x_1 \mbox{ lies between $0$ and $1$, }\\
  x_5 &=& x_4 + x_2 - x_3 \mbox{ lies between $0$ and $1$, }.
\end{eqnarray*}
This can be obtained from (\ref{twoequations}) by a permutation of the variables, 
so the contribution from $K_{ \{ \{ 1,6\} , \{ 2,4\} , \{ 3,5,7\} \} , u}$ must also be $\frac{9}{20}$, which proves the claim.

\section{The proof for Proposition~\ref{teo:first7moments}} \label{appendixkcounting}
Note first that multiplying both of sides of (\ref{cumequation}) with $c$ gives 
\begin{equation} \label{cumulantform}
  cM_n = \sum_{\rho\in{\cal P}(n)} K_{\rho , \omega} (cD)_{\rho},
\end{equation}
where we now can substitute the scaled moments (\ref{substequations1})-(\ref{substequations2}).
With ${\bf D}_1(N) = {\bf D}_2(N) = \cdots = {\bf D}_n(N)={\bf D}(N)$,
$D_{\rho}$ as defined in Definition~\ref{ddef} does only depend on the block cardinalities $|W_j|$,
so that we can group together the $K_{\rho , \omega}$ for $\rho$ with equal block cardinalities.
If we group the blocks of $\rho$ so that their cardinalities are in descending order, and
set
\[
  {\cal P}(n)_{r_1,r_2,...,r_k} =  \{ \rho = \{ W_1,...,W_k \} \in{\cal P}(n) | |W_i|=r_i\forall i \},
\]
where $r_1\geq r_2 \geq \cdots \geq r_k$,
and also write
\begin{equation} \label{cumform0}
  K_{r_1,r_2,...,r_k} = \sum_{\rho\in{\cal P}(n)_{r_1,r_2,...,r_k}} K_{\rho , \omega},
\end{equation}
(\ref{cumulantform}) can be written
\begin{equation} \label{cumulantform3}
  m_n = \sum_{\stackrel{r_1,...,r_k}{r_1+\cdots + r_k = n}} K_{r_1,r_2,...,r_k} \prod_{j=1}^k d_{r_j}.
\end{equation}
For the first $5$ moments this becomes
\begin{eqnarray}
  m_1 &=& K_1 d_1  \label{regressionform1} \\
  m_2 &=& K_2 d_2 + K_{1,1} d_1^2 \\
  m_3 &=& K_3 d_3 + K_{2,1} d_2 d_1^2 + K_{1,1,1} d_1^3 \\
  m_4 &=& K_4 d_4 + K_{3,1} d_3 d_1   + K_{2,2} d_2^2 + K_{2,1,1} d_2 d_1^2 + \nonumber \\
      & & K_{1,1,1,1} d_1^4 \\
  m_5 &=& K_5 d_5 + K_{4,1} d_4 d_1 + + K_{3,2} d_3 d_2 + \nonumber\\
      & & K_{3,1,1} d_3d_1^2 + K_{2,2,1} d_2^2d_1 + K_{2,1,1,1} d_2d_1^3 + \nonumber \\
      & & K_{1,1,1,1,1} d_1^5  \label{regressionform5}.
\end{eqnarray}
Thus, to prove Proposition~\ref{teo:first7moments}, we have to compute the $K_{r_1,r_2,...,r_k}$ by going through all partitions.  
We will have use for the following result, taken from~\cite{book:comblect}: 
\begin{lemma} \label{nccount}
  The number of noncrossing partitions in $NC(n)$ with $r_1$ blocks of length $1$, $r_2$ blocks of length $2$ and so on 
  (so that $r_1 + 2r_2+3r_3 +\cdots nr_n= n$) is 
  \[
    \frac{n!}{r_1!r_2!\cdots r_n! (n+1-r_1-r_2\cdots r_n)!}.
  \]
\end{lemma}
Using this and a similar formula for the number of partitions with prescribed block sizes, 
we obtain cardinalities for noncrossing partitions and the set of all partitions with a given block structure. 
These numbers are the used in the following calculations.  
For the proof of Proposition~\ref{teo:first7moments}, we need to compute (\ref{cumform0}) for all possible block cardinalities $(r_1,...,r_k)$, 
and insert these in (\ref{regressionform1})-(\ref{regressionform5}). 
The formulas for the three first moments are obvious, since all partitions of length $\leq 3$ are noncrossing. 
For the remaining computations, the following two observations save a lot of work: 
\begin{itemize}
  \item If $\rho_1\in {\cal P}(n_1)$, $\rho_2\in {\cal P}(n_2)$ with $n_1 < n_2$, and $\rho_1$ can be obtained from $\rho_2$ by omitting elements 
    $k$ in $\{ 1,...,n_2\}$ such that $k$ and $k+1$ are in the same block, then we must have that $K_{\rho_1 , u} = K_{\rho_2 , u}$. 
    This is straightforward to prove since it follows from the proof of Proposition~\ref{teo1} that $i_{k+1}$ can be chosen arbitrarily between $0$ and $N-1$ in such a case. 
  \item $K_{\rho_1 , u} = K_{\rho_2 , u}$ if the set of equations (\ref{xequations}) for $\rho_1$ can be obtained 
    by a permutation of the variables in the set of equations for $\rho_2$. 
    Since the rank of the matrix for (\ref{xequations}) equals the number of equations $-1$, we actually need only have that $|\rho_1|-1$ of the $|\rho_1 |$ 
    equations can be obtained from permutation of $|\rho_2 | - 1$ equations of the $|\rho_2 |$ equations in the equation system for $\rho_2$.
\end{itemize}

\subsection{The moment of fourth order}
The result is here obvious except for the case for the three partitions with block cardinalities $(2,2)$ 
(for all other block cardinalities, all partitions are noncrossing, so that $K_{r_1,r_2,...,r_k}$ is simply the number of noncrossing partitions 
with block cardinalities $(r_1,...,r_k)$. this number can be computed from Lemma~\ref{nccount}). 
Two of the partitions with blocks of cardinality $(2,2)$ are noncrossing, the third one is not. 
We see from Proposition~\ref{lemma:kcompute} that the total contribution is 
\[
\begin{array}{lll}
  K_{2,2} &=& 2+K_{ \{ \{ 1,3\} , \{ 2,4\} \} , u} \\
          &=& 2+\frac{2}{3} = \frac{8}{3}. 
\end{array}
\]
The formula for the fourth moment follows.

\subsection{The moment of fifth order}                
Here two cases require extra attention: 

\subsubsection{$\rho = \{ W_1 , W_2 \}$ with $| W_1 | = 3$, $| W_2 | =2$}
There are $10$ such partitions, and $5$ of them have crossings and contribute with $K_{ \{ \{ 1,3\} , \{ 2,4\} \} , u}$. 
The total contribution is therefore 
\[
\begin{array}{ll}
   & 5 + 5\times K_{ \{ \{ 1,3\} , \{ 2,4\} \} , u} \\
  =& 5 + 5\times \frac{2}{3} = \frac{25}{3}.
\end{array}
\]

\subsubsection{$\rho = \{ W_1 , W_2 , W_3 \}$ with $| W_1 | = | W_2 | =2$, $| W_3 | = 1$}
There are $15$ such partitions, of which $5$ have crossings. 
The total contribution is therefore 
\[
\begin{array}{ll}
   & 10 + 5\times K_{ \{ \{ 1,3\} , \{ 2,4\} \} , u} \\
  =& 10 + 5\times \frac{2}{3} = \frac{40}{3}.
\end{array}
\]
        
The computations for the sixth and seventh order moments are similar, but the details are skipped. 
These are more tedious in the sense that one has to count the number of partitions with a given block structure, 
and identify each partition with one of the coefficients listed in Proposition~\ref{lemma:kcompute}.

\section{The proof of Proposition~\ref{secondorderexps}} \label{appendixsecondorderexps}
$C_{i,j}({\bf D}(N) {\bf V}^H {\bf V})$ is computed as in Appendix~\ref{appendixteo0}. 
Since some terms in $E\left[ tr_L \left( {\bf A}^i \right) tr_L \left( {\bf A}^j \right) \right]$ cancel those in 
$E\left[ tr_L \left( {\bf A}^i \right) \right] E\left[ tr_L \left( {\bf A}^j \right) \right]$, 
we can restrict to summing over partitions of $1,2,...,i+j$ where at least one block contains elements from both $[1,...,i]$ and $[i+1,...,i+j]$. 
We denote this set by ${\cal P}(i,j)$, and set $n=i+j$. 
In our new calculations,(\ref{limitmoment4}) now instead takes the form
\begin{eqnarray}
  & & L\sum_{\rho\in{\cal P}(i,j)} \sum_{\begin{array}{c} \scriptsize (j_1,...,j_n) \\ \scriptsize\mbox{giving rise to }\rho \end{array}} \sum_{(i_1,...,i_n)} \nonumber \\
  & & \hspace{1cm} N^{|\rho |-i-j-1} L^{-1} c^{|\rho |-1} L^{-|\rho |} \nonumber \\
  & & \hspace{1cm} \times E \left( \prod_{k=1}^n \left( e^{j (\omega_{b(k-1)} -\omega_{b(k)}) i_k} \right) \right) \nonumber \\
  & & \hspace{1cm} \times {\bf D}_1(N)(j_1,j_1) \times\cdots\times {\bf D}_n(N)(j_n,j_n), \label{limitmoment4new}
\end{eqnarray}
where the normalizing factor $L$ from Definition~\ref{covtwotraces} has been included. 
Simplifying this as in Appendix~\ref{appendixteo0}, and restricting to uniform phase distribution, we obtain 
\[
  \lim_{L\rightarrow\infty} LC_{i,j}({\bf D}(N) {\bf V}^H {\bf V}) = \sum_{\rho\in{\cal P}(i,j)} c^{|\rho |-1} K_{2,\rho,u} D_{\rho},
\]
where $K_{2,\rho,u}$ is the volume of the solution set of 
\begin{equation} \label{xequations2}
  \sum_{k\in W_j} x_{\sigma^{-1}(k-1)} = \sum_{k\in W_j} x_k, 
\end{equation}
where $\sigma$ is the permutation which shifts $[1,i]$ and $[i+1,...,i+j]$ to the right cyclically so that the result is contained within the same interval. 
Thus, when the normalizing factor $L$ is included, we see that the second order moments exist. 

$C_{2,2}({\bf D}(N) {\bf V}^H {\bf V})$ in (\ref{vfluctuations2}) is computed by noting that 
$K_{2,\{ \{ 1,3 \} , \{ 2,4\} \},u}$ and $K_{2,\{ \{ 1,4 \} , \{ 2,3\} \},u}$ both equal $\frac{2}{3}$, 
and that there are $9$ other partitions in ${\cal P}(2,2)$, and $K_{2,\pi,u}=1$ for all these $\pi$ 
(all these values are computed as in Appendix~\ref{appendixkcompute}). 
By adding up for the different block cardinalities we get that 
\[
  c\lim_{L\rightarrow\infty} L C_{2,2}({\bf D}(N) {\bf V}^H {\bf V}) = d_4 + 4d_3d_1  \frac{4}{3}d_2^2 + 4d_2d_1^2,
\]
and using the substitution (\ref{substequations3}) we arrive at the desired result.

\section{The proof of Theorem~\ref{teo:exact4moments}} \label{appendixteoexactdist}
In order to get the exact expressions in Theorem~\ref{teo:exact4moments}, 
we now need to keep track of the $K_{\rho , u , N}$ defined by (\ref{kpindef}), not only the limits $K_{\rho , u}$ 
(if we had not assumed $\omega = u$, the calculations for $K_{\rho , \omega , N}$ would be much more cumbersome). 
When $\rho$ is a partition of $\{ 1,...,n \}$ and $n\leq 4$, 
we have that $K_{\rho , u , N} = K_{\rho , u} = 1$ when $\rho\neq \{ \{ 1,3 \} , \{ 2,4 \} \}$. 
We also have that 
\begin{equation} \label{exactkformula}
  K_{ \{ \{ 1,3 \} , \{ 2,4 \} \} , u , N} = \frac{2}{3} + \frac{1}{3N^2},
\end{equation}
where we have used that $\sum_{i=1}^N i^2 = \frac{N}{3} (N+1)(N+\frac{1}{2})$~\cite{rottmann}. 
We also need the exact expression for the quantity
\[
T_{\rho} =\sum_{\stackrel{(j_1,...,j_n)}{\mbox{giving rise to }\rho}} L^{-|\rho |} {\bf D}_1(N)(j_1,j_1) \times\cdots\times {\bf D}_n(N)(j_n,j_n)
\]
from (\ref{limitmoment5}) (i.e. we can not add (\ref{addterms1}) to obtain the approximation (\ref{limitform}) here). 
Setting $D_n^{(N,L)} =  tr_L\left( {\bf D}^n(N) \right)$, and $D_{\rho}^{(N,L)} = \prod_{i=1}^k D_{W_i}^{(N,L)}$, 
we see that 
\begin{equation} \label{recursive}
  T_{\rho} = D_{\rho}^{(N,L)} - \sum_{\rho ' > \rho} L^{|\rho '| - |\rho |} T_{\rho '} ,
\end{equation}
which can be used recursively to express the $T_{\rho}$ in terms of the $D_{\rho}^{(N,L)}$. 
We obtain the following formulas for $n=4$:
\begin{eqnarray}
  T_{ \{ \{ 1,2,3,4 \} \} } &=& D_4^{(N,L)} \label{formula1} \\
  T_{ \{ \{ 1,2,3 \} , \{ 4 \} \} } &=& D_3^{(N,L)} D_1^{(N,L)} - L^{-1} D_4^{(N,L)} \\
  T_{ \{ \{ 1,2 \} , \{ 3,4 \} \} } &=& (D_2^{(N,L)})^2 - L^{-1} D_4^{(N,L)} \\
  T_{ \{ \{ 1,2 \} , \{ 3 \} , \{ 4 \} \} } &=& D_2^{(N,L)} (D_1^{(N,L)})^2 \nonumber \\
                                            & & - 2 L^{-1} ( D_3^{(N,L)} D_1^{(N,L)} \nonumber \\
                                            & & - L^{-1} D_4^{(N,L)} ) \nonumber \\
                                            & & - L^{-1} \left( (D_2^{(N,L)})^2 - L^{-1} D_4^{(N,L)} \right) \nonumber \\
                                            & & - L^{-2} D_4^{(N,L)} \nonumber \\
  &=& D_2^{(N,L)} (D_1^{(N,L)})^2 \nonumber \\
  & & - L^{-1} (D_2^{(N,L)})^2 \nonumber \\
  & & - 2 L^{-1} D_3^{(N,L)} D_1^{(N,L)} \nonumber \\
  & & + 2 L^{-2} D_4^{(N,L)} \\
  T_{ \{ \{ 1 \} , \{ 2 \} , \{ 3 \} , \{ 4 \} \} } &=& (D_1^{(N,L)})^4 \nonumber \\
                                                    & & - 6 L^{-1} ( D_2^{(N,L)} (D_1^{(N,L)})^2 \nonumber \\
                                                    & & - L^{-1} (D_2^{(N,L)})^2 \nonumber \\
                                                    & & - 2 L^{-1} D_3^{(N,L)} D_1^{(N,L)} \nonumber \\
                                                    & & + 2 L^{-2} D_4^{(N,L)} ) \nonumber \\
                                                    & & - 3 L^{-2} (D_2^{(N,L)})^2 + 3 L^{-3} D_4^{(N,L)} \nonumber \\
                                                    & & - 4 L^{-2} D_3^{(N,L)} D_1^{(N,L)} \nonumber \\
                                                    & & + 4 L^{-3} D_4^{(N,L)} - L^{-3} D_4^{(N,L)} \nonumber \\
  &=& - 6L^{-3} D_4^{(N,L)} \nonumber \\
  & & + L^{-2} ( 8 D_3^{(N,L)} D_1^{(N,L)} \nonumber \\
  & & + 3 (D_2^{(N,L)})^2 ) \nonumber \\
  & & - 6 L^{-1} D_2^{(N,L)} (D_1^{(N,L)})^2 + \nonumber \\
  & & (D_1^{(N,L)})^4. \label{formula5}
\end{eqnarray}
For $n=3$ and $n=2$ the formulas are 
\begin{eqnarray}
  T_{ \{ \{ 1,2,3 \} \} }                 &=& D_3^{(N,L)} \label{newformula1} \\
  T_{ \{ \{ 1,2 \} , \{ 3 \} \} }         &=& D_1^{(N,L)} D_2^{(N,L)} - L^{-1} D_3^{(N,L)} \\
  T_{ \{ \{ 1 \} , \{ 2 \} , \{ 3 \} \} } &=& (D_1^{(N,L)})^3 - 3 L^{-1} D_1^{(N,L)} D_2^{(N,L)} \nonumber \\
                                          & & +2 L^{-2} D_3^{(N,L)} \\
  T_{ \{ \{ 1,2 \} \} }                   &=& D_2^{(N,L)} \\
  T_{ \{ \{ 1 \} , \{ 2 \} \} }           &=& (D_1^{(N,L)})^2 - L^{-1} D_2^{(N,L)}. \label{newformula5}
\end{eqnarray}
It is clear that (\ref{formula1})-(\ref{formula5}) and (\ref{newformula1})-(\ref{newformula5}) cover all possibilities when it comes to partition block sizes. 
Using (\ref{substequations1})-(\ref{substequations2}), and putting (\ref{exactkformula}), (\ref{formula1})-(\ref{formula5}), and (\ref{newformula1})-(\ref{newformula5}) 
into (\ref{limitmoment5}) we get the expressions in Theorem~\ref{teo:exact4moments} after some calculations.

If we are only interested in first order approximations rather than exact expressions, (\ref{recursive}) gives us 
\[
  T_{\rho} \approx D_{\rho} - \sum_{ \stackrel{\rho ' > \rho}{|\rho | - |\rho '| = 1} } L^{-1} D_{\rho '}, 
\]
which is easier to compute. 
Also, we need only first order approximations to $K_{\rho , u , N}$, which is much easier to compute than the exact expression. 
For (\ref{exactkformula}), 
\[
  K_{ \{ \{ 1,3 \} , \{ 2,4 \} \} , u , N} \approx \frac{2}{3} 
\]
is already a first order approximation. 
Inserting the approximations in (\ref{limitmoment5}) gives a first order approximation of the moments.

\section{The proof of Proposition~\ref{propunbounded}} \label{appendixpropunbounded}
We only state the proof for the case $c=1$. 
In~\cite{paper:brycdembojiang} it is stated that the asymptotic $2n$-moment ($m_{2n}$) of certain Hankel and Toeplitz matrices can be expressed in 
terms of volumes of solution sets of equations on the form (\ref{xequations}), with $\rho$ restricted to partitions with all blocks of length $2$. 
Rephrased in our language of Vandermonde mixed moment expansion coefficients, this means that
\begin{equation} \label{neweq2}
  m_{2n} = \sum_{\begin{array}{c} \scriptsize \rho\in{\cal P}(2n) \\ \scriptsize \rho\mbox{ has two elements in each block} \end{array}} K_{\rho,u}
\end{equation}
In the language of~\cite{paper:brycdembojiang}, the formula is not stated exactly like this, but rather in terms of 
volumes of solution sets of equations of the form (\ref{xequations}). This translates to (\ref{neweq2}), 
since we in Appendix~\ref{appendixteo1} interpreted $K_{\rho,u}$ as such volumes. 
In Proposition A.1 in~\cite{paper:brycdembojiang}, unbounded support was proved by showing that $(m_{2n})^{1/n}\rightarrow\infty$. 
Again denoting the asymptotic moments of Vandermonde matrices with uniform phase distribution by $V_n$, we have that $m_{2n}\leq V_{2n}$, 
since we sum over a greater class of partitions than in (\ref{neweq2}) when computing the Vandermonde moments. 
This means that $(V_{2n})^{1/n}\rightarrow\infty$ also, so that the 
asymptotic mean eigenvalue distribution of the Vandermonde matrices have unbounded support also.

\section{The proof of Theorem~\ref{teo:generaldist}} \label{appendixgeneraldist}
We will use the fact that
\begin{equation} \label{uniformexp}
\begin{array}{ll}
    K_{\rho , u , N} =& \frac{1}{(2\pi)^{|\rho |}N^{n+1-|\rho |}} \times \\
                      & \int_{(0,2\pi)^{|\rho |}} \prod_{k=1}^n \frac{1-e^{j N (x_{b(k-1)} - x_{b(k)})}}{1-e^{j (x_{b(k-1)} - x_{b(k)})}} \\
                      & dx_1\cdots dx_{|\rho |},
\end{array}
\end{equation}
where integration is w.r.t. Lebesgue measure.

For $\rho = 1_n$ Theorem~\ref{teo:generaldist} is trivial. We will thus assume that $\rho\neq 1_n$ in the following. 
We first prove that $\lim_{N\rightarrow\infty} K_{\rho , \omega , N}$ exists whenever $p_{\omega}$ is continuous. 
To simplify notation, define
\begin{eqnarray*}
  F(\omega) &=& \prod_{k=1}^n \frac{1-e^{j N (\omega_{b(k-1)} -\omega_{b(k)})}}{1-e^{j (\omega_{b(k-1)} -\omega_{b(k)})}} \\
            &=& \prod_{k=1}^n \frac{\sin\left( N (\omega_{b(k-1)} -\omega_{b(k)}) / 2\right)}{\sin\left( (\omega_{b(k-1)} -\omega_{b(k)}) / 2 \right)},
\end{eqnarray*}
and set $\omega = (\omega_1,...,\omega_{|\rho |})$ and $d\omega = d\omega_1\cdots d\omega_{|\rho |}$. 
Since $\omega$ is continuous, there exists a $p_{max}$ such that $p_{\omega}(\omega_i)\leq p_{max}$ for all $\omega_i$. Then we have that 
\[
\begin{array}{ll}
  |K_{\rho , \omega , N}| & \leq \frac{p_{max}^{|\rho |}}{N^{m+1-|\rho |}} \\
                          & \times \int_{[0,2\pi)^{|\rho |}} 
                                  \prod_{k=1}^n \left| \frac{\sin\left( N (x_{b(k-1)} - x_{b(k)}) / 2\right)}{\sin\left( (x_{b(k)} - x_{b(k+1)}) / 2 \right)} \right| dx ,
\end{array}
\]
where we have converted to Lebesgue measure, and where we have also written $dx=dx_1\cdots dx_{|\rho |}$. Consider first the set 
\[
  U = \{ \omega | |x_{b(k-1)} - x_{b(k)}| \leq \pi \forall k \} .
\]
When $\frac{2\pi}{N} \leq |\omega_{b(k-1)} - \omega_{b(k)}| \leq \pi$, it is clear that 
\begin{equation} \label{rightside}
  \left| \frac{\sin\left( N (x_{b(k-1)} - x_{b(k)}) / 2\right)}{\sin\left( (x_{b(k-1)} - x_{b(k)}) / 2 \right)} \right| 
  \leq 
  \left| \frac{4}{x_{b(k-1)} - x_{b(k)}} \right| ,
\end{equation}
since $\left| \sin\left( N (x_{b(k-1)} - x_{b(k)}) / 2\right) \right| \leq 1$, and since $|\sin(x)| \geq |\frac{x}{2}|$ when $|x| \leq \frac{\pi}{2}$. 
When $|x_{b(k-1)} - x_{b(k)}| \leq \frac{2\pi}{N}$ we have that
\begin{equation} \label{rightside2}
  \left| \frac{\sin\left( N (x_{b(k-1)} - x_{b(k)}) / 2\right)}{\sin\left( (x_{b(k-1)} - x_{b(k)}) / 2 \right)} \right| \leq N.
\end{equation}
Let $k_1,...,k_{|\rho |}\in\Z$, and assume that $k_{|\rho |}=0$. 
By using the triangle inequality, it is clear that on the set 
\[
  D_{k_1,...,k_{|\rho |-1}}
  = 
  \{ \omega | \left| x_i - \frac{2k_i\pi}{N} \right| \leq \frac{\pi}{N} \forall 1\leq i\leq |\rho | \} , 
\]
when $|k_r-k_s|\geq 2$ for all $r,s$, the $i$'th factor in $F(x)$ is bounded by $\frac{4N}{\left( |k_{b(r-1)} - k_{b(r)} | - 1 \right) \pi}$ due to (\ref{rightside}). 
Also, when $|k_r-k_s| < 2$ for some $r,s$, the corresponding factors in $F(x)$ are bounded by $N$ on $D_{k_1,...,k_{|\rho |}}$ due to (\ref{rightside2}).
Note also that the volume of $D_{k_1,...,k_{|\rho |-1}}$ is $(2\pi)^{|\rho |-1}N^{1-|\rho |}$. 
By adding some more terms (to compensate for the different behaviour for $|k_r-k_s|\geq 2$ and $|k_r-k_s| < 2$), 
we have that we can find a constant $D$ that 
\begin{equation} \label{expr}
\begin{array}{ll}
  \frac{1}{N^{n+1-|\rho |}} \int_U |F(x)| dx \\
  \leq \frac{1}{N^{n+1-|\rho |}} N^n \\
  \times \sum_{ \stackrel{0\leq k_1,...,k{|\rho |-1} < N}{\mbox{all }k_i\mbox{ different}} } \left( \prod_{r=1}^n \frac{D}{ |k_{b(r-1)} - k_{b(r)}|} \right) 2\pi (2\pi)^{|\rho |-1} N^{1-|\rho |} \\
  = (2\pi)^{|\rho |} D^n \sum_{ \stackrel{0\leq k_1,...,k{|\rho |-1} < N}{\mbox{all }k_i\mbox{ different}} }\prod_{r=1}^n \frac{1}{ |k_{b(r-1)} - k_{b(r)}| } ,  
\end{array}
\end{equation}
where we have integrated w.r.t. $x_{|\rho |}$ also (i.e. $k_{|\rho |}$ is kept constant in (\ref{expr})). 
A similar analysis as for $U$ applies for the complement set 
\[
  V = \{ \omega | \pi \leq |x_{b(k-1)} - x_{b(k)}| \leq 2\pi \mbox{ for some } k \} ,
\]
so that we can find a constant $C$ such that
\begin{equation} \label{expr0}
\begin{array}{l}
  \frac{1}{N^{n+1-|\rho |}} \int_{[0,2\pi)^{|\rho |}} |F(x)| dx \\
  \leq C \sum_{ \stackrel{0\leq k_1,...,k{|\rho |-1} < N}{\mbox{all }k_i\mbox{ different}} } \prod_{r=1}^n \frac{1}{|k_{b(r-1)} - k_{b(r)}|} ,
\end{array}
\end{equation}
It is clear this sum converges: 
First of all, this is only needed to prove for $\rho = 0_n$, since the summands for $\rho\neq 0_n$ is only a subset of the summands for $\rho = 0_n$. 

Secondly, for $\rho = 0_n$, (\ref{expr0}) can be bounded by considering convolutions of the following function with itself:
\begin{equation} \label{fdef}
  f(x) = \left\{ \begin{array}{ll} \frac{1}{|x|} & \mbox{ for } |x| > 1 \\ 0 & \mbox{ for } |x| \leq 1 \end{array} \right.
\end{equation}
The assumption that $f(x) = 0$ in a neighbourhood of zero is due to the fact that the $k_i$ are all different. 
Note that $|f(x)| \leq \frac{1}{|x|^{1-\epsilon}}$ for any $0< \epsilon < 1$. Also, the $n-2$-fold convolution 
(we wait with the $n-1$'th convolution till the end) of $\frac{1}{|x|^{1-\epsilon}}$ with itself exist outside $0$ 
whenever $0 < (n-2)\epsilon < 1$, and is on the form $r\frac{1}{|x|^{1-(n-2)\epsilon}}$ for some constant $r$~\cite{rottmann}. 
Therefore, (\ref{expr0}) is bounded by
\begin{eqnarray*}
  \int_{|x|>1} r \frac{1}{|x|^{1-(n-2)\epsilon}} \frac{1}{|x|} dx &=& \int_{|x|>1} r \frac{1}{|x|^{2-(n-2)\epsilon}} dx \\
                                                                  &=& \frac{2r}{(n-2)\epsilon - 1}.
\end{eqnarray*}
This proves that the entire sum (\ref{expr0}) is bounded,  
and thus also the statement on the existence of the limit $K(\rho, \omega)$ in Theorem~\ref{teo:generaldist} when the density is continuous. 

For the rest of the proof of Theorem~\ref{teo:generaldist} , we first record the following result:

\begin{lemma} \label{appendixlemma}
For any $\epsilon > 0$,
\begin{equation}
  \lim_{N\rightarrow\infty} \frac{1}{N^{n+1-|\rho |}} \int_{B_{\epsilon , r}} F(\omega) d\omega = 0,
\end{equation}
where
\[
  B_{\epsilon , r} = \{ (\omega_1,...,\omega_{|\rho |}) | | \omega_{b(r-1)} -\omega_{b(r)} | > \epsilon \} .
\]
\end{lemma}

{\bf Proof:} 
The set $B_{\epsilon , r}$ corresponds to those $k_1,...,k_{|\rho |}$ in (\ref{expr0}) for which $|k_{b(r-1)} - k_{b(r)}| > \frac{N}{2\pi}\epsilon$. 
Thus, for large $N$, we sum over $k_1,...,k_{|\rho |}$ in (\ref{expr0}) for which $|k_{b(r-1)} - k_{b(r)}|$ is arbitrarily large. 
By the convergence of the Fourier integral of $\frac{1}{|x|}$, it is clear that this converges to zero. 
\sluttmerke

Define 
\[
  B_{\epsilon} = \{ (\omega_1,...,\omega_{|\rho |}) | | \omega_i -\omega_j | > \epsilon \mbox{ for some } i,j \} .
\]
If $\omega\in B_{\epsilon}$, there must exist an $r$ so that $| \omega_{b(r-1)} -\omega_{b(r)} | > \frac{2\epsilon}{n}$, so that $\omega\in B_{r, 2\epsilon /n}$.
This means that 
\[
  B_{\epsilon} \subset \cup_r B_{r, 2\epsilon /n},
\]
so that by Lemma~\ref{appendixlemma} also
\[
  \lim_{N\rightarrow\infty} \frac{1}{N^{n+1-|\rho |}} \int_{B_{\epsilon}} F(\omega) d\omega = 0.
\]
This means that in the integral for $K_{\rho , \omega , N}$, we need only integrate over the $\omega$ which are arbitrarily close to the diagonal, 
(where $\omega_1 = \cdots = \omega_{|\rho |}$). We thus have
\[
\begin{array}{l}
  K_{\rho , \omega} = \lim_{N\rightarrow\infty} \frac{1}{N^{n+1-|\rho |}} \int_{[0,2\pi)^{|\rho |}} F(x) \prod_{r=1}^{|\rho |} p_{\omega}(x_r) dx \\
                    = \lim_{N\rightarrow\infty} \frac{1}{N^{n+1-|\rho |}} \int_{[0,2\pi)^{|\rho |}} F(x) p_{\omega}(x_{|\rho |})^{|\rho |} dx \\
                    = \lim_{N\rightarrow\infty} \frac{1}{N^{n+1-|\rho |}} \int_0^{2\pi} p_{\omega}(x_{|\rho |})^{|\rho |} \\
                      \hspace{1cm} \left( \int_{[0,2\pi)^{|\rho |-1}} F(x) dx_1\cdots dx_{|\rho |-1} \right) \\
                      \hspace{1cm} dx_{|\rho |}.
\end{array}
\]
We used here that the density is continuous. Using that 
\begin{equation} \label{haveusefor}
\begin{array}{l}
  \lim_{N\rightarrow\infty} \frac{1}{N^{n+1-|\rho |}} \int_{[0,2\pi )^{|\rho |-1}} F(x) dx_1\cdots dx_{|\rho |-1} \\
  = (2\pi)^{|\rho |-1} K_{\rho , u}
\end{array}
\end{equation}
when $x_{|\rho |}$ is kept fixed at an arbitrary value (this is straightforward by using the methods from the proof of Proposition~\ref{teo1} and (\ref{uniformexp})), 
we get that the above equals
\begin{eqnarray*}
  K_{\rho , u} (2\pi)^{|\rho |-1} \int_0^{2\pi} p_{\omega}(x_{|\rho |})^{|\rho |} dx_{|\rho |},
\end{eqnarray*}
which is what we had to show.

\section{The proof of Proposition~\ref{teo:uniformmin}} \label{appendixuniformmin}
Proposition~\ref{teo:uniformmin} will follow directly if we can prove the following result:

\begin{lemma}
Let $\omega_k$ ($1\leq k\leq n$) be the uniform distribution on $[\frac{2\pi (k-1)}{n},\frac{2\pi k}{n}]$
and define $\omega_{\lambda_1,...,\lambda_n}$ ($0\leq\lambda_i\leq 1,\lambda_1+\cdots +\lambda_n=1$) as the phase distribution with density 
$p_{\omega_{\lambda_1,...,\lambda_n}}=\lambda_1 p_{\omega_1}+\cdots +\lambda_n p_{\omega_n}$. Then
\[
  K_{\rho,\omega_{\frac{1}{n},...,\frac{1}{n}}} \leq K_{\rho,\omega_{\lambda_1,...,\lambda_n}}.
\]
\end{lemma}

{\bf Proof:} 
This follows immediately by noting that
\begin{eqnarray*}
  & &     K_{\rho,\omega_{\lambda_1,...,\lambda_n}} \\
  &=&     K_{\rho , u} (2\pi)^{|\rho |-1} \left( \int_0^{2\pi} p_{\omega_{\lambda_1,...,\lambda_n}}(x)^{|\rho |} dx \right) \\
  &=&     K_{\rho , u} (2\pi)^{|\rho |-1} \\
  & &     \times \int_0^{2\pi} (\lambda_1 p_{\omega_1}(x) + \cdots + \lambda_n p_{\omega_n}(x))^{|\rho |} dx \\
  &=&     K_{\rho , u} (2\pi)^{|\rho |-1} \times \\
  & &     ( (\lambda_1)^{|\rho |} \int_0^{2\pi} p_{\omega_1}(x)^{|\rho |} dx + \cdots \\
  & &      +(\lambda_n)^{|\rho |} \int_0^{2\pi} p_{\omega_n}(x)^{|\rho |} dx ) \\
  &=&     K_{\rho , u} (2\pi)^{|\rho |-1} \times\\
  & &     ( (\lambda_1)^{|\rho |} \int_0^{2\pi} p_{\omega_1}(x)^{|\rho |} dx + \cdots \\
  & &      +(\lambda_n)^{|\rho |} \int_0^{2\pi} p_{\omega_1}(x)^{|\rho |} dx ) \\
  &=&     K_{\rho , u} (2\pi)^{|\rho |-1} \left( (\lambda_1)^{|\rho |} + \cdots + (\lambda_n)^{|\rho |} \right) \\
  & &     \times \int_0^{2\pi} p_{\omega_1}(x) dx \\
  &\geq & K_{\rho , u} (2\pi)^{|\rho |-1} \left( \left(\frac{1}{n}\right)^{|\rho |} + \cdots + \left(\frac{1}{n}\right)^{|\rho |} \right) \\
  & &     \times \int_0^{2\pi} p_{\omega_1}(x)^{|\rho |} dx \\
  &=&     K_{\rho,\omega_{\frac{1}{n},...,\frac{1}{n}}},
\end{eqnarray*}
where we have used that $x_1^{|\rho |}+\cdots x_n^{|\rho |}$ constrained to $x_1+\cdots + x_n = 1$ achieves its minimum for $x_1=\cdots = x_n=\frac{1}{n}$. 
\sluttmerke

\section{The proof of Theorem~\ref{generalinfinitedensity1}} \label{appendixgeneralinfinitedensity1}
The contribution in the integral $K_{\rho , \omega , N}$ comes only from when the  $\omega_i$ coincide with the atoms of $p$. 
Actually, we evaluate $\frac{1-e^{jN\omega}}{1-e^{j\omega}}$ in points on the form $\omega = \alpha_i -\alpha_j$. 
This evaluates to $N^n p_i^n$ when all $\omega_i$ are chosen equal to the same atom $\alpha_j$. 
Since $\lim_{N\rightarrow\infty} \frac{1-e^{jN\omega}}{N\left( 1-e^{j\omega} \right)} = 0$ for any fixed $\omega\neq 0$, 
$\lim_{N\rightarrow\infty} K_{\rho , \omega , N} N^{-n} = 0$ when $\omega$ is chosen from nonequal atoms. 
(\ref{limitmoment4}) (with additional $1/N$-factors) thus becomes
\begin{equation} \label{ashere}
\begin{array}{l}
  \sum_{\rho\in{\cal P}(n)}                                   \\
  \sum_{\stackrel{(j_1,...,j_n)}{\mbox{giving rise to }\rho}} \\
  \sum_{(i_1,...,i_n)}                                        \\
  \hspace{1cm} N^{|\rho |-2n-1} c^{|\rho |-1} L^{-|\rho |} \\
  \hspace{1cm} \left( \sum_i N^n p_i^n + a_{\rho , N} N^n) \right) \\
  \hspace{1cm} {\bf D}_1(N)(j_1,j_1) {\bf D}_2(N)(j_2,j_2) \\
  \hspace{1cm} \cdots \times {\bf D}_n(N)(j_n,j_n),
\end{array}
\end{equation}
where $\lim_{N\rightarrow\infty} a_{\rho , N} = 0$. 
Multiplying both sides with $N$ and letting $N$ go to infinity gives
\[
  \lim_{N\rightarrow\infty} \sum_{\rho\in{\cal P}(n)} 
    N^{|\rho |-n} c^{|\rho |-1} 
    \left( \sum_i p_i^n + a_{\rho , N} \right)
    D_{\rho}.
\]
It is clear that this converges to $0$ when $\rho\neq 0_n$ (since $|\rho | < n$ in this case), 
so that the limit is 
\[
  c^{n-1} \left( \sum_i p_i^n \right) \alpha_{0_n} = c^{n-1} p^{(n)} \lim_{N\rightarrow\infty} \prod_{i=1}^n tr_L\left( {\bf D}_i(N) \right),
\]
which proves the claim

\section{The proof of Theorem~\ref{generalinfinitedensity2}} \label{appendixgeneralinfinitedensity2}
We need the following identity~\cite{rottmann}:
\[
  \int_0^{\infty} x^{-s} e^{jnx} dx = \frac{\Gamma(1-s)}{|n|^{1-s}} e^{\frac{j sgn(n) (1-s)\pi}{2}},
\]
where $sgn(x) = 1$ if $x > 0$, $sgn(x) = -1$ if $x < 0$, and $0$ otherwise.
From this it follows that
\begin{equation} \label{rewriteexp}
\begin{array}{l}
  \int_{-\infty}^{\infty} p_i |x-\alpha_i|^{-s} e^{jnx} dx = \\
  2p_i e^{jn\alpha_i} \frac{\Gamma(1-s)}{|n|^{1-s}} \cos\left( \frac{(1-s)\pi}{2} \right).
\end{array}
\end{equation}
Note that the measure with density $p$, has the same asymptotics near $\alpha_i$ as the measure with density $p_i |x-\alpha_i|^{-s}$ 
on 
\[ \left( -\left( \frac{1-s}{2p_i} \right)^{\frac{1}{1-s}}, \left( \frac{1-s}{2p_i} \right)^{\frac{1}{1-s}} \right) .\] 
As in the proof in Appendix~\ref{appendixgeneralinfinitedensity1}, 
the integral for the expansion coefficients is dominated by the behaviour near the points $(\alpha_i,...,\alpha_i)$.
To see this, note that 
the behaviour near the singular points on the diagonal is $O \left( s(|\rho | - n) - 1 \right)$ when polynomic growth of order $s$ 
of the density near the singular points is assumed. This is very much related to (\ref{expr0}) in Appendix~\ref{appendixgeneraldist}, 
since $K_{\rho , \omega}$ here in a similar way can be bounded by (taking into account new powers of $N$)
\begin{equation} \label{expr2}
\begin{array}{l}
  C \frac{1}{N^{n+ns+1-|\rho |}} N^n N^{-|\rho |} N^{|\rho | s} \\
  \times \sum_{ \stackrel{0\leq k_1,...,k{|\rho |} < N}{\mbox{all }k_i\mbox{ different}} } \prod_{r=1}^n \frac{1}{|k_{b(r-1)} - k_{b(r)}|} \prod_{t=1}^{|\rho |} k_t^{-s}.
\end{array}
\end{equation}
In (\ref{expr2}), the $N^n$-factor appears in exactly the same way as in the proof of Theorem~\ref{teo:generaldist} in Appendix~\ref{appendixgeneraldist}, 
$N^{-|\rho |}$ appears as a volume in $\R^{|\rho |}$, 
and $N^{|\rho | s}$ comes from evaluation of the density in the points $x_i = \frac{2k_i\pi}{N}$, $1\leq i\leq |\rho |$). 
Since $\frac{1}{|x|^s}$ has a bounded integral around $0$, and since the sum still converges (it is dominated by (\ref{expr0})), (\ref{expr2}) is 
\[
  O \left( s(|\rho | - n) - 1\right).
\] 
This has it's highest order when $|\rho | = n$, so that we can restrict to looking at $0_n$. 
Note also that we may just as well assume that $p_{\omega}(x)$ is identical to $p_i|x-\omega_i|^{-s}$ at an interval around $\omega_i$, since 
$\lim_{x\rightarrow\alpha_i} |x-\alpha_i|^s p_{\omega}(x) = p_i$ implies that 
\begin{equation} \label{extra}
  p_{\omega}(x) = p_i|x-\omega_i|^{-s} + k(x)|x-\omega_i|^{-s}
\end{equation}
where $\lim_{x\rightarrow\omega_i} k(x) = 0$. It is straightforward to see that the contribution of the second part in (\ref{extra}) to (\ref{expr2}) 
vanishes as $N\rightarrow\infty$, so that we may just as well assume that $p_{\omega}(x)$ is identical to $p_i|x-\omega_i|^{-s}$ at an interval around $\omega_i$, 
as claimed. Also, since 
\[
  \lim_{n\rightarrow\infty} \int_{|x|>\epsilon} x^{-s} e^{jnx} dx = 0
\]
for all $\epsilon > 0$, 
and since the contributions from large $n$ dominate in (\ref{newlimitmoment4}) below (since $\sum_n |n|^{-s}$ diverges), 
it is clear that we can restrict to an interval around $\omega_i$ when computing the limit also 
(since $p_{\omega}$ is continuous outside the singularity points, this follows from Theorem~\ref{teo:generaldist}, and due to the additional $\frac{1}{N^s}$-factor 
added to (\ref{vandermonde})). 
After restricting to $0_n$, multiplying both sides with $N$, summing over all singularity points, and using (\ref{rewriteexp}), we obtain the approximation 
\begin{eqnarray}
  & & \sum_{(i_1,...,i_n)} \sum_{a} \nonumber \\
  & & \hspace{1cm} N^{-ns} c^{n-1} \nonumber \\
  & & \hspace{1cm} \times \left( 2 p_a \Gamma(1-s) \cos\left( \frac{(1-s)\pi}{2} \right) \right)^n \nonumber \\
  & & \hspace{1cm} \times \prod_{k=1}^n \frac{ e^{j(i_{k-1} - i_k)\alpha_a} }{\left| i_{k-1} - i_k \right|^{1-s}} \nonumber \\
  & & \hspace{1cm} \times tr_L ({\bf D}_1(N)) \times\cdots\times tr_L ({\bf D}_n(N)) \label{newlimitmoment4}
\end{eqnarray}
to (\ref{limitmoment4}). Since $\prod_{k=1}^{n} e^{j(i_{k-1} - i_k)\alpha_a} = 1$, we recognize
\[
\begin{array}{lll}
  q^{(n,N)} &=& \left( 2 \Gamma(1-s) \cos\left( \frac{(1-s)\pi}{2} \right) \right)^n \left( \sum_a p_{a}^n \right) \times \\                                                   
          & & \sum_{(i_1,...,i_n)} N^{-ns} \prod_{k=1}^n \frac{1}{\left| i_{k-1} - i_k \right|^{1-s}},
\end{array}
\]
as a factor in (\ref{newlimitmoment4}) such that the limit of (\ref{newlimitmoment4}) as $N\rightarrow\infty$ can be written
\[
  c^{n-1} \lim_{N\rightarrow\infty} q^{(n,N)} \lim_{N\rightarrow\infty} \prod_{i=1}^n tr_L\left( {\bf D}_i(N) \right).
\]
It therefore suffices to prove that $\lim_{N\rightarrow\infty} q^{(n,N)} = q^{(n)}$. To see this, write 
\begin{eqnarray*}
  \frac{N^{-s}}{ \left| i_{k-1} - i_k \right|^{1-s}} &=& \frac{1}{N} \frac{1}{ \left( \frac{1}{N} \right)^{1-s} \left| i_{k-1} - i_k \right|^{1-s} } \\
                                                       &=& \frac{1}{N} \frac{1}{ \left| \frac{i_{k-1}}{N} - \frac{i_k}{N} \right|^{1-s} }.
\end{eqnarray*}
Summing over all $1\leq i_1,...,i_n\leq N$, it is clear from this that $q^{(n,N)}$ can be viewed as a Riemann sum which converges to $q^{(n)}$ as $N\rightarrow\infty$.

\section{The proof of Theorem~\ref{teo2} and Corollary~\ref{kornew}} \label{appendix:manymatrices}
{\bf Proof of Theorem~\ref{teo2}:} we define $S_j$ to be the blocks of $\sigma$, i.e.
\[
  S_j = \{ k | i_k = j\}.
\]
Note that Theorem~\ref{teo:generaldist} guarantees that the limit $K_{\rho , \omega} = \lim_{N\rightarrow\infty} K_{\rho , \omega , N}$ exists.
The partition $\rho$ simply is a grouping of random variables into independent groups.
It is therefore impossible for a block in $\rho$ to contain elements
from both $S_1$ and $S_2$, so that any block is contained in either $S_1$ or $S_2$.
As a consequence, $\rho\leq\sigma$.
\sluttmerke

Until now,
we have not treated mixed moments of the form
\[ {\bf D}_1(N) {\bf V}_{i_2} {\bf V}_{i_2}^H {\bf D}_2(N) {\bf V}_{i_3} {\bf V}_{i_3}^H \cdots \times {\bf D}_n(N) {\bf V}_{i_1} {\bf V}_{i_1}^H , \]
which are the same as the mixed moments of Theorem~\ref{teo2} except
for the position of the ${\bf D}_i(N)$. We will not go into depths
on this, but only remark that this case can be treated in the same
vein as generalized Vandermonde matrices by replacing the density
$p_f$ (or $p_{\lambda}$ in case of continuous generalized Vandermonde
matrices) with functions $p_{D_i}(x)$ defined by $p_{D_i}(x) = {\bf
D}_i(N)(\lfloor Lx \rfloor,\lfloor Lx \rfloor)$ for $0\leq x\leq 1$. This
also covers the case of mixed moments of independent, generalized
Vandermonde matrices (and, in fact, there are no restrictions on the
horizontal and vertical phase densities $p_{\omega_i}$ and
$p_{\lambda_j}$ for each matrix. They may all be different). The
proof for this is straightforward.

{\bf Proof of Corollary~\ref{kornew}:} 
this follows in the same way as Proposition~\ref{teo:first7moments} is proved from Proposition~\ref{lemma:kcompute}, by only considering $\rho$ which are less than $\sigma$,
and also by using Theorem~\ref{teo:generaldist}.
$\sigma$ are for the listed moments $\{ \{ 1 \} , \{ 2 \} \}$, $\{ \{ 1,3 \} , \{ 2,4 \} \}$, and $\{ \{ 1,3,5 \} , \{ 2,4,6 \} \}$, respectively.
\sluttmerke

\section{The proofs of Proposition~\ref{propasymptotic} and~\ref{propexact}} \label{mixedvandgauss}
The moments $E\left[ tr_n \left( {\bf W}^i \right) \right]$ will be related to the moments $P_i$ through three convolution stages:
\begin{enumerate}
  \item relating the moments of ${\bf W}$ with the moments of
    \begin{equation} \label{stage1}
      {\bf \Gamma} = {\bf V} {\bf P}^{\frac{1}{2}} \left( \frac{1}{K} {\bf S} {\bf S}^H \right) {\bf P}^{\frac{1}{2}} {\bf V}^H,
    \end{equation}
    from which we easily get the moments of
    \begin{equation} \label{stage1b}
      {\bf \tilde{S}} = \left( \frac{1}{K} {\bf S} {\bf S}^H \right) {\bf P}^{\frac{1}{2}} {\bf V}^H {\bf V} {\bf P}^{\frac{1}{2}},
    \end{equation}
  \item relating the moments of ${\bf S}$ with the moments of
    \begin{equation} \label{stage2}
      {\bf T} = {\bf P} {\bf V}^H {\bf V},
    \end{equation}
  \item relating the moments of ${\bf T}$ with the moments of ${\bf P}$.
\end{enumerate}
For the first stage, the moments of ${\bf \hat{W}}$ and ${\bf \Gamma}$ relate through the formulas
\begin{eqnarray}
  E\left[ tr_n \left( {\bf W}   \right) \right] &=& E\left[ tr_N\left({\bf \Gamma}\right) \right] + \sigma^2 \label{apply11}\\
  E\left[ tr_n \left( {\bf W}^2 \right) \right] &=& E\left[ tr_N\left({\bf \Gamma}^2\right) \right] \nonumber \\
                                                & & + 2{\sigma}^2 (1+c_1) E\left[ tr_N\left({\bf \Gamma}\right) \right] \nonumber \\
                                                & & + {\sigma}^4 (1 + c_1)\\
  E\left[ tr_n \left( {\bf W}^3 \right) \right] &=& E\left[ tr_N\left({\bf \Gamma}^3\right) \right] \nonumber \\
                                                & & + 3{\sigma}^2(1+c_1) E\left[ tr_N\left({\bf \Gamma}^2\right) \right] \nonumber \\
                                                & & + 3{\sigma}^2 c_1 E\left[ \left(tr_N\left({\bf \Gamma}\right)\right)^2 \right] \nonumber\\
                                                & & + 3{\sigma}^4 \left( c_1^2+3c_1+1+\frac{1}{K^2}\right) E\left[ tr_N\left({\bf \Gamma}\right) \right] \nonumber \\
                                                & & + {\sigma}^6 \left( c_1^2+3c_1+1+\frac{1}{K^2}\right), \label{apply13},
\end{eqnarray}
which are obtained by replacing ${\bf R}$ in~\cite{eurecom:channelcapacity} by ${\bf V}{\bf P}^{\frac{1}{2}}{\bf S}$,
with $c=c_1=\frac{N}{K}$. For the second part of the first stage, note that
\begin{eqnarray}
  E\left[ tr_N\left({\bf \Gamma}^k\right) \right]              &=& c_2   E\left[ tr_L\left({\bf \tilde{S}}^k\right) \right] \label{transform1} \\
  E\left[ \left(tr_N\left({\bf \Gamma}\right)\right)^k \right] &=& c_2^k E\left[ \left(tr_L\left({\bf \tilde{S}}\right)\right)^k \right] , \label{transform2}
\end{eqnarray}
where $c_2=\frac{L}{N}$.
We can now apply Theorem~\ref{teo:exact4moments} to obtain
\begin{eqnarray}
  c_3 E\left[ tr_L\left({\bf \tilde{S}}\right) \right]            &=& c_3 E\left[ tr_L\left({\bf T}\right) \right] \label{apply21} \\
  c_3 E\left[ tr_L\left({\bf \tilde{S}}^2\right) \right]          &=& c_3 E\left[ tr_L\left({\bf T}^2\right) \right] \nonumber \\
                                                          & & + c_3^2 E\left[ \left(tr_L\left({\bf T}\right)\right)^2 \right] \\
  c_3 E\left[ tr_L\left({\bf \tilde{S}}^3\right) \right]          &=& \left( 1 + K^{-2} \right) c_3 E\left[ tr_L\left({\bf T}^3\right) \right] \nonumber \\
                                                          & & + 3 c_3^2 E\left[ \left(tr_L{\bf T}\right) tr_L\left({\bf T}^2\right) \right] \nonumber \\
                                                          & & + c_3^3 E\left[ \left(tr_L\left({\bf T}\right)\right)^3 \right] \\
  E\left[ \left(tr_L\left({\bf \tilde{S}}\right)\right)^2 \right] &=& E\left[ \left(tr_L\left({\bf T}\right)\right)^2 \right] \nonumber \\
                                                          & & + \frac{1}{KL} E\left[ tr_L\left({\bf T}^2\right) \right] , \label{apply24}
\end{eqnarray}
where $c_3=\frac{L}{K}$, and ${\bf T} = {\bf P} {\bf V}^H {\bf V}$.
(\ref{apply11})-(\ref{apply13}), (\ref{transform1})-(\ref{transform2}), and (\ref{apply21})-(\ref{apply24}) can be
combined to
\begin{eqnarray}
  E\left[ tr_n \left( {\bf W}   \right) \right] &=& c_2 E\left[ tr_L\left({\bf T}\right) \right] + \sigma^2 \label{combined11} \\
  E\left[ tr_n \left( {\bf W}^2 \right) \right] &=& c_2 E\left[ tr_L\left({\bf T}^2\right) \right] + c_2c_3 E\left[ \left(tr_L\left({\bf T}\right)\right)^2 \right] \nonumber \\
                                                & & + 2 \sigma^2 (c_2+c_3) E\left[ tr_L\left({\bf T}\right) \right] + \sigma^4(1+c_1) \\
  E\left[ tr_n \left( {\bf W}^3 \right) \right] &=& c_2 \left( 1 + \frac{1}{K^2} \right) E\left[ tr_L\left({\bf T}^3\right) \right] \nonumber \\
                                                & & + 3c_2c_3 E\left[ \left(tr_L\left({\bf T}\right)\right) \left(tr_L\left({\bf T}^2\right)\right) \right] \nonumber \\
                                                & & + c_2c_3^2 E\left[ \left(tr_L\left({\bf T}\right)\right)^3 \right] \nonumber \\
                                                & & + 3\sigma^2\left( (1+c_1)c_2 + \frac{c_1c_2^2}{KL} \right) E\left[ tr_L\left({\bf T}^2\right) \right] \nonumber \\
                                                & & + 3\sigma^2 c_3(c_3+2c_2) E\left[ \left(tr_L\left({\bf T}\right)\right)^2 \right] \nonumber \\
                                                & & + 3\sigma^4 \left( c_1^2 +3c_1 + 1 + \frac{1}{K^2} \right) c_2 E\left[ tr_L\left({\bf T}\right) \right] \nonumber\\
                                                & & + \sigma^6\left( c_1^2 + 3c_1+1+\frac{1}{K^2}\right). \label{combined13}
\end{eqnarray}
Up to now, all formulas have provided exact expressions for the expectations.
For the next step, exact expressions for the expectations are only known when the phase distributions are uniform,
in which case the formulas are given by Theorem~\ref{teo:exact4moments}:
\begin{eqnarray}
  c_2E\left[ tr_L\left({\bf T}\right) \right]   &=& c_2tr_L({\bf P}) \label{apply31} \\
  c_2E\left[ tr_L\left({\bf T}^2\right) \right] &=& \left( 1 - N^{-1} \right) c_2tr_L({\bf P}^2) \nonumber \\
                                                & & + c_2^2(tr_L({\bf P}))^2 \\
  c_2E\left[ tr_L\left({\bf T}^3\right) \right] &=& \left( 1  - 3N^{-1}  + 2 N^{-2} \right) c_2tr_L({\bf P}^3) \nonumber \\
                                                & & + 3 \left( 1 - N^{-1} \right) c_2^2 tr_L({\bf P}) tr_L({\bf P}^2) \nonumber \\
                                                & & + c_2^3(tr_L({\bf P}))^3 \\
  E\left[ \left(tr_L\left({\bf T}\right)\right)^2 \right] &=& tr_L({\bf P})^2 \\
  E\left[ \left(tr_L\left({\bf T}\right)\right)^3 \right] &=& tr_L({\bf P})^3
\end{eqnarray}
\begin{equation} \label{apply36}
\begin{array}{l}
  E\left[ \left(tr_L\left({\bf T}\right)\right) \left(tr_L\left({\bf T}^2\right)\right) \right] \\
  = \left( 1 - N^{-1} \right) tr_L(P)tr_L({\bf P}^2) + c_2(tr_L({\bf P}))^3.
\end{array}
\end{equation}
If the phase distribution $\omega$ is not uniform, Theorem~\ref{teo0} and Theorem~\ref{teo:generaldist} gives the following approximation:
\begin{eqnarray}
  c_2E\left[ tr_L\left({\bf T}\right) \right]   &=& c_2 tr_L({\bf P}) \label{apply351} \\
  c_2E\left[ tr_L\left({\bf T}^2\right) \right] &\approx& c_2 tr_L({\bf P}^2) + c_2^2I_2(tr_L({\bf P}))^2 \\
  c_2E\left[ tr_L\left({\bf T}^3\right) \right] &\approx& c_2 tr_L({\bf P}^3) + 3 c_2^2 I_2 tr_L({\bf P}) tr_L({\bf P}^2) \nonumber \\
                                                & & + c_2^3 I_3 (tr_L({\bf P}))^3 \\
  E\left[ \left(tr_L\left({\bf T}\right)\right)^2 \right] &=& (tr_L {\bf P})^2 \\
  E\left[ \left(tr_L\left({\bf T}\right)\right)^3 \right] &=& (tr_L {\bf P})^3
\end{eqnarray}
\begin{equation} \label{apply356}
\begin{array}{l}
  E\left[ \left(tr_L\left({\bf T}\right)\right) \left(tr_L\left({\bf T}^2\right)\right) \right] \\
  \approx tr_L({\bf P})tr_L({\bf P}^2) + c_2I_2(tr_L({\bf P}))^3,
\end{array}
\end{equation}
where the approximation is $O(N^{-1})$,
and where $I_k$ is defined by (\ref{needthis}).

Proposition~\ref{propexact} is proved by combining (\ref{combined11})-(\ref{combined13}) with (\ref{apply31})-(\ref{apply36}), while
Proposition~\ref{propasymptotic} is proved by combining (\ref{combined11})-(\ref{combined13}) with (\ref{apply351})-(\ref{apply356}).
Proposition~\ref{propsamplingdist} is proved by first observing that the roles of $L$ and $N$ are interchanged,
since the Vandermonde matrix is replaced by its transpose.
This means that we obtain the formulas (\ref{combined11})-(\ref{combined13}), with $c_1$ and $c_3$ interchanged, and $c_2$ replaced with $\frac{1}{c_2}$.
The matrix ${\bf T}$ is now instead ${\bf V}{\bf V}^H$, and these can be scaled to obtain the moments of ${\bf V}^H{\bf V}$.
Finally the integrals $I_n$ or the angle $\alpha$ can be estimated from these moments, using (\ref{apply351})-(\ref{apply356})
with the moments of ${\bf P}$ replaced with $1$ (since no additional power matrix is included in the model).

Matlab code for implementing the steps (\ref{apply11})-(\ref{apply13}), (\ref{apply21})-(\ref{apply24}),
and (\ref{apply31})-(\ref{apply36}) can be found in~\cite{eurecom:vandermondeimpl}.

\section*{Acknowledgment}
The authors would like to thank the anonymous reviewers for their insightful and valuable comments, which have helped improve the quality of the paper. 
They would also like to thank the Associate Editor Prof. G. Taricco for a very professional processing of the manuscript.



\noindent
{\bf \O yvind Ryan} was born in Oslo, Norway. 
He studied mathematics at the University of Oslo, where he received the M.Sc and the Ph.D. degrees in 1993 and 1997, respectively. 

From 1997 to 2004, he worked as a consultant and product developer in various information technology 
projects. From 2004 to 2007, he was a postdoctoral fellow at the Institute of Informatics at the University of Oslo.
He is currently employed as a researcher at the Centre of Mathematics for Applications at the University of Oslo.
His research interests are applications of free probability theory and 
random matrices to the fields of wireless communication, finance, and 
information theory.

\noindent
{\bf Merouane Debbah} was born in Madrid, Spain. 
He entered the Ecole Normale Supérieure de Cachan (France) in 1996 where he received the M.Sc and the Ph.D. degrees respectively in 1999 and 2002. 

From 1999 to 2002, he worked for Motorola Labs on Wireless Local Area Networks and prospective fourth generation systems. 
From 2002 until 2003, he was appointed Senior Researcher at the Vienna Research Center for Telecommunications (ftw.), 
Vienna, Austria working on MIMO wireless channel modeling issues. 
From 2003 until 2007, he joined the Mobile Communications department of the Institute Eurecom (Sophia Antipolis, France) 
as an Assistant Professor. He is presently a Professor at Supelec (Gif-sur-Yvette, France), holder of the Alcatel-Lucent Chair on flexible radio. 
His research interests are in information theory, signal processing and wireless communications.

\end{document}